\documentclass[12pt]{article}
\usepackage{amssymb}
\usepackage{amsmath}
\usepackage{youngtab}
\usepackage{cite}
\usepackage[]{hyperref}
\input xy
\xyoption{all}

\numberwithin{equation}{section}

\begin{document}

\begin{center}

{\large\bf A proposal for nonabelian mirrors}

\vspace{0.2in}

Wei Gu, Eric Sharpe

Dep't of Physics\\
Virginia Tech\\
850 West Campus Dr.\\
Blacksburg, VA  24061\\

{\tt weig8@vt.edu},
{\tt ersharpe@vt.edu}

$\,$

\end{center}

In this paper we propose a systematic construction of mirrors  
of nonabelian
two dimensional (2,2) supersymmetric gauge theories.  
Specifically, we propose a construction of B-twisted Landau-Ginzburg
orbifolds whose correlation functions match those of A-twisted
supersymmetric gauge theories, and whose critical loci reproduce
quantum cohomology and Coulomb branch relations in A-twisted gauge
theories, generalizing the Hori-Vafa mirror construction.
We check this proposal in a wide variety of examples.  
For instance, we construct mirrors corresponding to Grassmannians and
two-step flag manifolds, as well as complete intersections therein, and
explicitly check predictions for 
correlation functions and quantum cohomology rings, as well as other
properties.
We also consider mirrors to examples of gauge theories
with $U(k)$, $U(k_1) \times U(k_2)$, $SU(k)$,
$SO(2k)$, $SO(2k+1)$, and $Sp(2k)$ gauge groups and a variety of matter 
representations, and compare to results in the literature for
the original two dimensional gauge theories.
Finally, we perform consistency checks of conjectures of 
Aharony et al that a two dimensional (2,2) supersymmetric pure $SU(k)$
gauge theory flows to a theory of $k-1$ free twisted chiral multiplets, and 
also consider the analogous question in pure $SO(3)$ theories.
For one discrete theta angle, the $SO(3)$
theory behaves the same as the $SU(2)$ theory; for the other, supersymmetry
is broken.  We also perform consistency checks of analogous statements
in pure supersymmetric $SO$ and $Sp$ gauge theories in two dimensions.

\begin{flushleft}
June 2018
\end{flushleft}

\newpage

\tableofcontents

\newpage

\section{Introduction}

Mirror symmetry has had a long and influential history in string theory,
leading for example to some of the original computations of Gromov-Witten
invariants, and has led to several generalizations such as homological
mirror symmetry and (0,2) mirror symmetry.  One of the unsolved problems
of the original mirror symmetry program was to systematically
understand mirrors for theories
constructed as nonabelian (2,2) supersymmetric gauge theories in two
dimensions.  In this paper we propose a
possible construction of such nonabelian mirrors, generalizing the
results of \cite{Hori:2000kt,Morrison:1995yh}.  
(That said, we only claim to have a proposal for
a mirror construction; we do not claim to have a proof of that proposal.)

The word `mirror' is sometimes used to mean a variety of 
things, so let us take a moment to clarify what we will mean in this
paper, beginning with A-twisted gauge theories.  
From \cite{Morrison:1994fr}[section 3,6], in an A-twisted
two-dimensional (2,2) supersymmetric
gauge theory, there are (at least) two classes of BRST-closed
observables:
\begin{itemize}
\item adjoint-valued scalars $\sigma$, the scalars in the (2,2) vector
multiplet, 
\item gauge-equivariantly-closed differential forms.
\end{itemize}
The former are typically most useful on semiclassical Coulomb branches,
the latter on semiclassical Higgs branches.  
That said, in phases that RG flow to nonlinear sigma models on geometries,
the $\sigma$'s are expected to flow to elements of $H^{1,1}$ (or rather
the images of the restriction of elements of $H^{1,1}$ of an ambient
space), and so one expects some sort of connection to equivariantly-closed
differential forms, a relation which is encoded in
the equations of motion \cite{edver}[equ'n (4.26)].

Given an A-twisted (2,2) supersymmetric 
theory, our proposal gives a B-twisted Landau-Ginzburg
orbifold which should have the same correlation functions and
Coulomb branch relations as the A-twisted theory.  The scalars $\sigma$
appear explicitly in our construction, just as they did in \cite{Hori:2000kt}.
In the case of
gauge theories corresponding to geometries such as Grassmannians and
flag manifolds, this means that the mirror will correctly encapsulate the
quantum cohomology relations.
In addition to mirrors of theories with some sort of geometric interpretation,
our construction is also defined for more general two-dimensional
(2,2) supersymmetric gauge theories, which we will illustrate in examples
with a variety of gauge groups and matter content.

We will show explicitly that our proposal satisfies a wide variety
of consistency tests, reproducing quantum cohomology/Coulomb branch
relations, correlation functions, and even the Coulomb branch excluded
loci, all from algebraic computations in the (classical) B model mirror,
in numerous examples.
All that said, there is an important subtlety:
the proposed mirror Landau-Ginzburg theories
typically will have poles in their superpotentials, corresponding to
nonabelian symmetry enhancements in the A-twisted gauge theories.  
(Other two-dimensional (2,2) supersymmetric
theories with superpotentials containing poles have also recently been
discussed in \cite{Berglund:2016yqo,Berglund:2016nvh}, in describing
the generalized CICY constructions of \cite{Anderson:2015iia}.
See also \cite{Ghoshal:1993qt} for an example of a two-dimensional
Landau-Ginzburg model with a
superpotential with a pole, in a different context, and
\cite{Aharony:1997bx}[section 6.1] for a three-dimensional theory
with a superpotential with a pole.)
We will argue that the locations of the poles, which match excluded loci
in A model computations, are dynamically excluded in the B model,
so that they do not impede understanding the essentially classical
physics of our B model topological field
theory computations.
In any event, as our computations are extremely
successful
at reproducing known A-twisted gauge theory results, we leave a more
complete understanding of the corresponding physical untwisted
quantum field theory to future work.

We begin in 
section~\ref{sect:prop} by describing the proposal for nonabelian
mirrors.
Briefly, the basic idea is that
we construct the mirror to the nonabelian gauge theory by taking the
mirror to a corresponding abelian gauge theory, consisting of a Weyl-group
orbifold of a $U(1)^r$ gauge theory where $r$ is the rank of the original
gauge group, with the same matter as the original gauge theory
(decomposed into $U(1)^r$ representations), together with some additional
superfields with the same indices as the
W bosons of the original theory.  In section~\ref{sect:justification},
we describe some of the motivations for this proposal,
we suggest a possible idea for a proof, and finally we
argue that in the B model,
correlation functions of this proposed mirror will match A model computations
in general.

In the rest of the paper, we check the resulting proposal in a variety
of examples.
In section~\ref{sect:grassmannian}, we apply these methods to construct
the mirror of (the GLSM for) a Grassmannian $G(k,n)$.  In gauge theoretic
terms, this is a $U(k)$ gauge theory with $n$ chiral superfields in the
fundamental representation.  We check the number of vacua, predicted
Coulomb
branch relations and quantum cohomology ring, and correlation
functions in the case of $G(2,n)$, and verify that all of these match
those of the original gauge theory.  We also discuss how the mathematical
relation $G(k,n) \cong G(n-k,n)$ is realized as a two-dimensional
Seiberg duality in the mirror, and explicitly compare our mirror to proposed
structures of Eguchi-Hori-Xiong \cite{ehx}, Rietsch \cite{r1,r2,r3}, 
Hori-Vafa \cite{Hori:2000kt}, and
Gomis-Lee \cite{Gomis:2012wy}.
In section~\ref{sect:two-step} we perform analogous checks of mirror
proposed here for a two-step flag manifold $F(k_1,k_2,n)$, corresponding
to a $U(k_1) \times U(k_2)$ gauge theory, comparing the predicted
Coulomb branch
relations (quantum cohomology ring) to that obtained
in the original A-twisted GLSM.

In section~\ref{sect:adj} we perform analogous computations
in a $U(k)$ gauge theory with matter in fundamental representations as
well as a superfield in the adjoint representation.
In section~\ref{sect:symmetric}
we perform analogous computations in a $U(k)$ gauge
theory with matter in fundamental representations as well as an
$m$-symmetric-tensor representation.  In section~\ref{sect:suk-twisted}
we perform analogous computations and comparisons in an $SU(k)$ gauge
theory with matter in the fundamental representation as well as
twisted masses, checking against results for such theories
in \cite{Hori:2006dk}[section 3].  In section~\ref{sect:so2k}
we perform analogous computations and comparisons for the proposed
mirror of an $SO(2k)$ gauge theory with matter in vector representations
and with twisted masses, 
checking against results for such A-twisted gauge theories
in \cite{Hori:2011pd}[section 4].  In section~\ref{sect:so2kp1} we 
perform analogous computations and comparisons in our proposed mirror
to an $SO(2k+1)$ gauge theory with matter in vector representations and
with twisted masses, checking against results for such A-twisted gauge theories
in \cite{Hori:2011pd}[section 4].  The results for this case are very
similar to those of the $SO(2k)$ gauge theories above, but there are
a few new quirks, such as the fact that the last component of each
vector is decoupled, and we see the appropriate corresponding structure
in the mirror.  In section~\ref{sect:sp2k} we perform analogous computations
and comparisons in our proposed mirror to an $Sp(2k) = USp(2k)$ gauge
theory with matter in fundamental representations and with twisted
masses, checking against results for such A-twisted gauge theoeires
in \cite{Hori:2011pd}[section 5].

In section~\ref{sect:pure}, we analyze mirrors to pure $SU(k)$ gauge
theories.  It was proposed in \cite{Aharony:2016jki} that these theories flow to
theories of $k-1$ free twisted chiral superfields, and we recover that result
from the mirror (at least to the extent possible in topological field
theory computations).  We also examine the same behavior in
$SO(3)$ theories, and discover that the result depends upon the value
of the discrete theta angle.  For one discrete theta angle, the result
is the same as for $SU(2)$; for the other, our mirror has no critical
loci, no supersymmetric vacua, which we interpret as supersymmetry breaking
in the A-twisted gauge theory.  The $SU(2)$ and $SO(3)$ results also
interrelate via nonabelian decomposition \cite{Sharpe:2014tca}, 
which says that an $SU(2)$
gauge theory with center-invariant matter should decompose
into a disjoint union of $SO(3)$ theories with the same matter and
varying discrete theta angles, which we discuss.
In section~\ref{sect:puresosp}, we perform the analogous computations
for pure supersymmetric $SO$ and $Sp$ gauge theories in two dimensions.
Our TFT computations are consistent with an analogous conjecture -- 
that for the right discrete theta angle, the theories flow in the IR
to a theory of free twisted chiral superfields, as many as the rank of the
gauge group.  That said, in our TFT computations we do not check,
for example, metrics, so we only state conjectures and do not claim proofs
of physical results for these theories, merely consistency checks of
some aspects.

To this point, we have primarily considered mirrors of theories with
vanishing superpotential.  In section~\ref{sect:hyp}, we compute the
mirror of a theory with a superpotential, a $U(k)$ gauge theory
corresponding to the GLSM for the Calabi-Yau hypersurface $G(2,4)[4]$.
This is primarily
an exercise in manipulating fields with nonzero R symmetries.
We compare to results in {\it e.g.} \cite{Hori:2006dk}.
We also consider the case of mirrors to complete intersection
Calabi-Yau's in $G(k,n)$, and compare to results for correlation
functions in \cite{Closset:2015rna}[section 8].

Regarding mirror geometries,
to be clear, the proposal in this paper yields a B-twisted Landau-Ginzburg model
that captures correlation functions of the original A model, but we do not
know if that same Landau-Ginzburg model also correlates to mirror geometries.
In the examples discussed above, we are not aware of any birational 
transformations from known mirror geometries to some with GLSMs in which
our proposal appears as a phase, possibly with some fields integrated out,
but neither can we exclude the possibility.
We do briefly discuss possible connections to geometry in 
section~\ref{sect:comments-mirror}, 
and in particular describe a central charge computation
there that is consistent with an interpretation of a geometric mirror.
We leave any further 
connection, if one exists, between the Landau-Ginzburg models
here and mirror geometries to future work.

In appendix~\ref{app:rietsch}, we explicitly describe
Eguchi-Hori-Xiong's and Rietsch's proposed mirrors to Grassmannians,
which we compare to the results of our proposal
in section~\ref{sect:compare-mirrors}.
Finally, in appendix~\ref{app:r-charges}, we provide some technical
notes on mirrors to fields with nonzero R charges.

Although much of this paper is spent justifying our proposal by
checking its predictions against known results, we do in addition use
it to make some new predictions, such as, for one example,
for excluded loci and
Coulomb branch (quantum cohomology) relations for gauge theories with
symmetric tensor matter. Perhaps the most interesting new results are
in sections~\ref{sect:pure} and \ref{sect:puresosp}, where we
refine predictions for IR behavior
of pure two-dimensional supersymmetric $SU$ theories 
and extend those predictions to pure $SO$ and $Sp$ theories.

Finally, 
we should emphasize to the reader that we are proposing a construction of
a mirror topological Landau-Ginzburg model.  Although it is our hope that
this construction can be understood as a topological
twist of a duality of physical theories, we have not performed any checks
beyond topological field theories, and in fact, as we do not specify
a K\"ahler potential, we have not given sufficient data to
specify untwisted theories.  It would certainly be interesting to pursue
such a direction.  One starting point
would be to compute and compare hemisphere
partition functions, which may be an efficient means to check for 
possible physical dualities.  Again, our primary
intent in this paper is merely to propose
a duality of topological field theories, and we leave questions of
physical dualities for future work.

\section{Proposal}
\label{sect:prop}

Consider an A-twisted (2,2) GLSM with gauge group $G$ (of dimension $n$ and
rank $r$) and matter in
some representation ${\cal R}$ (of dimension $N$).  
For simplicity, in this paper we will assume $G$ is connected.
For the moment, we will assume that the A-twisted gauge theory has no
superpotential, and we will consider generalizations later in this section.
We propose that
the mirror is an orbifold of a Landau-Ginzburg
model.  We will describe the Landau-Ginzburg model first,
then the orbifold.  The Landau-Ginzburg model
has the following matter fields:
\begin{itemize}
\item $r$ (twisted) chiral superfields
$\sigma_a$, corresponding to a choice of Cartan subalgebra of the
Lie algebra of $G$,
\item $N$ (twisted) chiral superfields $Y_i$, each of imaginary periodicity
$2 \pi i$ as in \cite{Hori:2000kt}[section 3.1], which we will discuss
further in a few paragraphs,
\item $n-r$ (twisted) chiral superfields $X_{\tilde{\mu}}$,
\end{itemize}
with superpotential\footnote{
If we want to be careful about QFT scales, the second line should be written
\begin{displaymath}
 \sum_{i=1}^N \mu \exp\left(-Y_i\right) \: + \:
\sum_{\tilde{\mu}=1}^{n-r} \mu X_{\tilde{\mu}},
\end{displaymath}
where $\mu$ is a pertinent mass scale.
}
\begin{eqnarray}
W & = & \sum_{a=1}^r \sigma_a \left( 
\sum_{i=1}^N \rho_i^a Y_i \: - \: \sum_{\tilde{\mu}=1}^{n-r} \alpha_{\tilde{\mu}}^a 
\ln X_{\tilde{\mu}}
\: - \: t_a \right) 
\nonumber \\
& & \: + \: \sum_{i=1}^N \exp\left(-Y_i\right) \: + \:
\sum_{\tilde{\mu}=1}^{n-r} X_{\tilde{\mu}}, 
\label{eq:proposal-w}
\end{eqnarray}
where the $\rho_i^a$ are the weight vectors for the representation
${\cal R}$, and the $\alpha_{\tilde{\mu}}^a$ are root vectors for the
Lie algebra of $G$.  (We will sometimes write\footnote{
Taking into account QFT mass scales, $Z_{\tilde{\mu}} = - \ln \left( \mu X_{\tilde{\mu}} \right)$.
}
 $X_{\tilde{\mu}}$ in terms of
$Z_{\tilde{\mu}} = - \ln X_{\tilde{\mu}}$ for convenience, but $X_{\tilde{\mu}}$
is the fundamental\footnote{
We use the term `fundamental field' to implicitly indicate the form of 
the path integral
integration measure.  In the present case, in the B model, the integration
measure (over constant zero modes) has the form
\begin{displaymath}
\int \left( \prod_a d^2 \sigma_a \right)
\left( \prod_i d^2 Y_i \right)
\left( \prod_{\tilde{\mu}} d^2 X_{\tilde{\mu}} \right).
\end{displaymath}
Later in this section when we discuss mirrors to fields with R-charges,
we will say that different fields are `fundamental', for example,
for a mirror to a field of R-charge two, the fundamental field would be
$\exp( - Y)$.  In such a case, instead of $d^2 Y$, one would have
$d^2 \exp( - Y)$.  
} field.)   
(In passing, in later sections, we will slightly modify our index
notation:  $i$ will be broken into flavor and color components,
and $\tilde{\mu}$ will more explicitly reflect the adjoint representation,
in the form of $\tilde{\mu} \mapsto \mu \nu$, for $\mu$, $\nu$ 
ranging over the same values as $a$..
However, for the moment, this convention is very efficient for outlining
the proposal.)  
For reasons we will discuss in section~\ref{sect:gkn:excluded}, 
we believe that the loci $\{ X_{\tilde{\mu}} = 0 \}$ are dynamically
excluded by {\it e.g.} diverging potentials.
(This will lead to an algebraic derivation
of the excluded locus in A model Coulomb branch computations.)
Finally, for reasons of brevity, we will often omit the term `twisted'
when describing the $\sigma$, $Y$, and $X$ superfields; 
the reader should add it as needed
from context.

Strictly speaking, the $\sigma_a$ should be understood as curvatures
of vector multiplets of the original A-twisted gauge theory:
$\sigma_a \propto \overline{D}_+ D_- V_a$ for vector multiplets $V_a$,
just as in \cite{Hori:2000kt}.
This means the $\sigma$ terms in the superpotential
above encode theta angle terms such as
$\theta F_{z \overline{z}}$, which tie into periodicities of the
$Y$ fields (to which we will return next).  The reader should also
note that our notation is slightly nonstandard:  whereas other papers
use $\Sigma$, we use $\sigma$ to denote both the twisted chiral
superfield (the curvature of $V$) as well as the lowest component of the
superfield.  (As these $\sigma$'s often occur
inside and next to summation symbols, we feel our slightly nonstandard notation
will improve readability.)
In the limit that the gauge coupling of the original theory becomes infinite,
the $\sigma_a$ become Lagrange multipliers, from the form of the
kinetic terms \cite{Hori:2000kt}[equ'n (3.69)].  As a result,
we will often speak of integrating them out.
(On occasion, we will utilize the fact that we are in a TFT to integrate
out other fields as well.)

We have not carefully specified to which two-dimensional (2,2) supersymmetric
theories the ansatz above should apply.  Certainly we feel it should apply
to theories with isolated vacua and theories describing compact CFTs, and
we have checked numerous examples of this form.  In addition, later we will
also see it reproduces results for non-regular theories (in the sense of
\cite{Hori:2011pd}), as well as results for theories that flow to
free twisted 
chiral multiplets.  In any event, as we have not provided a proof of
the ansatz above, we can not completely nail down a range of validity.

To clarify the $Y$ periodicities, 
\begin{displaymath}
Y_i \: \sim \: Y_i \: + \: 2 \pi i,
\end{displaymath}
so that the $Y$'s take values in a 
torus of the form ${\mathbb C}^{{\rm dim}\,{\cal R}} / 2 \pi i {\mathbb Z}$,
and the superpotential terms $\exp(-Y_i)$ are well-defined.
Then, schematically, the contraction $\rho Y$ has periodicity 
$2 \pi i M$
for $M$ the weight lattice.  For abelian cases, this is merely the usual
affine shift by $2 \pi i$ that appeared in \cite{Hori:2000kt}, 
independently for each $Y$,
but may be a little more complicated in nonabelian cases.
Furthermore, in our conventions, the weight lattice is normalized
so that the theta angle periodicities of the original gauge theory
are of the form $2 \pi M$, since phases picked up by $\rho Y$
are absorbed into theta angles.
Finally, note that
in the first line of the superpotential above, the log branch cut ambiguity
effective generates shifts of weight lattice periodicities by roots.

So far we have described the Landau-Ginzburg model.  The proposed
mirror is an orbifold of the theory above, by the Weyl group $W$,
acting on the $\sigma_a$, $Y_i$, and $X_{\mu}$.  (The action can be
essentially inferred from the quantum numbers, and we will describe it
in explicit detail in examples.)

The superpotential above, written in terms of root and weight vectors, is 
invariant under this Weyl group action simply
because the Weyl group permutes root vectors into other root vectors
and weight vectors into other weight vectors for any finite-dimensional
representation, see {\it e.g.} \cite{jacobson}[chapter VIII.1],
\cite{fultonharris}[chapter 14.1].
As a result, the Weyl group maps $Z_{\tilde{\mu}}$'s to other 
$Z_{\tilde{\mu}}$'s,
and $Y_i$'s to other $Y_i$'s, consistently with changes in
$\rho^a_i$ and $\alpha^a_{\tilde{\mu}}$.
Furthermore, we take each Weyl group reflection to also act in the
same way on the $\sigma$'s as on the root and weight lattices, so that
the combinations
\begin{displaymath}
\sum_a \sigma_a \alpha^a_{\tilde{\mu}}, \: \: \:
\sum_a \sigma_a \rho^a_i
\end{displaymath}
are permuted at the same time and in the same way as the 
$X_{\tilde{\mu}}$ and $Y_i$.  This guarantees that the superpotential
remains invariant under the Weyl group, a fact we shall also
check explicitly in examples.

In practice, in the examples in this paper, the Weyl group will act
by permutations and sign flips, and so it will be straightforward
to check that, so long as the $\sigma$'s are also permuted and sign-flipped,
the superpotential is invariant.
In addition, in some of the examples we shall compute,
we shall also see alternate representations of the superpotential
above, in which the $\sigma_a$ terms above involve nontrivial
matrix multiplications, rather than just root and weight
vectors.  In such cases, we will check explicitly that the
superpotential is again invariant under the Weyl group action.

We require the mirror Landau-Ginzburg model admit a B twist,
which constrains the orbifold.  After all, to define the B twist in
a closed string theory,
the orbifold must be such that the square of the holomorphic top-form
is invariant \cite{Sharpe:2006qd}.  
(It is sometimes said that the B model is only
defined for Calabi-Yau's, but as discussed in \cite{Sharpe:2006qd}, 
the Calabi-Yau
condition for existence of the closed string B model
can be slightly weakened.)  In the present case, each element of the
Weyl group acts by exchanging some of the fields, possibly with signs.
Under such (signed) interchanges, a holomorphic top-form will change by
at most a sign; a square of the holomorphic top-form will be invariant.
Therefore, this Weyl orbifold will always be compatible with the B twist.

In our proposal, we have deliberately not specified the K\"ahler
potential.  As we are working with topologically-twisted theories,
and the space of $\sigma$s, $Y$s, and $X$s is topologically trivial,
the K\"ahler potential is essentially irrelevant.  One suspects
that in a physical, untwisted, nonabelian mirror, the K\"ahler
potential terms would reflect nonabelian T-duality, just as the
kinetic terms in the Hori-Vafa proposal \cite{Hori:2000kt} reflected
abelian duality.  We do briefly outline an idea of how one might go about
proving this proposal in section~\ref{sect:just:hints}, 
and in that section, we make
a few tentative suggestions for a possible form of the K\"ahler potential.
It would be interesting to pursue this in future
work.

Our proposal only refers to the Lie algebra of the A model gauge theory,
not the gauge group.  Different gauge groups with the same Lie algebra
can encode different nonperturbative physics, see {\it e.g.}
\cite{Pantev:2005rh,Pantev:2005zs,Hellerman:2006zs,Caldararu:2007tc,Sharpe:2014tca}.  
Here, we conjecture that the different Lie groups with the same
Lie algebra (and matter content) are described in the mirror by
rescalings of the mirror roots $\alpha^a_{\tilde{\mu}}$  and matter weights
$\rho^a_i$.  Integrating out the $\sigma$'s would then
result in gerbe structures as discussed in analogous abelian cases in
\cite{Pantev:2005zs}.  (See {\it e.g.} \cite{Tachikawa:2013kta}[section 4.5]
for related
observations in other theories.)  We will study this matter to a limited
extent
later in section~\ref{sect:pure}, 
but aside from that will leave such considerations
for future work.

By computing the critical locus along $Y$'s and $Z$'s, we also
find the operator mirror map, in the sense of \cite{Gu:2017nye}:
\begin{eqnarray}
\exp(-Y_i) & = & \sum_{a=1}^r \sigma_a \rho_i^a, 
\label{eq:op-mirror-1}\\
X_{\tilde{\mu}}  & = & \sum_{a=1}^r \sigma_a \alpha_{\tilde{\mu}}^a .
\label{eq:op-mirror-2}
\end{eqnarray}
We interpret the right-hand side as defining A model Coulomb branch
operators, which this map shows us how to relate to B model operators.

In principle, to make the ansatz above useful for general cases,
one would like to be able to evaluate Landau-Ginzburg correlation
functions on general orbifolds.  Many Landau-Ginzburg computations
are known, especially massless spectrum computations in conformal models
\cite{Intriligator:1990ua,Kachru:1993pg,Bertolini:2017lcz},
and more recently \cite{Bertolini:2018now},
but correlation function computations on orbifolds are not, to our
knowledge, understood in complete generality.  On the other hand, in many simple
cases we can get by with less.  In particular, in the examples
in this paper, 
the critical points of the
superpotential are not located at orbifold fixed points. 
(This is essentially because of the assumption that $X_{\tilde{\mu}} \neq 0$
mentioned earlier.  One of the effects of this assumption is to make the
superpotential well-defined -- although it has poles where any $X_{\tilde{\mu}}$
vanishes, it becomes ill-defined when multiple $X_{\tilde{\mu}}$ vanish,
an issue which we will return to in section~\ref{sect:regularization},
where we will discuss this as a regularization issue.  In any event,
since the Weyl group will interchange
the $\sigma_a$, the orbifold fixed-point locus will lie where some
$X_{\tilde{\mu}}$ vanish.
This also corresponds to one of the
conditions for Bethe vacua, discussed in {\it e.g.} 
\cite{Closset:2017vvl}[section 2.1].)
Rescaling
the worldsheet metric in the B-twisted theory, one quickly finds
that the bosonic contribution to the path integral is of the form
\cite{Guffin:2008kt}[section 2.2], \cite{Vafa:1990mu}
\begin{displaymath}
\lim_{\lambda \rightarrow \infty}
\int_X d \phi \exp\left( - \sum_i | \lambda \partial_i W |^2 \right),
\end{displaymath}
and so vanishes unless the critical locus intersects the fixed-point
locus.  As a result, since in this paper we are computing {\it e.g.}
correlation functions of
untwisted operators on genus zero worldsheets,
we are able to consistently omit
contributions from twisted sectors in the computations presented here.

As a consistency check, let us specialize to the case that $G = U(1)^r$.
In this case, there is no Weyl 
orbifold, there are no fields $X_{\tilde{\mu}}$, and
the mirror is defined by the fields $\sigma_a$ and $Y_i$ with
superpotential
\begin{displaymath}
W \: = \:
\sum_{a=1}^r \sigma_a \left(
\sum_{i=1}^N Q_i^a Y_i \: - \: t_a \right)
\: + \: \sum_{i=1}^N \exp\left(-Y_i\right),
\end{displaymath}
since the weight vectors $\rho_i^a$ reduce to the charge matrix $Q_i^a$.
This is precisely the mirror of an abelian GLSM discussed
in \cite{Hori:2000kt}, as expected.

Let us now return to the nonabelian theory.
If the fields $\phi_i$ of the original A model have 
twisted masses $\tilde{m}_i$, then the mirror proposal is the same
orbifold but with a different superpotential, given by
\begin{eqnarray}
W & = & \sum_{a=1}^r \sigma_a \left( 
\sum_{i=1}^N \rho_i^a Y_i \: + \: 
\sum_{\tilde{\mu}=1}^{n-r} \alpha_{\tilde{\mu}}^a Z_{\tilde{\mu}}
\: - \: t_a \right) 
\nonumber \\
& & \: - \:
\sum_{i=1}^N  \tilde{m}_i 
\left( Y_i \: - \: \sum_a \rho_i^a t_a \right) 
\nonumber \\
& & \: + \: \sum_{i=1}^N \exp\left(-Y_i\right) \: + \:
\sum_{\tilde{\mu}=1}^{n-r} X_{\tilde{\mu}} .
\label{eq:proposal-w:mass}
\end{eqnarray}

Computing the critical locus along the $Y$'s and $X$'s yields the
operator mirror map including twisted masses and R-charges:
\begin{eqnarray}
\exp(-Y_i) & = & - \tilde{m}_i  + \sum_{a=1}^r \sigma_a \rho_i^a, 
\label{eq:op-mirror-mass-1}
\\
X_{\tilde{\mu}}  & = & \sum_{a=1}^r \sigma_a \alpha_{\tilde{\mu}}^a .
\label{eq:op-mirror-mass-2}
\end{eqnarray}

We can also formally derive quantum cohomology relations in a similar
fashion.  The critical locus for $\sigma_a$ is
\begin{equation}
\sum_{i=1}^N \rho_i^a Y_i \: + \: \sum_{\tilde{\mu}=1}^{n-r} \alpha^a_{
\tilde{\mu}} Z_{\tilde{\mu}} \: = \: t_a,
\end{equation}
and exponentiating gives
\begin{equation}
\left[ \prod_i \left( \exp(-Y_i) \right)^{ \rho^a_i } \right]
\left[ \prod_{\tilde{\mu}} \left( X_{\tilde{\mu}}  \right)^{\alpha^a_{
\tilde{\mu}} } \right] \: = \: q_a.
\end{equation}
Applying the operator mirror map equations above, this becomes
\begin{equation}
\left[ \prod_i \left( \sum_b \sigma_b \rho^b_i - \tilde{m}_i 
\right)^{\rho^a_i} \right]
\left[ \prod_{\tilde{\mu}} \left( \sum_b \sigma_b \alpha^b_{\tilde{\mu}}
\right)^{\alpha^a_{\tilde{\mu}} } \right] \: = \: q_a.
\end{equation}
In practice, we will see later in section~\ref{sect:compare-corr-fn} that 
\begin{displaymath}
\left[ \prod_{\tilde{\mu}} \left( \sum_b \sigma_b \alpha^b_{\tilde{\mu}}
\right)^{\alpha^a_{\tilde{\mu}} } \right]
\end{displaymath}
is a $\sigma$-independent constant matching that discussed in
\cite{Hori:2013ika}[setion 10], so we can write the quantum cohomology
relations as either
\begin{equation}
\prod_i \left( \sum_b \sigma_b \rho^b_i - \tilde{m}_i 
\right)^{\rho^a_i} \: = \: \tilde{q}_a,
\end{equation}
or equivalently in the mirror
\begin{equation}
\prod_i \exp\left( - \rho_i^a Y_i \right) \: = \: \tilde{q}_a,
\end{equation}
where $\tilde{q}_a$ differs from $q_a$ by the constant discussed above.

As written above, our proposal is for the mirror to an A-twisted gauge
theory with no superpotential.  Let us now consider the case that the
A-twisted theory has a superpotential.  In this case, one must specify
nonzero R charges for the fields, so that the superpotential has
R charge two.  Furthermore, in order for the A twist to exist,
those R charges must be integral (see {\it e.g.} \cite{Guffin:2008kt},
\cite{Hori:2013ika}[section 3.4], \cite{Closset:2015ohf}[section 2.1]).  
(Technically, on Riemann surfaces
of nonzero genus, this requirement can be slightly relaxed, but in order
to have results valid for all genera, we will assume the most restrictive
form, namely the genus zero result that R charges are integral.)

Given an A-twisted gauge theory with superpotential and suitable
R charges, we can now define the mirror.  Both the A-twisted theory and
its mirror will be independent of the details of the (A model)
superpotential (which is BRST exact in the A model,
see {\it e.g.} \cite{Guffin:2008kt}[section 3.1]), 
though not independent of the R charges of the fields.
Our proposal is that the B-twisted mirror has exactly the same form as discussed
above -- same number of fields, same mirror superpotential -- but with
one minor quirk, that the choice of fundamental field changes.
Specifically, if a field $\phi_i$ of the A model has nonzero R-charge $r_i$,
then the fundamental field in the mirror is
\begin{displaymath}
X_i \: \equiv \: \exp(-(r_i/2) Y_i),
\end{displaymath}
and in the expressions above, we take $Y_i$ to mean
\begin{displaymath}
Y_i \: = \: - \frac{2}{r_i} \ln X_i.
\end{displaymath}
(We will outline the reason for this identification in
appendix~\ref{app:r-charges}).
Furthermore, in this case, ultimately because of the periodicity
of $Y_i$, the mirror theory with field $X_i$ has a cyclic orbifold
of order $2/r_i$ (which we assume to be an integer), for the same
reasons as discussed elsewhere in Hori-Vafa \cite{Hori:2000kt} mirrors.
(Of course, if the original field has $r_i=0$, then there is no change
in fundamental field, and so no orbifold in the mirror.)
The reader should also note that field redefinitions in the mirror
may introduce additional orbifolds, which is essentially what happens
in the Hori-Vafa mirror to the quintic, for example.

Note that for the A-twisted theory to exist, every $r_i$ must be an integer,
and for the orbifold in the B model mirror to be well-defined,
we must require $2/r_i$ (for nonzero $r_i$) to also be an integer.  
Also taking into account a positivity condition discussed in
\cite{Hori:2013ika}[section 3.4], 
this means we are effectively restricted to the choices
$r_i \in \{0, 1, 2\}$ in our proposal.  If a gauge theory has
a superpotential that is incompatible with such choices of R charges,
then either the A twist does not exist or our proposed mirror does not
apply.

In principle, in the language of the dictionary above,
mirrors to the W bosons act like mirrors to fields of
R charge two, and are the fields $X_{\tilde{\mu}}$ rather than
the $Z_{\tilde{\mu}}$.  Also, since $2/2=1$, there is no orbifold
(beyond the Weyl group orbifold) associated with the $X_{\tilde{\mu}}$
specifically.

Finally, we should mention that the axial R symmetry of the A model
theory appears here following the same pattern as in
\cite{Hori:2000kt}[equ'n (3.30)].  Specifically, 
under R$_{\rm axial}$,
\begin{equation}
Y_i \: \mapsto \: Y_i \: - \: 2 i \alpha,
\end{equation}
and
\begin{equation}
X_{\tilde{\mu}} \: \mapsto \: X_{\tilde{\mu}} \exp\left( + 2 i \alpha \right),
\end{equation}
so that for example the superpotential terms
\begin{displaymath}
\sum_i \exp\left( - Y_i \right) \: + \:
\sum_{\tilde{\mu}} X_{\tilde{\mu}}
\end{displaymath}
have charge $2$, as one would expect.
In that vein, note that 
\begin{equation}
\exp\left( - (r_i/2) Y_i \right) \: \mapsto \:
\exp\left( - (r_i/2) Y_i \right) \exp\left( + i r_i \alpha \right),
\end{equation}
as one would expect for a field of R charge $r_i$.
Similarly, the effect of the R charge on the $\sigma$ terms is
to generate a term
\begin{equation}
(- 2 i \alpha) \sum_a \sigma_a  \left(
\sum_i \rho^a_i + \sum_{\tilde{\mu}} \alpha^a_{\tilde{\mu}} \right)
\: = \:
(- 2 i \alpha) \sum_a \sigma_a  \left(
\sum_i \rho^a_i \right)
\end{equation}
(since the sum over $\alpha^a_{\tilde{\mu}}$ will vanish),
reflecting the fact that if the A model theory has an axial anomaly,
then an R$_{\rm axial}$ rotation will shift theta angles.

\section{Justification for the proposal}
\label{sect:justification}

The bulk of this paper will be spent checking examples,
which to our minds will be the best verification of the proposal,
but before working through those examples, we wanted to briefly describe
the origin of some of the details of the proposal above, as well
as perform some consistency tests, such as a general comparison of
correlation functions.

\subsection{General remarks}

At least for the authors, one of the motivations for
this work was to find a UV realization of factors of the form
\begin{displaymath}
\prod_{a < b} \left( \sigma_a - \sigma_b \right)
\end{displaymath}
appearing in integration measures, such as the 
Hori-Vafa conjecture for nonabelian mirrors 
in \cite{Hori:2000kt}[appendix A], and later in
expressions for supersymmetric partition functions of nonabelian
gauge theories in \cite{Benini:2012ui,Doroud:2012xw}. 
Later, \cite{Gomis:2012wy} studied $S^2$ partition
functions of Hori-Vafa mirrors, and in section 4 of that paper,
applied the same methods to predict the form of partition functions
of the mirror of a $U(k)$ gauge theory (with $k>1$)
corresponding to a Grassmannian, 
where again they found factors in
the integration measure of the same form (albeit squared\footnote{
In open string computations, one gets factors of 
$\prod_{a<b} (\sigma_a - \sigma_b)$,
whereas in closed string computations, one typically gets factors of
$\prod_{a<b} (\sigma_a - \sigma_b)^2$.  As this paper is focused on closed
string computations, we will see the latter.
}),
a result we will duplicate later.

We reproduce such factors via the fields $X_{\tilde{\mu}}$, the
mirrors to the W bosons.
The basic idea originates in an observation in
\cite{Halverson:2013eua}[section 2], which relates the partition function
of a nonabelian theory to that of an associated `Cartan theory,'
an abelian gauge theory in which the nonabelian gauge group is replaced
by its Cartan torus, and in addition to the chiral multiplets of the
nonabelian theory, one adds an additional set of chiral multiplets of
R charge two corresponding
to the nonzero roots of the Lie algebra.  It is briefly argued that the
$S^2$ partition function of the original nonabelian theory matches the
$S^2$ partition function of the associated Cartan theory.  In effect,
we are taking this observation a step further, by dualizing the 
associated Cartan theory in the sense of \cite{Hori:2000kt} to construct
this proposal for nonabelian mirrors.

We take the Weyl orbifold to get the right moduli space:
the Coulomb branch moduli space is not quite just the moduli space
of a $U(1)^r$ gauge theory, as one should also identify $\sigma$ fields
related by the Weyl group.  Note the Weyl group does not survive the
adjoint Higgsing; instead, we taking the orbifold so as to reproduce
the correct Coulomb branch.  This is analogous\footnote{
We would like to thank I.~Melnikov for providing this analogy.
} to the $c=1$ boson
at self-dual radius:  as one moves away from the self-dual point,
the $SU(2)$ is broken to $U(1)$ on both sides, and the Weyl orbifold allows
one to forget about radii that are smaller, since they are all Weyl equivalent
to larger radii.  Another example is the construction of the $u$ plane in
four-dimensional $N=2$ Seiberg-Witten theory.
In the present case, we will see for example in the
case of Grassmannians that to get the correct number of vacua,
one has to quotient by the Weyl group.

\subsection{Suggestions of a route towards a proof}
\label{sect:just:hints}

We do not claim to have a rigorous proof of the proposal of this paper,
but there is a simple idea for a proof.  Given a (2,2) supersymmetric
GLSM, imagine moving to a generic point on the Coulomb branch, described by a 
Weyl-group orbifold of an abelian gauge theory, with gauge group equal
to the Cartan of the original theory.  Now, apply abelian duality
to this abelian gauge theory\footnote{
In other words, to any nonabelian gauge theory we can associate
a toric variety or stack, defined by matter fields plus W bosons at a generic
point on the Coulomb branch.  Our proposal seems consistent with
abelian duality for that toric variety.
}.  One will T-dualize the original matter
fields (which become the $Y_i$) as well as the W bosons (which become
the $X_{\mu \nu}$).  

For later purposes, it will be instructive to fill in a few steps.
That said, we emphasize that we are not claiming we have a rigorous
demonstration.  Our goal here is merely to suggest a program, and to
investigate the form of a possible K\"ahler potential to justify
certain plausibility arguments elsewhere.

For ordinary matter fields $\Phi$, of charge $\rho^a$ under the
$a$th $U(1)$, T-duality in this context \cite{Hori:2000kt,Morrison:1995yh}
says that the field should be described by an `intermediate' Lagrangian
density of the form   \cite{Hori:2000kt}[equ'n (3.9)]
\begin{equation}
L_{\Phi} \: = \: \int d^4 \theta \, \left(
\exp\left( 2 \sum_a \rho^a V_a \: + \: B \right) \: - \: \frac{1}{2}
\left( Y + \overline{Y} \right) B \right) .
\end{equation}
Reviewing the analysis of \cite{Hori:2000kt}[section 3.1],
if one integrates over $Y$, one gets constraints
\begin{equation}
\overline{D}_+ D_- B \: = \: 0 \: = \: D_+ \overline{D}_- B,
\end{equation}
which are solved by taking
\begin{equation} \label{eq:just:pf:b1}
B \: = \: \Psi \: + \: \overline{\Psi}.
\end{equation}
Plugging back in, one finds
\begin{equation}
L_{\Phi} \: = \: \int d^4 \theta \, \exp\left(  2 \sum_a \rho^a V_a
\: + \: \Psi \: + \: \overline{\Psi} \right) \: = \:
\int d^4 \theta \, \overline{\Phi} \exp\left(  2 \sum_a \rho^a V_a \right)
\Phi,
\end{equation}
the original Lagrangian, for $\Phi = \exp(\Psi)$.

If one instead integrates over $B$ first, then one recovers the dual
theory, as follows.  Integrating out $B$ first yields
\begin{equation}  \label{eq:just:pf:b2}
B \: = \: -  2 \sum_a \rho^a V_a \: + \: \ln \left( \frac{ Y + \overline{Y}}{
2} \right),
\end{equation}
and plugging this in we find
\begin{eqnarray}
L_{\Phi} & = &
\int d^4 \theta \, \left( \frac{ Y + \overline{Y} }{2} \: + \:
 \left( Y + \overline{Y} \right) \sum_a \rho^a V_a
\: - \: \frac{1}{2} \left( Y + \overline{Y} \right) 
 \ln \left( \frac{ Y + \overline{Y}}{
2} \right) \right), 
\\
& = &
\int d^4 \theta \, \left(
 \left( Y + \overline{Y} \right) \sum_a \rho^a V_a
\: - \: \left( \frac{Y + \overline{Y} }{2} \right)  
 \ln \left( \frac{ Y + \overline{Y}}{
2} \right) \right).
\end{eqnarray}
Since $Y$ is a twisted chiral superfield, 
the first term can be written
\begin{equation}
\int d^4 \theta \, \left(   Y \sum_a \rho^a V_a
\right)
\: = \: \int d^2 \theta\, \sum_a \sigma_a \rho^a Y,
\end{equation}
where $\sigma_a = \overline{D}_+ D_- V_a$,
and so this term contributes to the superpotential.

Equating the two forms~(\ref{eq:just:pf:b1}), (\ref{eq:just:pf:b2}) for
$B$, one finds
\begin{equation}
Y + \overline{Y} \: = \: 2
\overline{\Phi} \exp\left(  2 \sum_a \rho^a V_a \right)
\Phi,
\end{equation}
and from the K\"ahler potential term above, we see that the metric
seen by the kinetic terms for $Y$ components is
\begin{equation}
ds^2 \: = \: \frac{ | dy |^2 }{ 2 (y + \overline{y} ) },
\end{equation}
where $y$ is the scalar part of $Y$.

So far this analysis is entirely standard.  Now, let us think about the
analogous analysis for T-duals of the W bosons.  Here, we take the
W bosons to be described by chiral superfields $W_{\tilde{\mu}}$
and the Lagrangian density
\begin{equation}  \label{eq:just:pf:worig}
L_W \: = \: \int d^4 \theta \, \overline{W}_{\tilde{\mu}}
\exp\left( 2 \sum_a \alpha^a_{\tilde{\mu}} V_a \right) W_{\tilde{\mu}}.
\end{equation}
Proceeding as before, we can consider the intermediate Lagrangian density
\begin{equation}
L_W \: = \:
\int d^4 \theta \, \left(
\exp\left( 2 \sum_a \alpha^a_{\tilde{\mu}} V_a \: + \: B_{\tilde{\mu}}
\right) \: - \: \frac{1}{2} \left( Z_{\tilde{\mu}} + \overline{
Z}_{\tilde{\mu}} \right) B_{\tilde{\mu}} 
\right).
\end{equation}
Our analysis will closely follow the pattern for $\Phi$, $Y$.
Integrating over the $Z_{\tilde{\mu}}$ recovers the original
Lagrangian density~(\ref{eq:just:pf:worig}).
Integrating out the $B_{\tilde{\mu}}$, one finds
\begin{equation}
L_W \: = \:
\int d^4 \theta \, \left(
\left( Z_{\tilde{\mu}} + \overline{Z}_{\tilde{\mu}} \right) \sum_a
\alpha^a_{\tilde{\mu}} V_a \: - \:
\left( \frac{ Z_{\tilde{\mu}} + \overline{Z}_{\tilde{\mu}} }{2} \right)
\ln \left( \frac{ Z_{\tilde{\mu}} + \overline{Z}_{\tilde{\mu}} }{2} \right)
\right).
\end{equation}
The first term can be rewritten as a superpotential
contribution.  The primary difference here is that we take
the fundamental field to be $X_{\tilde{\mu}} = \exp(-Z_{\tilde{\mu}})$.
In terms of $X_{\tilde{\mu}}$, the kinetic term takes the form
\begin{equation}
\int d^4 \theta \, 
 \left( \frac{ \ln X_{\tilde{\mu}} + \ln \overline{X}_{\tilde{\mu}} }{2}
\right)
\ln \left( -
\frac{ \ln X_{\tilde{\mu}} + \ln \overline{X}_{\tilde{\mu}} }{2}
\right) ,
\end{equation}
and from this K\"ahler potential it is straightforward to compute that the
metric for the kinetic terms has the form
\begin{equation}
ds^2 \: = \: \frac{ | dx |^2 }{ 2 | x |^2 \ln | x |^2 }.
\end{equation}

To resolve subtleties in renormalization, in \cite{Hori:2000kt}, 
it was noted that the kinetic terms were written in terms of a bare field
$Y_0$ related to a renormalized field by \cite{Hori:2000kt}[equ'n (3.23)]
\begin{equation}
Y_0 = \ln( \Lambda_{UV}/\mu ) + Y,
\end{equation}
and then in a suitable limit, the metric on the $Y$'s becomes flat.
The analogue here is to write $X_0 = (\mu/\Lambda_{UV}) X$,
so that the metric for the kinetic term becomes
\begin{equation}
\frac{ | dx |^2 }{ |x|^2 \left( - 2 \ln(\Lambda_{UV}/\mu) + \ln |x|^2 \right)}.
\end{equation}
Even in the analogous scaling limit however, this metric diverges as
$x \rightarrow 0$, suggesting that the kinetic terms dynamically forbid
$x=0$.

The take-away observation from the computation above is that the
proposed kinetic terms for the W-boson mirrors have singularities at
$X=0$.  Now, granted, 
a more rigorous analysis of
duality might well work along the lines of nonabelian T-duality rather than
abelian T-duality in a Cartan, and so yield different kinetic terms still,
see {\it e.g.} \cite{CaboBizet:2017fzc} for a pertinent discussion of
nonabelian T-duality.
Furthermore, these kinetic terms will receive quantum corrections,
that could even smooth out singularities of the form above, see
{\it e.g.}
\cite{Hori:2000kt,Aharony:1997bx,Aharony:2016jki,Adams:2001sv}.

In passing, let us point out a few other consistency checks.
As observed in \cite{Closset:2015rna}[appendic C.4], 
in the A-twisted gauge theory, supersymmetric W bosons contribute to
supersymmetric localization as chiral multiplets of R-charge two,
so that the mirror should be a twisted chiral multiplet (same as the
$X$ fields), and the R charge dictates that the mirror fields should appear
linearly in the superpotential (as the fundamental field is
$\exp(- (r/2) Y)$).  The mass of the $X$ fields themselves is a bit off:
\begin{equation}
\frac{\partial^2 W}{\partial X_{\tilde{\mu}} \partial X_{\tilde{\nu}}}
\: = \: \delta_{\tilde{\mu} \tilde{\nu}} \frac{
\sum_a \sigma_a \alpha^a_{\tilde{\mu}} }{ X_{\tilde{\mu}}^2 }
\: = \: \delta_{\tilde{\mu} \tilde{\nu}} \frac{1}{ \sum_a \sigma_a
\alpha^a_{\tilde{\mu}} }
\end{equation}
after applying the mirror map, whereas the mass of a W boson is instead
$\sum_a \sigma_a \alpha^a_{\tilde{\mu}}$.  On the other hand,
\begin{equation}
\frac{\partial^2 W}{\partial \ln X_{\tilde{\mu}} 
\partial X_{\tilde{\nu}} } \: = \: \delta_{\tilde{\mu} \tilde{\nu} }
X_{\tilde{\mu}} \: = \: \delta_{ \tilde{\mu} \tilde{\nu}}
\sum_a \sigma_a \alpha^a_{\tilde{\mu}},
\end{equation}
after applying the mirror map, exactly right to match the mass of the
W bosons, suggesting that the W boson mirrors are $\ln X_{\tilde{\mu}}$.

\subsection{Comparison of correlation functions}
\label{sect:compare-corr-fn}

In this section, we will give a formal outline of how (some)
A model correlation functions match B model correlation functions in the
proposed nonabelian mirror, by in the mirror 
formally integrating out the mirrors to the W bosons
and the matter fields, yielding a theory of $\sigma$'s only.  
We will give several versions of this comparison, of varying levels
of rigour.  We will focus exclusively on correlation functions
of Weyl-group-invariant untwisted-sector
operators, which together with the fact 
that the Weyl group
orbifold fixed points do not intersect superpotential critical points in
the examples in this paper, will enable us to largely gloss over
the Weyl group orbifold.

\subsubsection{First argument -- iterated integrations out}

Begin with the basic mirror proposal, the
Landau-Ginzburg orbifold described in
section~\ref{sect:prop},
with superpotential $W$ given in~(\ref{eq:proposal-w:mass}).
If none of the critical points intersect fixed points of the
Weyl group orbifold, then we can integrate out the $X_{\tilde{\mu}}$,
as we shall outline next.

First, it is straightforward to compute from
the superpotential~(\ref{eq:proposal-w:mass}) that
\begin{eqnarray}
\frac{\partial W}{\partial X_{\tilde{\mu}} } & = & 1 \: - \:
\frac{ \sum_a \sigma_a \alpha^a_{\tilde{\mu}} }{ X_{\tilde{\mu}} },
\\
\frac{\partial^2 W}{\partial X_{\tilde{\mu}} \partial X_{\tilde{\nu}} }
& = &
\delta_{\tilde{\mu} \tilde{\nu}} \frac{
\sum_a \sigma_a \alpha^a_{\tilde{\mu}} }{ X_{\tilde{\mu}}^2 },
\end{eqnarray}
a diagonal matrix of second derivatives.
Evaluating on the critical locus (and identifying the
field $\sigma_a$ with the mirror field $\sigma_a$,
reflecting their common origin),
\begin{equation}
\frac{\partial^2 W}{\partial X_{\tilde{\mu}}^2 } \: = \:
\frac{1}{\sum_a \sigma_a \alpha^a_{\tilde{\mu}}}.
\end{equation}
So long as the determinant of the matrix of second derivatives is nonvanishing,
the $X_{\tilde{\mu}}$ are massive, so it is consistent
to integrate them out.
(If the matrix of second derivatives were to have a zero
eigenvalue somewhere, integrating out the $X_{\tilde{\mu}}$ would,
of course, not be consistent.)

To integrate them out, we follow the same logic as
\cite{Guffin:2008kt}[section 2.2], \cite{Vafa:1990mu}.  
Briefly, the pertinent terms in the
Lagrangian are of the form
\begin{equation}
\sum_{\tilde{\mu}} | \partial_{\tilde{\mu}} W |^2 \: + \:
 \psi_+^{\tilde{\mu}} \psi_-^{\tilde{\nu}} 
\partial_{\tilde{\mu}} \partial_{\tilde{\nu}} W
\: + \: {\rm c.c.}
\end{equation}
Expanding the purely bosonic term about the critical locus
$X^o_{\tilde{\mu}}$,
given by
\begin{displaymath}
X^o_{\tilde{\mu}} \: = \:  \sum_a \sigma_a \alpha^a_{\tilde{\mu}} ,
\end{displaymath}
we write
\begin{equation}
\sum_{\tilde{\mu}} | \partial_{\tilde{\mu}} W |^2 \: = \: 0 \: + \:
\left. \left|  \partial_{\tilde{\nu}} \partial_{\tilde{\mu}} W \right|^2
 \right|_{X^o}  | \delta X_{\tilde{\nu}} |^2,
\end{equation}
suppressing higher-order terms as in \cite{Guffin:2008kt}[section 2.2].
Performing the Gaussian integral over $\delta X_{\tilde{\nu}}$ yields a factor
\begin{displaymath}
\frac{1}{H_X \overline{H}_X},
\end{displaymath}
for $H_X$ the determinant of the matrix of second derivatives with respect
to the $X_{\tilde{\mu}}$, meaning
\begin{equation}
H_X \: = \: \prod_{\tilde{\mu}} \frac{1}{ \sum_a \sigma_a \alpha^a_{\tilde{\mu}}}.
\end{equation}
Repeating the same for the Yukawa interactions
\begin{displaymath}
\psi_+^{\tilde{\mu}} \psi_-^{\tilde{\nu}} 
\partial_{\tilde{\mu}} \partial_{\tilde{\nu}} W
\: + \: {\rm c.c.}
\end{displaymath}
as in \cite{Guffin:2008kt}[section 2.2], \cite{Vafa:1990mu} 
yields another factor of
$\overline{H}_X H_X^g$ at genus $g$.  Putting these factors together
results in a net factor of $1/H_X^{1-g}$ in correlation
functions on a genus $g$ worldsheet.

Another effect of integrating out the $X_{\tilde{\mu}}$
should be to modify the superpotential~(\ref{eq:proposal-w:mass}), evaluating
the $X_{\tilde{\mu}}$ on the critical loci:
\begin{eqnarray}
W_0 & = & \sum_{a=1}^r \sigma_a \left( 
\sum_{i=1}^N \rho_i^a Y_i \: - \: \sum_{\tilde{\mu}=1}^{n-r} \alpha_{\tilde{\mu}}^a 
\ln \left( \sum_b \sigma_b \alpha^b_{\tilde{\mu}} \right)
\: - \: t_a \right) 
\nonumber \\
& & \: + \: \sum_{i=1}^N \exp\left(-Y_i\right) 
\: - \: \sum_{i=1}^N \tilde{m}_i Y_i
\: + \: 
\sum_{\tilde{\mu}}  \sum_a \sigma_a \alpha^a_{\tilde{\mu}} ,
\\
& = &
\sum_{a=1}^r \sigma_a \left[
\sum_{i=1}^N \rho_i^a Y_i \: - \:
\sum_{\tilde{\mu}=1}^{n-r} \alpha_{\tilde{\mu}}^a 
\left( \ln \left( \sum_b \sigma_b \alpha^b_{\tilde{\mu}} \right) \: - \: 1
\right)
\: - \: t_a \right] 
\nonumber \\
& & \: + \: \sum_{i=1}^N \exp\left(-Y_i\right) 
\: - \: \sum_{i=1}^N \tilde{m}_i Y_i
\label{eq:proposal-int-out}
\end{eqnarray}
(up to constant terms we have omitted).
The $\sigma ( \ln(\sigma)-1 )$ term in the $\sigma_a$ constraint,
originating from integrating out the $X_{\tilde{\mu}}$ fields,
reflects the shift of the FI parameter described in
\cite{Hori:2013ika}[section 10].  We can simplify this constant
by rewriting it as a sum over positive roots:
\begin{eqnarray}
\lefteqn{
\sum_{\tilde{\mu}=1}^{n-r} \alpha_{\tilde{\mu}}^a 
\left( \ln \left( \sum_b \sigma_b \alpha^b_{\tilde{\mu}} \right) \: - \: 1
\right)
} \nonumber \\
& = &
\sum_{\rm pos'} \alpha_{\tilde{\mu}}^a 
\ln \left( \sum_b \sigma_b \alpha^b_{\tilde{\mu}} \right) 
\: - \:
\sum_{\rm pos'} \alpha_{\tilde{\mu}}^a \left( \ln \left( \sum_b \sigma_b \alpha^b_{\tilde{\mu}} \right) - \pi i \right),
\\
& = &
\sum_{\rm pos'} i \pi  \alpha_{\tilde{\mu}}^a,
\label{eq:genl-corr:simp1}
\end{eqnarray}
giving a shift of the theta angle matching that given in
\cite{Hori:2013ika}[equ'n (10.9)].

Altogether, the effect of integrating out the
$X_{\tilde{\mu}}$ is to
add a factor of $\overline{H}_X H_X^g/(H_X \overline{H}_X) = 1/H_X^{1-g}$
to correlation functions (at genus $g$):
\begin{equation}
\langle {\cal O} \rangle \: = \: 
\int [ D Y_i ] [D \sigma_a] \, {\cal O} \left( \prod_{\tilde{\mu}}
\left( \sum_a \sigma_a \alpha^a_{\tilde{\mu}} \right) \right)^{1-g}  
\exp(-S_0),
\end{equation}
or more simply, for the case of isolated vacua (and no contributions
from orbifold twisted
sectors),
\begin{equation}
\langle {\cal O} \rangle \: = \: \frac{1}{|W|}
\sum_{\rm vacua} \frac{{\cal O}}{ (\det \partial^2 W_0)^{1-g} } 
\left(  \prod_{\tilde{\mu}}
\left( \sum_a \sigma_a \alpha^a_{\tilde{\mu}} \right) \right)^{1-g}.
\end{equation}
Note that we can rewrite the new factor above solely in terms
of the positive roots:
\begin{equation}
 \prod_{\tilde{\mu}}
\left( \sum_a \sigma_a \alpha^a_{\tilde{\mu}} \right) 
\: \propto \:
\prod_{\rm pos' roots} \left( \sum_a \sigma_a \alpha^a_{\tilde{\mu}} \right)^2.
\end{equation}

So far, we have glossed over the fact that there is a Weyl-group orbifold
present.  For genus zero computations, since $\det \partial^2 W_0$ and
\begin{displaymath}
\prod_{\tilde{\mu}}
\left( \sum_a \sigma_a \alpha^a_{\tilde{\mu}} \right) 
\end{displaymath}
are both invariant under the Weyl group, so long as ${\cal O}$ itself is
also Weyl group invariant, the effect of the Weyl group orbifold is solely
to contribute the overall factor of $1/|W|$, where $|W|$ is the order of the
Weyl group.  For genus $g > 0$, one should be more careful, as partition
functions now contain sums over twisted sectors.  However, the B model
localizes on constant maps, so therefore so long as no critical points
of the superpotential intersect the fixed point locus of the orbifold,
we do not expect any twisted sector contributions to correlation functions
of Weyl-group-invariant operators, even at genus $g > 0$.

Our analysis so far has been rather formal, but in fact, we
will see in section~\ref{sect:gkn:int-out} that the results are consistent with
concrete computations in the case of Grassmannians.

To review, so far we have argued that
correlation functions (on a genus $g$ worldsheet of
fixed complex structure) take
the form
\begin{equation}
\langle {\cal O} \rangle \: = \: \frac{1}{|W|}
\sum_{\rm vacua} \frac{{\cal O}}{ (\det \partial^2 W_0)^{1-g} } 
\left(  \prod_{\tilde{\mu}}
\left( \sum_a \sigma_a \alpha^a_{\tilde{\mu}} \right) \right)^{1-g}
\end{equation}
(for isolated vacua),
where $W_0$ is the superpotential~(\ref{eq:proposal-int-out}).

Next, we integrate out the $Y_i$ fields, in the same
fashion.
It is straightforward to compute
\begin{eqnarray*}
\frac{\partial W_0}{\partial Y_i} & = & \sum_a \sigma_a \rho^a_i
\: - \: \exp\left( - Y_i \right) \: - \: \tilde{m}_i,
\\
\frac{\partial^2 W_0}{\partial Y_i \partial Y_j} & = &
+ \delta_{ij} \exp\left( - Y_i \right).
\end{eqnarray*}
The critical points $Y^o_i$
for $Y_i$ follow from the derivative above as
\begin{equation}
\exp\left( - Y^o_i \right) \: = \: \sum_a \sigma_a \rho^a_i
\: - \: \tilde{m}_i .
\end{equation}
Integrating out the superfield $\delta Y_i = Y_i - Y^o_i$ as
in section~\ref{sect:gkn:int-out} results in correlation functions
with an extra factor of $1/H_Y^{1-g}$ for $H_Y$ the determinant of the
matrix of second derivatives with respect to $Y$'s, namely
\begin{displaymath}
H_Y \: = \: \prod_{i=1}^N \exp\left( - Y_i \right),
\end{displaymath}
and superpotential $W_{00}$
given by evaluating $W_0$ at $Y^o_i$, meaning
\begin{eqnarray}
W_{00} & = & - \sum_a \sum_i \sigma_a \rho^a_i \ln\left(  \sum_b
\sigma_b \rho^b_i \: - \: \tilde{m}_i\right) \: + \: 
\sum_a \sum_i \sigma_a \rho^a_i
\: - \: \sum_a \sigma_a t_a
\nonumber \\
& & 
\: - \: \sum_a \sum_{\tilde{\mu}} \sigma_a \alpha^a_{\tilde{\mu}}
\ln \left( \sum_b \sigma_b \alpha^b_{\tilde{\mu}} \right)
\: + \: 
\sum_{\tilde{\mu}} \sum_a \sigma_a \alpha^a_{\tilde{\mu}}
\nonumber \\
& & 
\: + \: \sum_i \tilde{m}_i \ln \left(
\sum_b
\sigma_b \rho^b_i \: - \: \tilde{m}_i\right) ,
\\
& = & - \sum_a \sum_i \sigma_a \rho^a_i \ln\left(  \sum_b
\sigma_b \rho^b_i \: - \: \tilde{m}_i\right) \: + \: 
\sum_a \sum_i \sigma_a \rho^a_i
\: - \: \sum_a \sigma_a t_a
\nonumber \\
& & 
- \sum_{\rm pos'} i \pi \alpha^a_{\tilde{\mu}} \sigma_a
\:  + \: \sum_i \tilde{m}_i \ln \left(
\sum_b
\sigma_b \rho^b_i \: - \: \tilde{m}_i\right),
\end{eqnarray}
where we have used the simplification~(\ref{eq:genl-corr:simp1}).

Concretely, this means correlation functions (for isolated vacua,
away from fixed points of the orbifold)
on a worldsheet of genus $g$ (and fixed complex structure) are given by
\begin{equation}
\langle {\cal O} \rangle \: = \: \frac{1}{|W|}
\sum_{\rm vacua} \frac{{\cal O}}{ (\det \partial^2 W_{00})^{1-g} } 
\left[  \prod_{\tilde{\mu}}
\left( \sum_a \sigma_a \alpha^a_{\tilde{\mu}} \right) \right]^{1-g}
\left[ \prod_{i=1}^N \left( \sum_a \sigma_a \rho^a_i  \: - \:
\tilde{m}_i \right) \right]^{g-1},
\label{eq:corr-fn:sig-only}
\end{equation}
where the matrix of second derivatives $\partial^2 W_{00}$ now consists
solely of derivatives with respect to $\sigma$'s.
We deal with the Weyl-group-orbifold in the same fashion as
in section~\ref{sect:gkn:int-out}:  since the B model localizes on
constant maps, and we assume that the critical points of the superpotential
do not intersect the fixed points of the orbifold, there are no
twisted sector contributions at any worldsheet genus.

Up to overall factors, the expression~(\ref{eq:corr-fn:sig-only})
B model correlation function for the
mirror to the A-twisted gauge theory,
matches \cite{Nekrasov:2014xaa}[section 4],
with $\Delta^2(\sigma)$ reproducing
\begin{displaymath}
\prod_{\tilde{\mu}}
\left( \sum_a \sigma_a \alpha^a_{\tilde{\mu}} \right) 
\end{displaymath}
and $\exp(-2 {\cal U}_0 )$ reproducing
\begin{displaymath}
\prod_{i=1}^N \left( \sum_a \sigma_a \rho^a_i  \: - \:
\tilde{m}_i \right) ,
\end{displaymath}
and $W_{00}$ matches the ``$W_0$'' given in
\cite{Nekrasov:2014xaa}[equ'ns (2.17), (2.19)].
Also up to factors, for genus zero worldsheets,
the expression~(\ref{eq:corr-fn:sig-only}) also
matches \cite{Closset:2015rna}[equ'n (4.77)] for an A-twisted (2,2)
supersymmetric gauge theory in two dimensions,
where $Z_0^{\rm 1-loop}$ encodes \cite{Closset:2015rna}
\begin{displaymath}
\left[ \prod_{\tilde{\mu}}
\left( \sum_a \sigma_a \alpha^a_{\tilde{\mu}} \right) \right]
\left[
\prod_{i=1}^N \left( \sum_a \sigma_a \rho^a_i  \: - \:
\tilde{m}_i \right) \right]^{-1}.
\end{displaymath}
See also \cite{Closset:2017vvl}[equ'n (2.37)].
In passing, a (0,2) supersymmetric version of the same A model result
is given in \cite{Closset:2015ohf}[equ'n (3.63)].

There is another formal argument to demonstrate that
correlation functions should match.
If we integrate out the mirrors to the W bosons, but not other fields,
then as shown in \cite{Gu:2017nye}[section 4.1],
$\det \partial^2 W_0$ matches the product of $Z_{\rm 1-loop}$ and
the Hessian that arise in A model computations, which together with the
factor of 
\begin{displaymath}
\prod_{\rm pos' roots} \left( \sum_a \sigma_a \alpha^a_{\tilde{\mu}} 
\right)^2
\end{displaymath}
in correlation functions arising from integrating out the $X$ fields,
implies
that B model
correlation functions match their A model counterparts, as expected.

Our computation of B model correlation functions glossed over
cross-terms in the superpotential such as $\partial^2 W/
\partial X_{\tilde{\mu}} \partial \sigma_a$.  In the next subsection,
we shall revisit this computation from another perspective,
taking into account those cross-terms, and derive the same result
for correlation functions that we have derived in this
subsection.

\subsubsection{Second argument}

In this section, we will give a different formal derivation of the
correlation functions, that will give the same result -- the
correlation functions in our B-twisted proposed mirror (of untwisted
sector states) will match conventional computations on Coulomb branches
of A-twisted gauge theories.
Instead of sequentially integrating out the $X_{\tilde{\mu}}$, then the
$Y_i$, let us formally consider instead a direct computation of
correlation functions, assuming that critical loci are isolated
(and distinct from fixed points of the orbifold).
(If critical loci are not isolated, one could suitably
deform the superpotential to make them isolated.)

Correlation functions are then of the form
\begin{displaymath}
\langle {\cal O} \rangle \: = \: 
\sum_{\rm vacua} \frac{ {\cal O} }{ H^{1-g} },
\end{displaymath}
where $H$ is the determinant of the matrix of second derivatives.
Write
\begin{equation}
H \: = \: \det \left[ \begin{array}{cc}
A & B \\
C & D
\end{array} \right],
\end{equation}
where $A$ is the submatrix of derivatives with respect to
$X_{\tilde{\mu}}$ and $Y_i$, 
\begin{eqnarray*}
\frac{ \partial^2 W}{\partial X_{\tilde{\mu}} \partial X_{\tilde{\nu}} }
& = & \delta_{\tilde{\mu} \tilde{\nu}} \left( \sum_c \sigma_c
\alpha^c_{\tilde{\mu}} \right)^{-1},
\\
\frac{ \partial^2 W}{\partial Y_i \partial Y_j} & = &
\delta_{ij} \left( \sum_c \sigma_c \rho^c_i \: - \: 
\tilde{m}_i \right),
\\
\frac{\partial^2 W}{\partial X_{\tilde{\mu}} \partial Y_i} & = &
0,
\end{eqnarray*}
(on the critical locus,)
$B = C^T$ are the matrices of
derivatives of the form
\begin{eqnarray*}
\frac{\partial^2 W}{\partial X_{\tilde{\mu}} \partial \sigma_a }
& = & - \alpha^a_{\tilde{\mu}}  \left( \sum_c \sigma_c
\alpha^c_{\tilde{\mu}} \right)^{-1},
\\
\frac{\partial^2 W}{\partial Y_i \partial \sigma_a } & = &
\rho^a_i,
\end{eqnarray*}
and $D$ is the matrix of second derivatives with respect to
$\sigma$'s.  Since $\sigma$ only appears linearly in the superpotential,
$D=0$.

Putting this together, from \cite{powell}, we can write
\begin{equation}
H \: = \: \left( \det A \right) \det \left( D - C A^{-1} B \right).
\end{equation}
The factor $\det A$ we have seen previously:  since $A$ is diagonal,
it is straightforward to see that
\begin{equation}
\det A \: = \: 
\left[ \prod_{\tilde{\mu}} \left( \sum_c \sigma_c
\alpha^c_{\tilde{\mu}}\right) \right]^{-1}
\left[ \prod_i \left(  \sum_c \sigma_c \rho^c_i \: - \: 
\tilde{m}_i \right)
\right].
\end{equation}

Note first that
\begin{equation}
\left( \frac{1}{\det A} \right)^{1-g}
\end{equation}
is the same factor that appears multiplying operators in
correlation functions in our previous
expression~(\ref{eq:corr-fn:sig-only}); we have duplicated it without
any extra factors, despite the fact that our previous analysis omitted
cross-terms such as $\partial^2 W/ \partial X_{\tilde{\mu}} 
\partial \sigma_a$.
The remaining factor,
\begin{displaymath}
 \det \left( D - C A^{-1} B \right),
\end{displaymath}
can be interpreted as the usual Hessian from some superpotential
we shall label $W_{\rm eff}$, which we shall see next will
coincide with the $W_{00}$ of the previous subsection.  

Proceeding carefully, since $C = B^T$,
$D-0$, and $A$ is symmetric, the quantity $C A^{-1} B$ is a symmetric
matrix, so we can define a function $W_{\rm eff}$ as follows:
\begin{equation}
\left( - C A^{-1} B \right)_{ab} \: = \: \frac{
\partial^2 W_{\rm eff} }{ \partial \sigma_a \partial \sigma_b }.
\end{equation}
Computing the matrix multiplication,
we find
\begin{equation}
\left( - C A^{-1} B \right)_{ab}
\: = \: - \sum_{\tilde{\mu}} \frac{ \alpha^a_{\tilde{\mu}}
\alpha^b_{\tilde{\mu}} }{
\sum_c \sigma_c \alpha^c_{\tilde{\mu}} }
\: - \:
\sum_i \frac{ \rho^a_i \rho^b_i }{
\sum_c \sigma_c \rho^c_i - \tilde{m}_i}.
\end{equation}
Curiously, it can be shown that for the superpotential $W_{00}$ computed
in the previous subsection,
\begin{equation}
\frac{\partial^2 W_{00} }{ \partial \sigma_a \partial \sigma_b}
\: = \:
- \sum_{\tilde{\mu}} \frac{ \alpha^a_{\tilde{\mu}}
\alpha^b_{\tilde{\mu}} }{
\sum_c \sigma_c \alpha^c_{\tilde{\mu}} }
\: - \:
\sum_i \frac{ \rho^a_i \rho^b_i }{
\sum_c \sigma_c \rho^c_i - \tilde{m}_i},
\end{equation}
the same as the result above, hence we can identify
\begin{equation}
W_{\rm eff} \: = \: W_{00}
\end{equation}
(up to irrelevant terms annihilated by the second derivative).

Phrased more simply, by more carefully taking into account
all fields and cross-terms, we reproduce the same result for
correlation functions derived in the previous subsection, which
itself matches results in the literature for A model correlation
functions.

In fact, there is a more general statement of this form that can
be made, that for B model correlation functions, sequential
`integrations-out' are equivalent to correlation function computations.  
Consider for simplicity a superpotential $W = W(x,y)$,
a function of two variables.  Assuming isolated critical points,
correlation functions are weighted by a factor of
\begin{eqnarray*}
\det\left[ \begin{array}{cc}
\frac{\partial^2 W}{\partial x^2} & \frac{\partial^2 W}{\partial x \partial y}
\\
\frac{\partial^2 W}{\partial y \partial x} & 
\frac{\partial^2 W}{\partial y^2}
\end{array} \right]
& = &
\left( \frac{\partial^2 W}{\partial x^2} \right)
\left( \frac{\partial^2 W}{\partial y^2} \right)
\: - \:
\left( \frac{\partial^2 W}{\partial x \partial y}\right)^2,
\\
& = &
\left( \frac{\partial^2 W}{\partial x^2} \right)
\left[ \frac{\partial^2 W}{\partial y^2}
\: - \:
\left( \frac{\partial^2 W}{\partial x \partial y}\right)^2
\left( \frac{\partial^2 W}{\partial x^2} \right)^{-1} \right],
\end{eqnarray*}
(mimicking the form of the result in \cite{powell}).
We claim, as an elementary result, that
\begin{equation}
 \frac{\partial^2 W}{\partial y^2}
\: - \:
\left( \frac{\partial^2 W}{\partial x \partial y}\right)^2
\left( \frac{\partial^2 W}{\partial x^2} \right)^{-1}
\: = \:  \frac{\partial^2 W_0}{\partial y^2},
\end{equation}
where $W_0 = W(x_0(y), y)$, for $x_0$ the critical loci of $W$ defined by
\begin{equation}
\left.
\frac{\partial W}{\partial x}
\right|_{x = x_0(y)} \: = \: 0.
\end{equation}
The trivial generalization to multiple variables establishes the
equivalence of the two arguments described in this section.

To demonstrate this, we compute:
\begin{eqnarray*}
\frac{\partial W_0}{\partial y} & = &
\frac{\partial W(x_0,y)}{\partial x_0} \frac{\partial x_0}{\partial y}
\: + \: \frac{\partial W(x_0,y)}{\partial y},
\\
\frac{\partial^2 W_0}{\partial y^2} & = &
\frac{\partial^2 W(x_0,y)}{\partial y^2}
 \: + \: 2 \frac{\partial^2 W}{\partial x_0
\partial y} \frac{\partial x_0}{\partial y} \: + \:
\frac{\partial^2 W}{\partial x_0^2} \left( \frac{\partial x_0}{\partial y}
\right)^2.
\end{eqnarray*}
From the fact that $\partial W(x_0,y)/\partial x_0 = 0$, we have that
\begin{equation}
\frac{\partial}{\partial y} \frac{\partial W(x_0,y)}{\partial x_0}
\: = \:
\frac{\partial^2 W}{\partial x_0^2} \frac{\partial x_0}{\partial y}
\: + \: \frac{\partial^2 W}{\partial x_0 \partial y} \: = \: 0,
\end{equation}
and plugging into the equation above we find
\begin{eqnarray}
\frac{\partial^2 W_0}{\partial y^2} & = &
\frac{\partial^2 W}{\partial y^2} \: + \: 
2 \frac{\partial^2 W}{\partial x_0
\partial y} \left( - \frac{\partial^2 W}{\partial x_0 \partial y} \right)
\left( \frac{\partial^2 W}{\partial x_0^2} \right)^{-1}
\nonumber \\
& & 
\: + \:
\frac{\partial^2 W}{\partial x_0^2} \left(  - \frac{\partial^2 W}{\partial x_0 \partial y} \right)^2
\left( \frac{\partial^2 W}{\partial x_0^2} \right)^{-2},
\\
& = &
\frac{\partial^2 W}{\partial y^2} \: - \:
\left( \frac{\partial^2 W(x_0,y)}{\partial x_0 \partial y} \right)^2
\left( \frac{\partial^2 W(x_0,y)}{\partial x_0^2} \right)^{-1},
\end{eqnarray}
as claimed, establishing the desired equivalence.

In passing, note in the argument above that since $\partial W_0/\partial y
= 0$, since $\partial W(x_0,y)/\partial x_0 = 0$, we also have that
$\partial W(x_0,y)/\partial y = 0$.

\section{Example:  Grassmannian $G(k,n)$}
\label{sect:grassmannian}

For our first example, we will compute the prediction for the mirror to
a Grassmannian.  Other proposals for this case also exist in the literature,
see {\it e.g.} \cite{Hori:2000kt,ehx,r1,r2,r3,Gomis:2012wy},
to which we will compare.

\subsection{Predicted mirror}

Here, the A-twisted gauge theory is a $U(k)$ gauge theory with
$n$ chiral superfields in the fundamental representation.
The resulting GLSM describes the Grassmannian $G(k,n)$ \cite{edver}.

The mirror is predicted to be
an $S_k$-orbifold of a Landau-Ginzburg model with matter fields
$Y_{ia}$ ($i \in \{1, \cdots n\}$, $a \in \{1, \cdots k \}$),
$X_{\mu \nu} = \exp(-Z_{\mu \nu})$, 
$\mu, \nu \in \{1, \cdots, k\}$, and superpotential
\begin{equation}   \label{eq:gkn:mirror-w} 
W \: = \: \sum_a \sigma_a \left( \sum_{i b} \rho^a_{i b} Y_{ib}
\: + \: \sum_{\mu \nu} \alpha^a_{\mu \nu} Z_{\mu \nu} \: - \: t \right)
\: + \:
\sum_{i a} \exp\left( - Y_{ia} \right) \: + \:
\sum_{\mu \neq \nu} X_{\mu \nu} ,
\end{equation}
where\footnote{
Let us illustrate the $\alpha$'s explicitly for the case of $U(3)$.  
Begin by describing the
Cartan subalgebra of $U(3)$ as
\begin{displaymath} 
\left[ \begin{array}{ccc} 
a & 0 & 0 \\ 
0 & b & 0 \\
0 & 0 & c \end{array} \right].
\end{displaymath}
Describe the W bosons as
\begin{displaymath}
\left[ \begin{array}{ccc}
0 & A_{12} & A_{13} \\
A_{21} & 0 & A_{23} \\
A_{31} & A_{32} & 0 
\end{array} \right].
\end{displaymath}
Under a gauge transformation in the Cartan,
\begin{displaymath}
\left[ \begin{array}{ccc}
a^{-1} & 0 & 0 \\
0 & b^{-1} & 0 \\
0 & 0 & c^{-1} \end{array} \right]
\left[ \begin{array}{ccc}
0 & A_{12} & A_{13} \\
A_{21} & 0 & A_{23} \\
A_{31} & A_{32} & 0 
\end{array} \right]
\left[ \begin{array}{ccc}
a & 0 & 0 \\
0 & b & 0 \\
0 & 0 & c \end{array} \right]
\: = \:
\left[ \begin{array}{ccc}
0 & a^{-1} A_{12} b & a^{-1} A_{13} c \\
b^{-1} A_{21} a & 0 & b^{-1} A_{23} c \\
c^{-1} A_{31} a & c^{-1} A_{32} b & 0
\end{array} \right].
\end{displaymath}
Thus, we see that
\begin{displaymath}
\alpha^a_{\mu \nu} = - \delta^a_{\mu} + \delta^a_{\nu}.
\end{displaymath}
}
\begin{displaymath}
\rho^a_{i b} \: = \: \delta^a_b, \: \: \:
\alpha^a_{\mu \nu} \: = \: - \delta^a_{\mu} + \delta^a_{\nu},
\end{displaymath}
$X_{\mu \nu} = \exp(-Z_{\mu \nu})$ is a fundamental field, 
and the $X_{\mu \nu}$, $Z_{\mu \nu}$ need not be (anti)symmetric but are only
defined for $\mu \neq \nu$, as the diagonal entries would correspond to
the elements of the Cartan subalgebra that we use to define constraints
via $\sigma$'s.
Between $X_{\mu \nu}$ and $Z_{\mu \nu}$, $X_{\mu \nu}$ is the
fundamental field, but it will sometimes be convenient to work with its
logarithm, so we retain $Z_{\mu \nu} \equiv - \ln X_{\mu \nu}$.

We orbifold the space of fields $\sigma_a$, $Y_{ia}$, and $Z_{\mu \nu}$ by the
Weyl group of the gauge group.  Now,
the Weyl group of $U(k)$ is the symmetric group on $k$ entries.
It acts on a Cartan torus by permuting $U(1)$ elements.
In the present case, that means the orbifold acts 
by permuting the $\sigma_a$, by making corresponding permutations of
the $Y_{ia}$ (acting on the $a$ index, leaving the $i$ fixed), and 
correspondingly on the $X_{\mu \nu}$ (associated with root vectors).  We will
see concrete examples in the next subsections.

In passing, the Weyl group $S_k$ acts by interchanging fields,
which will leave a holomorphic top-form on the space of
fields $\sigma_a$, $Y_{ia}$, and $X_{\mu \nu}$ invariant up to a sign.
(For example, for (two-dimensional)
Calabi-Yau surfaces $M$, namely $T^4$ and $K3$, $S_k$ leaves invariant the
holomorphic top-form on $M^k$ \cite{hh}.)
As discussed earlier and in \cite{Sharpe:2006qd}, this is sufficient for
the B twist to exist.  A more general orbifold might not be compatible
with the B twist, but as previously discussed, the Weyl orbifold is
always compatible with the B twist.

We begin working
in the untwisted sector of the orbifold.
(Later we will observe that only the untwisted sector is relevant.)
Integrating out the $\sigma_a$, we get constraints
\begin{displaymath}
\sum_i Y_{i a} \: - \: \sum_{\nu \neq a} \left( Z_{a \nu} - Z_{\nu a} \right)
\: - \: t \: = \: 0,
\end{displaymath}
which we use to eliminate $Y_{n a}$:
\begin{displaymath}
Y_{n a} \: = \: - \sum_{i=1}^{n-1} Y_{i a} \: + \:
\sum_{\nu \neq a} \left( Z_{a \nu} - Z_{\nu a} \right) \: + \: t.
\end{displaymath}
Define
\begin{eqnarray}
\Pi_a & = & \exp(-Y_{n a} ), \\
& = &
q \left( \prod_{i=1}^{n-1} \exp(+ Y_{i a} ) \right)
\left( \prod_{\mu \neq a} \frac{ X_{a \mu} }{ X_{\mu a} }
 \right),
\label{eq:gkn:pi-defn}
\end{eqnarray}
for $q = \exp(-t)$,
then the superpotential for the remaining fields, after applying
the constraint, reduces to
\begin{equation}  \label{eq:gkn:final-sup}
W \: = \: \sum_{i=1}^{n-1} \sum_{a=1}^k \exp\left( - Y_{ia} \right)
\: + \: \sum_{\mu \neq \nu} X_{\mu \nu}
\: + \: \sum_{a=1}^k \Pi_a.
\end{equation}

Note that since $\Pi_a$ contains factors of the form $1/X$, the superpotential
above has poles where $X_{\mu \nu} = 0$.  Such structures can arise
after integrating out fields in more nearly conventional Landau-Ginzburg
theories, as we shall review in section~\ref{sect:regularization}, 
and have also appeared in other discussions of two-dimensional
(2,2) supersymmetric theories {\it e.g.}
\cite{Berglund:2016yqo,Berglund:2016nvh,Ghoshal:1993qt,Aharony:1997bx}.
In the next section, we shall argue that physics excludes the
loci where $X_{\mu \nu} = 0$, and so at least insofar as our
semiclassical analysis of the B-twisted theory is concerned, the presence
of poles will not be an issue.

So far we have discussed the untwisted sector of the Weyl orbifold.
However, since none of the critical loci land at fixed points of the
orbifold, we do not expect any {\it e.g.} twisted sector contributions,
and so for the purposes of this paper, we will omit the possibility of
twisted sector contributions.

\subsection{Excluded loci}
\label{sect:gkn:excluded}

We saw in section~\ref{sect:compare-corr-fn} 
that when the $X$ fields are integrated
out, the integration measure is multiplied by a factor 
proportional to 
\begin{equation}
\prod_{\tilde{\mu}} \langle \sigma, \alpha_{\tilde{\mu}} \rangle
\: = \: 
\prod_{\tilde{\mu}} \left( \sum_a \sigma_a \alpha^a_{\tilde{\mu}} 
\right),
\end{equation}
which therefore suppresses contributions from vacua such that any
$\langle \sigma, \alpha_{\tilde{\mu}} \rangle$ vanish.

Thus, points where $\langle \sigma, \alpha_{\mu \nu} \rangle$ vanish,
necessarily do not contribute.  In (A-twisted) gauge theories in two
dimensions, this is a standard and well-known effect:  in Coulomb branch
computations, one must exclude certain loci.  For the case of the
Grassmannian, one excludes the loci where any $\sigma_a$s collide,
corresponding to the same loci discussed here, and also to loci where
there is semiclassically an enhanced nonabelian gauge symmetry.
In supersymmetric localization computations, the excluded loci appear
in the same fashion -- as the vanishing locus of a measure factor in
correlation functions (see {\it e.g.} \cite{Guo:2015caf}[section 2.2]).

Thus, the loci $\{ \langle \sigma, \alpha_{\tilde{\mu}}\rangle = 0 \}$ 
must be excluded.
It remains to understand this excluded locus phenomenon from the perspective
of the theory containing the $X_{\tilde{\mu}}$ fields, 
before they are integrated out.

It turns out that a nearly identical argument applies to the theory
containing $X_{\tilde{\mu}}$ fields.
To understand this fact,
we first need to utilize the operator
mirror map~(\ref{eq:op-mirror-2}), which says
\begin{equation} 
X_{\tilde{\mu}}
 \: =  \:
\sum_a \sigma_a \alpha^a_{\tilde{\mu}}
\: = \: \langle \sigma, \alpha_{\tilde{\mu}} \rangle. 
\end{equation}
Thus, the locus where $\langle \sigma, \alpha_{\tilde{\mu}} \rangle$
vanishes is the same as the locus where $X_{\tilde{\mu}}$ vanishes.

If one tracks through the integration-out, one can begin to
see a purely mechanical
reason for the zeroes of the measure:  the mass of $X_{\tilde{\mu}}$
is proportional to
\begin{equation}
\frac{1}{ \langle \sigma, \alpha_{\tilde{\mu}} \rangle },
\end{equation}
and so the $X_{\tilde{\mu}}$ become infinitely massive at the excluded
loci.  When one computes Hessian factors $H = \partial^2 W$ weighting
critical loci in correlation function computations, it turns out that,
just as the mass becomes infinite, so too does the Hessian, to which
the mass contributes.  (Indeed, it is difficult to see how the Hessian
could fail to become infinite in such cases.)  For example,
in section~\ref{sect:gkn:corr-fns} we will see that the Hessian for the
mirror to $G(2,n)$ has a factor
\begin{equation}
\frac{1}{ \left( \Pi_1 - \Pi_2 \right)^2 },
\end{equation}
where $\Pi_a$ is mirror to $\sigma_a$.  The critical loci where
$X_{\tilde{\mu}}$ vanish correspond in this case to the locus
where $\Pi_1 \rightarrow \Pi_2$, so we see that the Hessian diverges.
Since correlation functions are weighted by factors of $1/H$,
if $H$ diverges, then the critical locus in question cannot contribute
to correlation functions.

More generally, the result above is related by the operator mirror map
to results in supersymmetric localizations for Coulomb branch computations
with $\sigma$'s.  Specifically, we will see later that the operator
mirror map relates $\Pi_a$ to $\sigma_a$, so the Hessian above is
mirror to 
\begin{equation}
H \: \leftrightarrow \:
\frac{1}{ \left( \sigma_1 - \sigma_2 \right)^2} \: = \:
\frac{1}{\langle \sigma, \alpha_{12} \rangle^2},
\end{equation}
hence $1/H$ is mirror to the measure factor
\begin{equation}
\frac{1}{H} \: \leftrightarrow \:
\left( \sigma_1 - \sigma_2 \right)^2 \: = \:
\langle \sigma, \alpha_{12} \rangle^2.
\end{equation}
More generally, the operator mirror map directly relates the
vanishing measure factors to vanishing $1/H$ factors, and so we see
that critical loci along the excluded locus cannot contribute to
correlation functions, in both A-twisted theories of $\sigma$'s as well
as the proposed mirror, for essentially identical reasons in each case.

So far we have established at a mechanical level that critical loci
along the excluded locus (where the $X_{\tilde{\mu}}$ vanish) cannot
contribute.  Next, we shall outline less mechanical reasons in the physics
of the proposed mirror for why this exclusion should take place.
This matter will be somewhat subtle, as we shall see, but nevertheless even
without computing Hessians one can see several issues with the excluded
locus in the mirror theory that would suggest these loci should be excluded.

First,
focusing on the $X$ fields, 
the bosonic potential diverges where any
one $X_{\tilde{\mu}}$ vanishes -- so generically these points are
excluded dynamically.  One has to be slightly careful about higher codimension
loci, however.  
Because the superpotential contains
ratios of the form $X_{\mu \nu}/X_{\nu \mu}$, if multiple $X_{\tilde{\mu}}$
vanish, then the superpotential has $0/0$ factors, which are 
ill-defined (as we shall discuss further in section~\ref{sect:regularization}).
Since the critical loci are defined as the loci where 
$\nabla W$ vanishes, formally the bosonic potential
\begin{equation}
U \: = \: | \nabla W |^2
\end{equation}
also vanishes along critical loci, which appears to say that at higher
codimension loci such as critical loci, the bosonic potential will be 
finite, not infinite, where enough $X_{\tilde{\mu}}$ vanish.
That said, in typical examples in this paper, the critical locus consists
of isolated points, not a continuum, so working just in critical loci
themselves one cannot continuously approach a point where
all $X_{\tilde{\mu}}$ vanish, so one cannot reach such
points through a limit of
critical loci.

The fact that the bosonic potential diverges when any one
$X_{\tilde{\mu}}$ vanishes means that the bosonic potential diverges
generically when $X_{\tilde{\mu}}$ vanish.  As noted above, there are
higher-codimension loci where the superpotential and bosonic potential
are ambiguous.  It is natural to suspect that some regularization of the
quantum field theory 
may effectively 'smooth over' these higher-codimension ambiguities,
so that the quantum field theory sees a continuous (and infinite)
potential.  We will elaborate on this suspicion later in
section~\ref{sect:regularization}.

As this matter is extremely subtle, let us examine it from another perspective.
The superpotential is ambiguous at points where all $X_{\tilde{\mu}}$ vanish,
but we can still consider limits as one approaches such points.
Consider for example the mirror superpotential~(\ref{eq:gkn:final-sup}) for the
case of $G(2,n)$.  The critical locus equations are 
\begin{eqnarray}
\frac{\partial W}{\partial Y_{i1}} & = &
- \exp\left( - Y_{i1} \right) \: + \:
q \left( \prod_{j=1}^{n-1} \exp\left( + Y_{j1} \right) \right) \frac{X_{12}}{
X_{21}},
\\
\frac{\partial W}{\partial Y_{i2}} & = &
- \exp\left( - Y_{i2} \right) \: + \: 
q \left( \prod_{j=1}^{n-1} \exp\left( + Y_{j2} \right) \right) \frac{ X_{21} }{
X_{12} },
\\
\frac{\partial W}{\partial X_{12}} & = &
1 \: + \: 
q \left( \prod_{j=1}^{n-1} \exp\left( + Y_{j1} \right) \right) \frac{1}{X_{21}}
\: - \:
q \left( \prod_{j=1}^{n-1} \exp\left( + Y_{j2} \right) \right) \frac{ X_{21} }{
X_{12}^2 },
\\
\frac{\partial W}{\partial X_{21}} & = &
1 \: - \:
q \left( \prod_{j=1}^{n-1} \exp\left( + Y_{j1} \right) \right) \frac{ X_{12} }{
X_{21}^2 } \: + \:
q \left( \prod_{j=1}^{n-1} \exp\left( + Y_{j2} \right) \right) \frac{1}{
X_{12} }.
\end{eqnarray}
Because of the ratios in the last two equations, the limit of these
equations as one approaches a critical locus point can be a little
subtle.  Assuming that the first two derivatives vanish, we can
rewrite the last two equations in a more convenient form:
\begin{eqnarray}
\frac{\partial W}{\partial X_{12}} & = &
1 \: + \: \frac{ \exp\left( - Y_{i1} \right) - \exp\left( - Y_{i2} \right)
}{ X_{12} },
\\
\frac{\partial W}{\partial X_{21} } & = &
1 \: - \: \frac{  \exp\left( - Y_{i1} \right) - \exp\left( - Y_{i2} \right)
}{ X_{21} },
\end{eqnarray}
for any $i$.
If we are approaching a `typical' critical locus point, not on the
proposed excluded locus, then $X_{12, 21} \neq 0$ and
$\exp(-Y_{i1}) \neq \exp(-Y_{i2})$, so the limits of the derivatives
above are well-defined and vanish unambiguously at the critical point,
consistent with supersymmetry.
Now, consider instead a critical point on the excluded locus,
where $X_{\mu \nu}$ vanish and (as we shall see later from {\it e.g.} the
operator mirror map) $\exp(-Y_{i1}) = \exp(-Y_{i2})$.  Strictly speaking,
the derivatives above are not uniquely defined at this point, as
they have a term of the form $0/0$.  Suppose we approach this point
along a path such that
\begin{equation}
\exp\left( - Y_{i1}\right) - \exp\left( -Y_{i2} \right) \: = \:
\alpha X_{12} \: = \: - \alpha X_{21},
\end{equation}
for some constant $\alpha$.  Then the limit of the derivatives along
this path is easily computed to be
\begin{equation}
\lim \frac{\partial W}{\partial X_{12}} \: = \: 1 + \alpha
\: = \: \lim \frac{\partial W}{\partial X_{21}}.
\end{equation}
For most paths of this form, so long as $\alpha \neq -1$,
the limit of the derivatives is nonzero, and so appears incompatible
with supersymmetry.  This is an artifact of the critical locus in
the proposed excluded locus; for other critical loci, not in the
excluded locus, the limits of these derivatives are well-defined and
vanish.

More globally, understanding these excluded loci is  
one of the motivating factors behind this proposed mirror construction.
After all, a condition such as $\sigma_a \neq \sigma_b$ for $a\neq b$
is an example of an open condition, in the sense that it specifies
an open set, rather than a closed set.  To specify an open condition
in physics would seem to require either an integration measure that
vanishes at the excluded points, or a potential function that excludes
those points.  In effect, both arise here:  the proposed mirror
superpotential describes a bosonic potential that excludes these
points, and if we integrate out the pertinent fields to get a theory
of just $\sigma$'s, then as we have already seen, the result is
an integration measure which vanishes at the points.

In fact, one of the strengths of this proposed mirror is that it gives
a purely algebraic way to determine those excluded loci -- as the points
where the $X_{\mu \nu}$ vanish. Sometimes 
these A model exclusions have been empirical, see for
example \cite{Hori:2016txh}[footnote 4, p. 26], so in such cases, the
analysis here gives one a more systematic means of understanding the
A model excluded loci.

Later in this paper we will check in numerous examples beyond
Grassmannians that the excluded loci predicted in this fashion by
the proposed mirror superpotential, match the excluded loci that are
believed to arise on the A model side.  In fact, in every example
we could find in the literature,
the excluded loci determined in gauge theory Coulomb branch analyses
match those determined by the loci $\{ X_{\tilde{\mu}} = 0 \}$.

\subsection{Check: Number of vacua}
\label{sect:gkn:vacua}

The Euler characteristic of the Grassmannian $G(k,n)$ is
\begin{displaymath}
\left( \begin{array}{c}
n \\ k \end{array} \right).
\end{displaymath}
In this section we will check that the proposed B model
mirror has this number of vacua.

The critical locus of the superpotential~(\ref{eq:gkn:final-sup})
is defined by
\begin{eqnarray}
\frac{\partial W}{\partial Y_{ia}}: & &
\exp\left( - Y_{ia} \right) \: = \: \Pi_a, \\
\frac{\partial W}{\partial X_{\mu \nu}}: & &
X_{\mu \nu}  \: = \: - \Pi_{\mu} + \Pi_{\nu}.
\label{eq:gkn:crit2}
\end{eqnarray}

Plugging into the definition~(\ref{eq:gkn:pi-defn}) of $\Pi_a$, we find
\begin{displaymath}
\Pi_a \: = \: q \left( \frac{1}{\Pi_a} \right)^{n-1}
\left( \prod_{\mu \neq a} \frac{
- \Pi_a + \Pi_{\mu} 
}{
-\Pi_{\mu} + \Pi_a
}
\right)
\: = \:
q (-)^{k-1} (\Pi_a)^{1-n},
\end{displaymath}
hence
\begin{equation}  
(\Pi_a)^n \: = \: (-)^{k-1} q.
\label{eq:gkn:vac:main}
\end{equation}
As discussed in section~\ref{sect:gkn:excluded}, the $X_{\mu \nu}$ do
not vanish, 
which means that the $\Pi_a$
are all distinct.

Since the $\Pi_a$ are distinct, and
from~(\ref{eq:gkn:vac:main}), each is an $n$th root of $(-)^{k-1}q$, 
there are therefore
\begin{displaymath}
n (n-1) \cdots (n-k+1)
\end{displaymath}
different vacua, before taking into account the Weyl group orbifold.

Finally, we need to take into account the $S_k$ orbifold.
The Weyl group orbifold acts by exchanging $Y_{ia}$ with different
values of $a$, hence exchanges different $\Pi_a$.
(It also exchanges the $\sigma_a$'s with one another, and interrelates
the $X_{\mu \nu}$, though for the moment that is less relevant.)
Thus, in the untwisted sector, there are
\begin{displaymath}
\frac{
n (n-1) \cdots (n-k+1)
}{
k!
} \: = \:
\left( \begin{array}{c}
n \\ k \end{array} \right)
\end{displaymath}
critical loci or vacua.

The fixed-point locus of the Weyl orbifold lies along
loci where some of the $\Pi_a$ coincide; since the critical locus
requires all $\Pi_a$ distinct, we see that none of the critical loci
can lie at fixed points of the Weyl orbifold group action.
As a result, we
do not expect any contributions from twisted sectors, as discussed
previously in section~\ref{sect:prop}.  

Thus, we find that the proposed mirror has
\begin{displaymath}
\left( \begin{array}{c}
n \\ k \end{array} \right)
\end{displaymath}
vacua, matching the number of vacua of the original A-twisted GLSM
for $G(k,n)$.

In passing, the details of this computation closely match the
details of the analogous computation in the A-twisted GLSM for $G(k,n)$,
where one counts solutions of $( \sigma_a )^n = (-)^{k-1} q$, subject
to the excluded-locus
constraint that $\sigma_a \neq \sigma_b$ if $a \neq b$.
If one did try to include vacua where the $\Pi_a$ are not distinct,
including vacua on the excluded loci, 
then at minimum
the computation would no longer closely match the A model computation,
and furthermore (modulo the possibility of extra twisted sector vacua
contributing with sufficient signs), it is not at all clear that the resulting
Witten index would necessarily match that of the Grassmannian.

\subsection{Compare B ring to A ring}

Let us first compare against the operator mirror 
maps~(\ref{eq:op-mirror-1}), (\ref{eq:op-mirror-2}).
For the case of the Grassmannian, these operator mirror maps predict
\begin{eqnarray*}
\exp\left( - Y_{ia} \right) & = & \sum_b \sigma_b \rho^a_{ib}
\: = \: \sigma_a, 
\\
X_{\mu \nu}
 & = & 
\sum_a \sigma_a \alpha^a_{\mu \nu} \: = \: - \sigma_{\mu} + \sigma_{\nu}.
\end{eqnarray*}
On the critical locus, we computed
\begin{eqnarray*}
\exp\left( - Y_{ia} \right) & = & \Pi_a, \\
X_{\mu \nu}
 & = & - \Pi_{\mu} + \Pi_{\nu}.
\end{eqnarray*}
Thus, we see the operator mirror map is completely consistent with our
computations for the critical locus, and in particular, the mirror
map identifies
\begin{equation}  \label{eq:gkn:mirror}
\sigma_a \: \leftrightarrow \: \Pi_a.
\end{equation}

The equation~(\ref{eq:gkn:vac:main}) above is the mirror of the A model Coulomb
branch statement  
\begin{equation}  \label{eq:qc:sig}
(\sigma_a)^n \: = \: (-)^{k-1} q,
\end{equation}
which determines the quantum
cohomology ring of the Grassmannian, which is of the form
(see {\it e.g.} \cite{buch1,buch2,tamvakisrev,bertram1,edver,st})
\begin{equation}  \label{eq:qcring}
{\mathbb C}[x_1, \cdots, x_{n-k}] / \langle D_{k+1}, \cdots, D_{n-1},
D_n + (-)^n \tilde{q} \rangle,
\end{equation}
for some constant $\tilde{q} \propto q$,
where 
\begin{displaymath}
D_m \: = \: \det \left( x_{1+j-i} \right)_{1 \leq i, j \leq m},
\end{displaymath}
in conventions in which $x_m=0$ if $m < 0$ or $m > n-k$, and $x_0=1$.

In particular, the chiral ring of this B model theory is finite-dimensional,
and matches that of the A-twisted theory.  Although the superpotential
has poles, in this instance (and for the other theories in this paper),
the chiral ring remains finite.  See for example \cite{Ghoshal:1993qt} for
a discussion of a different Landau-Ginzburg theory with a superpotential
pole, not of the form discussed in this paper, where the chiral ring
is not finite.

For an explicit derivation of the quantum cohomology
ring of the Grassmannian
from the Coulomb branch relations~(\ref{eq:gkn:vac:main}),
see {\it e.g.} 
\cite{Guo:2015caf}[section 3.3].
In fact, at this point we could stop and observe
that the critical loci, defined by the equation above, satisfy
the same form as the critical loci of the one-loop effective twisted
superpotential on the Coulomb branch in the original A-twisted GLSM,
including the orbifold by the Weyl
group action, hence the B model shares the quantum cohomology ring
of the A model.  

Let us briefly outline the idea of how the quantum cohomology ring is
derived from the Coulomb branch relations, sketching
\cite{Guo:2015caf}[section 3.3].  First, we identify each
$x_i$ with a Schur polynomial $s_{\lambda}(\sigma)$ in the variables
$\sigma_1, \cdots, \sigma_k$, associated to a Young tableau $\lambda$ with
$i$ horizontal boxes.  These are symmetric polynomials, invariant
under the (Weyl-)orbifold group.
For example, for $k=2$,
\begin{eqnarray*}
x_1 & = & s_{\tiny\yng(1)}(\sigma) \: = \: \sigma_1 + \sigma_2,
\\
x_2 & = & s_{\tiny\yng(2)}(\sigma) \: = \: 
\sigma_1^2 + \sigma_1 \sigma_2 + \sigma_2^2,
\\
x_3 & = & s_{\tiny\yng(3)}(\sigma) \: = \:
\sigma_1^3 + \sigma_1^2 \sigma_2 + \sigma_1 \sigma_2^2 + \sigma_2^3.
\end{eqnarray*}
Without using the relation~(\ref{eq:qc:sig}), it is straightforward
to verify that $D_m = 0$ for $m > k$, simply as an algebraic consequence
of the expressions
for $x_i$ in terms of $\sigma_a$'s, and this is the origin of most
of the relations in the quantum cohomology ring~(\ref{eq:qcring}) in this
language.

The relation~(\ref{eq:qc:sig}) modifies relations involving $n$th powers
of $\sigma$'s.
For example, for $k=2$, it is straightforward to check that
\begin{equation}
x_4 - x_3 \sigma_1 \: = \: (\sigma_2)^4,
\end{equation}
again using algebraic properties of the expansions in terms of 
$\sigma_a$'s.
Consider the case $n=4$, for which we know $x_3=0$.  The algebraic relation
above then implies $x_4 = (\sigma_2)^4 \propto q$, giving the
desired relation.  Other cases follow similarly.

\subsection{Correlation functions in $G(2,n)$}
\label{sect:gkn:corr-fns}

As another consistency test,
we will now outline correlation function computations in the 
proposed B-twisted Landau-Ginzburg
orbifold mirror to $G(2,n)$ for various values of $n$, 
and compare them to results for correlation functions in the
original A-twisted gauge theory.

Before wading into the details of the computations, it may be helpful
to first very briefly review the analogous computations for the B-twisted
mirror to ${\mathbb P}^n$.  This is a Landau-Ginzburg model 
with superpotential of the form
\begin{equation}
W \: = \: \exp(-Y_1) + \cdots + \exp(-Y_n) + q \prod_{i=1}^n \exp(+ Y_i).
\end{equation}
The critical locus is defined by $\exp(-Y_i)^{n+1} = q$ for all $i$.
Genus zero correlation functions have the form
\begin{equation}
\langle f \rangle \: = \: \sum_{\rm vacua} \frac{f}{H},
\end{equation}
where $H$ is the determinant of the matrix of second derivatives
of the superpotential $W$, and we identify vacua with the
critical locus.  In the present case,
if we define $X = \exp(-Y_i)$, so that the critical locus is
$X^{n+1} = q$,
then $H = \det \partial^2 W = (n+1) X^n$.
Correlation functions then take the form of a sum over $(n+1)$th roots of
unity:
\begin{equation}
\langle X^k \rangle \: \propto \: \sum_{\rm vacua} \frac{
X^k }{ X^n },
\end{equation}
and so will be nonzero if $k = n + m(n+1)$, corresponding to cases
in which the summand is a multiple of $X^{n+1} = q$.  
For other values of $k$, the sum vanishes, as the corresponding sum
over roots of unity vanishes.  This result matches the form of
genus zero A model correlation
functions on ${\mathbb P}^n$, and we will see that computations in
the mirror to $G(2,n)$ have a similar flavor.

Now, let us return to the mirror of $G(2,n)$.  
As stressed previously, since we are computing correlation functions
of untwisted sector operators, and critical loci do not overlap 
orbifold fixed points, the sole effect of the orbifold
will be to multiply the correlation function by a factor of $1/|W|$,
for $|W|$ the order of the Weyl group.

From the superpotential~(\ref{eq:gkn:final-sup}) (after integrating out
$\sigma_1$), 
we have the following derivatives:
\begin{eqnarray*}
\frac{\partial W}{\partial Y_{ia} } & = &
- \exp\left( - Y_{ia} \right) \: + \: \Pi_a
\mbox{   for }i < n, 
\\
\frac{\partial W}{\partial X_{\mu \nu}} & = &
1 \: + \: \frac{\Pi_{\mu} }{X_{\mu \nu} } \: - \:
\frac{\Pi_{\nu}}{ X_{\mu \nu} } \mbox{     for } \mu \neq \nu,
\end{eqnarray*}
\begin{eqnarray*}
\frac{\partial^2 W}{\partial Y_{jb} \partial Y_{ia} } & = &
\delta_{ij} \delta_{ab} \exp\left( - Y_{ia} \right) \: + \:
\delta_{ab} \Pi_a,
\\
\frac{\partial^2 W}{\partial X_{\mu \nu} \partial Y_{ia}} & = &
\delta_{a \mu} \frac{ \Pi_{\mu} }{ X_{\mu \nu} } \: - \:
\delta_{a \nu} \frac{ \Pi_{\nu} }{ X_{\mu \nu} },
\\
\frac{\partial^2 W}{\partial X_{\rho \sigma} \partial X_{\mu \nu} } & = &
\delta_{\mu \rho} \delta_{\nu \sigma} \frac{ - \Pi_{\mu} + \Pi_{\nu} }{
X_{\rho \sigma} X_{\mu \nu} }
\\
& & \: + \: \left( \delta_{\rho \mu} - \delta_{\sigma \mu} \right)
\frac{ \Pi_{\mu} }{ X_{\rho \sigma} X_{\mu \nu} }
\: - \: \left( \delta_{\rho \nu} - \delta_{\sigma \nu} \right)
\frac{\Pi_{\nu} }{X_{\rho \sigma} X_{\mu \nu} }.
\end{eqnarray*}

Clearly, correlation functions will be nontrivial.
Let us now specialize to Grassmannians $G(2,n)$.
It is straightforward to compute\footnote{
Some potentially useful identities can be found in {\it e.g.}
\cite{powell}.
}:
\begin{itemize}
\item for $G(2,3)$, 
\begin{displaymath}
H \: \equiv \: \det\left( \partial^2 W \right) \: = \:
- 9 \frac{ (\Pi_1)^2 ( \Pi_2)^2 }{ (\Pi_1 - \Pi_2)^2 },
\end{displaymath}
\item for $G(2,4)$,
\begin{displaymath}
H \: \equiv \: \det\left( \partial^2 W \right) \: = \:
-16 \frac{ (\Pi_1)^3 (\Pi_2)^3 }{ (\Pi_1 - \Pi_2)^2 },
\end{displaymath}
\item for $G(2,5)$,
\begin{displaymath}
H \: \equiv \: \det\left( \partial^2 W \right) \: = \:
- 25 \frac{ (\Pi_1)^4 ( \Pi_2)^4 }{ (\Pi_1 - \Pi_2)^2 }.
\end{displaymath}
\end{itemize}
All derivatives above are evaluated on the critical locus.
From the results above, we conjecture that for $G(2,n)$ for general $n \geq 3$,
\begin{equation}
H \: \equiv \: \det\left( \partial^2 W \right) \: = \:
- n^2 \frac{ (\Pi_1)^{n-1} (\Pi_2)^{n-1} }{ (\Pi_1 - \Pi_2)^2 }.
\end{equation}

Let us compute correlation functions in the mirrors to $G(2,n)$ for
$3 \leq n \leq 5$, and compare to the correlation functions computed for the
corresponding A-twisted gauge theories in
\cite{Guo:2015caf}.  Note that given the quantum cohomology relations,
once we establish that the classical correlation functions match
(up to an overall scale), all the remaining correlation functions are
guaranteed to match.

Reference \cite{Guo:2015caf} considers correlation functions A-twisted
gauge theories corresponding to 
$G(2,3)$ (see \cite{Guo:2015caf}[section 4.2]), 
$G(2,4)$ (see \cite{Guo:2015caf}[section 4.3]),
and $G(2,5)$ (see \cite{Guo:2015caf}[section 4.4]).
In each case, for $G(2,n)$, the nonzero classical ($q=0$) correlation 
functions are
\begin{equation}
\langle \sigma_1^{n-1} \sigma_2^{n-3} \rangle, \: \: \:
\langle \sigma_1^{n-2} \sigma_2^{n-2} \rangle, \: \: \:
\langle \sigma_1^{n-3} \sigma_2^{n-1} \rangle.
\end{equation}
All other correlation functions of products of $\sigma$'s of degree
$2n-4$ vanish.  The three nonzero classical correlation functions are related as
\begin{equation} \label{eq:g2n:corr-fn-relns}
\langle \sigma_1^{n-1} \sigma_2^{n-3} \rangle
\: = \:
\langle \sigma_1^{n-3} \sigma_2^{n-1} \rangle, \: \: \:
\langle \sigma_1^{n-2} \sigma_2^{n-2} \rangle \: = \:
-2 \langle \sigma_1^{n-1} \sigma_2^{n-3} \rangle
\: = \: 
-2 \langle \sigma_1^{n-3} \sigma_2^{n-1} \rangle,
\end{equation}
so that
\begin{equation} \label{eq:g2n:corr-fn-relns-2}
\langle \left(  \sigma_1^{n-1} \sigma_2^{n-3} \: + \:
 \sigma_1^{n-2} \sigma_2^{n-2} \: + \:
 \sigma_1^{n-3} \sigma_2^{n-1} \right) \rangle \: = \: 0.
\end{equation}
Although the overall normalization is not essential,
in reference~\cite{Guo:2015caf}, we list here the normalized values
in the normalization convention of that
paper:
\begin{equation}
\langle \sigma_1^{n-2} \sigma_2^{n-2} \rangle \: = \: \frac{2}{2!}, \: \: \:
\langle \sigma_1^{n-1} \sigma_2^{n-3} \rangle \: = \: - \frac{1}{2!} \: = \:
\langle \sigma_1^{n-3} \sigma_2^{n-1} \rangle.
\end{equation}
Not only will our mirror's correlation functions have the same
ratios, in fact their normalized values will be identical.

From the operator mirror map~(\ref{eq:gkn:mirror}), in the mirror we should
make corresponding statements about correlation functions of
products of $\Pi_1$ and $\Pi_2$.  As explained earlier, 
correlation functions in the Landau-Ginzburg orbifold mirror to
$G(2,n)$ take the
form
\begin{equation}
\langle \Pi_1^k \Pi_2^{\ell} \rangle \: = \: \frac{1}{2!}
\sum_{\rm vacua} \frac{ \Pi_1^k \Pi_2^{\ell} }{ H }
\: = \:
- \frac{1}{2!} \frac{1}{n^2}
\sum_{\rm vacua} \frac{ (\Pi_1 - \Pi_2)^2 \Pi_1^k \Pi_2^{\ell} }{
\Pi_1^{n-1} \Pi_2^{n-2} }.
\end{equation}
Note that because of the $(\Pi_1 - \Pi_2)$ factors in the numerator,
we no longer need to restrict to vacua described by distinct $\Pi_a$,
since cases in which they coincide do not contribute;
instead, we can replace the sum over vacua with a sum over two sets of
$n$th roots of unity, corresponding (up to scale) with separate solutions for
$\Pi_1$ and $\Pi_2$.

For example, let us compute
\begin{eqnarray*}
\langle \Pi_1^{n-2} \Pi_2^{n-2} \rangle & = &
- \frac{1}{2! n^2} \sum_{\rm vacua}
\frac{ (\Pi_1 - \Pi_2)^2 \Pi_1^{n-2} \Pi_2^{n-2} }{ \Pi_1^{n-1} 
\Pi_2^{n-1} },
\\
 & = &
- \frac{1}{2! n^2} \left( - \frac{1}{q} \right)^2\sum_{\rm vacua}
\Pi_1 \Pi_2  (\Pi_1 - \Pi_2)^2 \Pi_1^{n-2} \Pi_2^{n-2},
\\
& = & - \frac{1}{2! n^2 q^2} \sum_{\rm vacua}
\left( \Pi_1^2 - 2 \Pi_1 \Pi_2 + \Pi_2^2 \right) \Pi_1^{n-1} 
\Pi_2^{n-1}, 
\\
& = & - \frac{1}{2! n^2 q^2} \sum_{\rm vacua}
(-2) ( \Pi_1 \Pi_2 ) \Pi_1^{n-1} 
\Pi_2^{n-1}, 
\\
& = & - \frac{1}{2! n^2 q^2} \sum_{\rm vacua}
(-2) (-q)^2,
\\
& = &  \frac{2 n^2}{2! n^2} \: = \: \frac{2}{2!},
\end{eqnarray*}
where we have used the relations $\Pi_1^n = - q = \Pi_2^n$.
Reasoning similarly, it is straightforward to demonstrate that
\begin{displaymath}
\langle \Pi_1^{n-1} \Pi_2^{n-3} \rangle \: = \:
- \frac{n^2}{2! n^2} \: = \: - \frac{1}{2!} \: = \:
\langle \Pi_1^{n-3} \Pi_2^{n-1} \rangle,
\end{displaymath}
which immediately obey the analogues of the 
relations~(\ref{eq:g2n:corr-fn-relns}), (\ref{eq:g2n:corr-fn-relns-2})
for $\Pi_{1, 2}$ in place of $\sigma_{1,2}$, and in fact even has the
same overall normalization.  Using similar reasoning, it
is also trivial to verify that all other correlation functions of products
of $\Pi$'s of degree $2n-4$ vanish.

Thus, we see that the classical genus zero correlation
functions in the proposed B-twisted mirror to $G(2,n)$ 
match those of the original A-twisted theory, and since the quantum
cohomology relations match, we immediately have that all genus zero
correlation functions match.

\subsection{Regularizing the discontinuity}
\label{sect:regularization}

When we integrate out the $\sigma$ fields, the resulting superpotential
has a somewhat odd form.  For Grassmannians, the
superpotential~(\ref{eq:gkn:final-sup}) contains factors of the
form 
\begin{equation}  \label{eq:gkn:bad-fact} 
\frac{ X_+ }{ X_- }.
\end{equation}
a property shared by most of the mirror superpotentials in this paper.
Ordinarily one does not consider
superpotentials with poles at all, but as previously discussed,
we are interpreting these in an effective field theory sense, much as
\cite{Berglund:2016yqo,Berglund:2016nvh} 
(see also \cite{Ghoshal:1993qt,Aharony:1997bx}), 
which should suffice for an
essentially classical analysis of the B-twisted theory.
To further confuse matters, as we shall review shortly, the toy
superpotential above is discontinuous at $X_+ = 0 = X_-$, making its
analysis potentially more confusing.  
As discussed earlier in section~\ref{sect:gkn:excluded},
we believe that the locus $\{X_- = 0 \}$ (or its analogue in our
actual models) is dynamically excluded, so the matter is somewhat moot,
but is still of interest as a matter of principle.

In this section, we will consider this matter in more detail.
We will discuss potential regularizations of these discontinuities, presumably
corresponding physically to extra degrees of freedom present in the
physical theory but largely irrelevant in the topological field theory.
Unfortunately, we will not develop a fully satisfactory regularization
scheme, but as our B model topological field theory computations seem
to give consistent results, we leave the search for a `correct' regularization
to future work along with proposals for mirrors to physical untwisted QFTs.

Perhaps the first thing to observe is that poles in superpotentials are 
actually easy to generate from manifestly consistent theories.
Consider for example a topological B model describing three
chiral superfields $x_1$, $x_2$, $x_3$, and a polynomial superpotential
\begin{equation}
W \: = \: x_1 x_3 \: + \: x_2^2 x_3^2  \: + \: x_1 \: + \: x_2.
\end{equation}
As the superpotential is polynomial, we do not expect any physical oddities
in this theory.
Since we are working in the B model, we ought to be able to integrate out
the $x_3$ field.  Omitting measure factors discussed earlier,
this results in the superpotential
\begin{equation}
W \: = \: - \frac{ x_1^2 }{4  x_2^2} \: + \: x_1 \: + \: x_2.
\end{equation}
This has a pole where $\{x_2 = 0\}$, reflecting the fact that the
vev of $x_2$ acts as a mass for $x_3$, and so when $x_2$ vanishes,
$x_3$ is a massless field, so integrating it out is problematic.
Thus, in this example, the pole indicates that the low-energy theory is
missing information about fields that have become massless, in the
same spirit as {\it e.g.} the interpretation of conifold singularities
in moduli spaces \cite{Strominger:1995cz}.

Returning to the superpotential factors above, the
factor~(\ref{eq:gkn:bad-fact}) is not a continuous function at the
point $X_+ = 0 = X_-$, simply because the limit is not uniquely determined.
Consider for example a path to the origin of the form $X_+ = t X_-$.
Along that path,
\begin{displaymath}
\lim_{ X_- \rightarrow 0 } \frac{ X_+ }{ X_- }
\: = \: \lim_{ X_- \rightarrow 0 } \frac{ t X_- }{ X_- } \: = \: t,
\end{displaymath}
for any $t$.  As this limit is not independent of path, 
the function $X_+/X_-$ does not have a well-defined limit
at the point $X_+ = 0 = X_-$,
and so is discontinuous at that point.

The reader might now ask how one should make sense of a Landau-Ginzburg
theory with a superpotential that is discontinuous at a point.
First, as a practical matter, in section~\ref{sect:gkn:excluded}
we discussed why
the locus where any $X_{\tilde{\mu}}$ vanishes is dynamically excluded,
so as
a practical matter we do not directly encounter these discontinuities
in our classical analysis of the B model.  That said, there is still
a question of principle concerning how one would understand a QFT with
such a discontinuity.

Briefly, we propose to
interpret this discontinuity as a signal of extra degrees of
freedom, not visible in the B model topological field theory, that would
be important to specify in a physical theory.  Without specifying the
physical mirror, we cannot uniquely specify the missing information,
but we can at least discuss potential regularizations, leaving a thorough
analysis (in a physical theory) for future work.

Mathematically, there is a standard mechanism to deal with a
function such as $X_+/X_-$:  perform a blowup of the space of
$X_+$, $X_-$ at the point $X_+ = 0 = X_-$.  The blowup separates the
divisors of poles and zeroes, leading to a well-defined function.
In more detail, we replace the original coordinate patch
${\mathbb C}[X_+, X_-]$ with two new coordinate patches, call them
$U_A$ and $U_B$:
\begin{equation}
U_A: \: {\mathbb C}[X, z], \: \: \:
U_B: \: {\mathbb C}[z^{-1}, Y],
\end{equation}
where the original coordinates are related to the coordinates on each
patch as follows:
\begin{equation}
(X_+,X_-) \: = \: (X, Xz) \: = \: (Y z^{-1}, Y),
\end{equation}
or
\begin{equation}
X \: = \: X_+, \: \: \: Y \: = \: X_- \: = \: X_+ z .
\end{equation}
In any event, we see
\begin{equation}
\frac{ X_+ }{ X_- } \: = \:
\frac{1}{z}.
\end{equation}
On the blowup, the ratio $X_+/X_-$ is now well-defined:  the divisor of
poles at $\{ X_- = 0\}$ runs through one pole of the $S^2$ in the blowup,
whereas the divisor of zeroes at $\{X_- = 0 \}$ runs through the other
pole of the $S^2$ in the blowup.  The blowup has separated the intersection
point, inserting a ${\mathbb P}^1$ between the two divisors so that they
no longer intersect.

Unfortunately, the blowup of ${\mathbb C}^2$ at the origin,
as described above, is not Calabi-Yau,
and so would not be compatible with the B twist.  On general principles,
we expect the mirror of an A-twistable theory to always be B-twistable,
so whatever extra degrees of freedom arise at those loci
should be compatible with a Calabi-Yau condition.

Although the blowup of ${\mathbb C}^2$ at the origin is not
Calabi-Yau, there is a related construction it may be tempting to
perform instead.
Instead of inserting a ${\mathbb P}^1$, making the geometry locally
look like the total space of ${\cal O}(-1) \rightarrow {\mathbb P}^1$,
we can instead replace the local geometry with the total space of
${\cal O}^{\times}(-1) \rightarrow {\mathbb P}^1$, meaning the line bundle
with the zero section deleted.
Since
\begin{displaymath}
K_{{\mathbb P}^1} \: = \: {\cal O}(-2) \: = \: {\cal O}(-1)^{\otimes 2},
\end{displaymath}
the total space of ${\cal O}^{\times}(-1) \rightarrow {\mathbb P}^1$
is Calabi-Yau \cite{Pantev:2005wj}[appendix A].
We could describe this with local coordinates of the form
\begin{equation}
U_A: \: {\mathbb C}[e^A, z], \: \: \:
U_B: \: {\mathbb C}[z^{-1}, e^B],
\end{equation}
following the same pattern as above, with apparent geodesic incompleteness
cured with suitable kinetic terms that make it dynamically prohibitive
to approach the gaps.  Furthermore, when applied to examples elsewhere in this
text, it is straightforward to check that this procedure has the desired
effect of removing the excluded loci.  Unfortunately, it can be shown that this
alternative blowup is isomorphic to ${\mathbb C}^2 - 0$:  it is not really
a blowup at all, instead merely an omission of the origin of 
${\mathbb C}^2$, and taking as fundamental fields the coordinates used
above, the correlation functions do not come out correctly.

For completeness, note that
we would also need to extend the Weyl orbifold over this blowup.
For the simple toy models above, a Weyl orbifold could only act as
$X_+ \leftrightarrow X_-$, which would be implemented by mapping
$z \leftrightarrow z^{-1}$.

Another approach would be to utilize the orbifold.  After all,
Calabi-Yau resolutions of orbifold singularities are well-known.
However, here again we run into a hitch.  In this toy model, the orbifold
acts as $[{\mathbb C}^2 / S_2]$, where the $S_2$ exchanges the two factors, so
that
\begin{equation}
[{\mathbb C}^2 / S_2] \: = \: {\mathbb C} \times 
[ {\mathbb C} / {\mathbb Z}_2 ],
\end{equation}
where the ${\mathbb Z}_2$ acts by sign flips.
Although this orbifold is B-twistable \cite{Sharpe:2006qd},
it is not Calabi-Yau.  
In theories without superpotentials,
such orbifold have been studied in {\it e.g.} \cite{Adams:2001sv},
where it is described that
the tip of the orbifold flattens dynamically; however, it is not clear
what happens in the present context.  A bit more generally for cases such as
$G(2,n)$, the orbifold takes the form
\begin{equation}
[ {\mathbb C}^{2\ell} / S_2 ] \: = \: {\mathbb C}^{\ell} \times
[ {\mathbb C}^{\ell} / {\mathbb Z}_2 ].
\end{equation}
When $\ell$ is odd, again this is B-twistable but not Calabi-Yau,
so we do not expect a Calabi-Yau resolution.  When $\ell$ is even,
it is Calabi-Yau but Calabi-Yau resolutions may still not exist.
For example, ${\mathbb C}^4/{\mathbb Z}_2$ is an example of a terminal
singularity, which is Calabi-Yau but admits no Calabi-Yau resolution.

Finally, one could imagine deforming the superpotential or the
bosonic potential, doing computations, and then taking a limit as the
deformation is removed.  We have tried this in a few examples, but have
not succeeded in finding a generally useful deformation.  We leave
this approach for future work.

To review, to make the superpotential well-defined, we have tried
several approaches -- blowups of the space, introduction of additional fields
-- none of which was
satisfactory.
If one could find a satisfactory method, satisfying the Calabi-Yau property,
then presumably one could apply the fact that the B model is independent of
K\"ahler moduli to argue that the B model is independent of the blowup.
In such a case, the topological field theory computations in this paper
would be unaffected, but presumably a physical theory would see the
difference.  We leave the correct regularization of these
discontinuities to future work.
In any event, since these discontinuities lie along a dynamically excluded
locus, understanding this issue is not essential for our analysis.

\subsection{Integrating out $X$ fields}
\label{sect:gkn:int-out}

So far we have carried through the analysis including the mirrors
$X_{\mu \nu}$ to the W bosons.  
As an alternative method of analysis of this model, one can for
example integrate out the $X_{\mu \nu}$, and perform computations in
the resulting simpler Landau-Ginzburg model.
We described this formally in section~\ref{sect:compare-corr-fn};
in this section, we will compare results more explicitly for the
case of Grassmannians, and will verify in concrete computations
that the proposal of section~\ref{sect:compare-corr-fn} does indeed
work.

First, in order to be able to sensibly integrate out the $X_{\mu \nu}$,
they must be massive.  It is straightforward to compute from the
superpotential~(\ref{eq:proposal-w:mass}) that for the special case of
$G(k,n)$, on the critical locus, the matrix of second derivatives
takes the form
\begin{equation}
\frac{\partial^2 W }{ \partial X_{\tilde{\mu}} \partial X_{\tilde{\nu}} }
\: = \:
\delta_{\tilde{\mu} \tilde{\nu} }
\frac{1}{-\sigma_{\mu} + \sigma_{\nu}},
\end{equation}
where we have expanded $\tilde{\mu}$ into the $\mu \nu$ index convention
that we have used for $G(k,n)$.
Furthermore, for this case,
from the constraints discussed in section~\ref{sect:gkn:vacua}, 
we know that the $\sigma_a$ are
distinct, hence the $X_{\tilde{\mu}} = X_{\mu \nu}$ 
are always massive, so it is consistent
to integrate them out.
(If the matrix of second derivatives were to have a zero
eigenvalue somewhere, integrating out the $X_{\tilde{\mu}}$ would,
of course, not be consistent.)

As discussed in section~\ref{sect:compare-corr-fn},
one effect of integrating out the $X_{\tilde{\mu}} = X_{\mu \nu}$
should be to change the superpotential~(\ref{eq:proposal-w:mass}) 
to the form~(\ref{eq:proposal-int-out}), which in the present case
is
\begin{eqnarray}
W_0 & = & \sum_{a=1}^r \sigma_a \left( 
\sum_{i=1}  Y_{ia} \: - \: \sum_{\mu \neq \nu} \left(- \delta^a_{\mu} + \delta^a_{\nu}
\right) 
\ln \left( - \sigma_{\mu} + \sigma_{\nu} \right)
\: - \: t \right) 
\nonumber \\
& & \: + \: \sum_{i,a} \exp\left(-Y_{ia}\right) 
\: + \: 
\sum_{\mu \neq \nu} \left( - \sigma_{\mu} + \sigma_{\nu} \right), 
\\
& = &
\sum_{a=1}^r \sigma_a \left(  
\sum_{i=1}^n  Y_{ia} \: - \: ( t - (k-1) \pi i) \right) 
\: + \: \sum_{i,a} \exp\left(-Y_{ia}\right) ,
\label{eq:gkn:int-out:sup1}
\end{eqnarray}
(up to constant terms we have omitted),
where we have used the identities
\begin{eqnarray*}
\sum_{\mu \neq \nu}  \left( - \sigma_{\mu} + \sigma_{\nu} \right)
& = & 0,
\\ 
\sum_{\mu \neq \nu} \left(- \delta^a_{\mu} + \delta^a_{\nu}
\right) 
\ln \left( - \sigma_{\mu} + \sigma_{\nu} \right)
& = & \sum_{\nu \neq a} \ln \left( \frac{ - \sigma_{\nu} + \sigma_a }{
- \sigma_a + \sigma_{\nu} } \right),
\\
& = &
\sum_{\nu \neq a} \ln(-) \: = \: (k-1)\pi i.
\end{eqnarray*}
The other effect of integrating out the $X_{\mu \nu}$ is to
add a factor of 
\begin{displaymath}
\overline{H}_X H_X^g/(H_X \overline{H}_X) \: = \: 1/H_X^{1-g}
\end{displaymath}
to correlation functions (at genus $g$), which for $G(k,n)$ take
the form
\begin{equation}
\langle {\cal O} \rangle \: = \: \frac{1}{|W|}
\sum_{\rm vacua} \frac{{\cal O}}{ (\det \partial^2 W_0)^{1-g}} 
\left(  \prod_{\mu < \nu}
\left( \sigma_{\mu} - \sigma_{\nu} \right)^2  \right)^{1-g}.
\end{equation}

Now, let us compare the formal results we have outlined above from
section~\ref{sect:compare-corr-fn} to concrete results obtained
by actually computing the Hessians of the Landau-Ginzburg models
for the mirror of $G(k,n)$ both with and without the $X_{\tilde{\mu}} = 
X_{\mu \nu}$.  To that end, in equation~(\ref{eq:gkn:int-out:sup1}), 
if we integrate out the $\sigma_a$, we get the constraint
\begin{displaymath}
\sum_{i=1}^n  Y_{ia} \: = \: \tilde{t},
\end{displaymath}
for 
\begin{equation}
\tilde{t} \: = \: t \: - \: (k-1) \pi i,
\end{equation}
which we can use to eliminate $Y_{na}$:
\begin{equation}
Y_{na} \: = \: \tilde{t} \: - \: \sum_{i=1}^{n-1} Y_{ia}.
\end{equation}
Plugging back into the superpotential~(\ref{eq:gkn:int-out:sup1}),
we get
\begin{equation}
W_0 \: = \:
\sum_{i=1}^{n-1} \sum_a \exp\left( - Y_{ia} \right)
\: + \: \sum_a \tilde{q} \prod_{i=1}^{n-1} \exp\left( + Y_{ia} \right),
\label{eq:gkn:int-out:sup2}
\end{equation}
for 
\begin{equation}
\tilde{q} \: = \: \exp\left( - \tilde{t} \right) \: = \:
(-)^{k-1} \exp\left( - t \right) \: = \:
(-)^{k-1} q.
\end{equation}
With an eye towards correlation functions, we compute
\begin{eqnarray*}
\frac{\partial W_0}{\partial Y_{ia} } & = &
- \exp\left( - Y_{ia} \right) \: + \: \tilde{q} \prod_{i=1}^{n-1} 
\exp\left( + Y_{ja} \right),
\\
\frac{\partial^2 W_0}{\partial Y_{ia} \partial Y_{jb} } & = &
+ \delta_{ij} \delta_{ab} \exp\left( - Y_{ia} \right) \: + \:
\delta_{ab} \tilde{q} \prod_{k=1}^{n-1} \exp\left( + Y_{ka} \right).
\end{eqnarray*}
The operator mirror map derived from~(\ref{eq:gkn:int-out:sup1}) is simply
\begin{equation}
\exp\left( - Y_{ia} \right) \: \leftrightarrow \: \sigma_a,
\end{equation}
for $i \in \{1, \cdots, n\}$.  Self-consistency for $i=n$ requires
that
\begin{displaymath}
\exp\left( - Y_{na} \right) \: = \: \tilde{q} \prod_{i=1}^{n-1} \exp\left(
+ Y_{ia} \right)
\end{displaymath}
match $\exp\left( - Y_{ia} \right)$ on the critical locus
for $i < n$, which implies
\begin{equation}
\left( \sigma_a \right)^n \: = \: \tilde{q}.
\end{equation}

In the $G(2,n)$ example, the determinant of the matrix
of second derivatives of the superpotential~(\ref{eq:gkn:int-out:sup2})
with respect to the $Y_{ia}$ ($i < n$) can be computed to be
\begin{equation}
H \: = \: 16 \sigma_1^3 \sigma_2^3,
\end{equation}
whereas the determinant of the matrix of second derivatives with respect
to both the $Y_{ia}$ and the $X_{\mu \nu}$ that we computed previously
in section~\ref{sect:gkn:corr-fns} is
\begin{equation}
H' \: = \: - 16 \frac{ \sigma_1^3 \sigma_2^3 }{ (\sigma_1 - \sigma_2)^2}.
\end{equation}
Up to a sign, the only difference between $H$ (after integrating out
$X_{\mu \nu}$) and $H'$ (before integrating out $X_{\mu \nu}$) is
a factor of $(\sigma_1 - \sigma_2)^2$, which the formal argument above
explains perfectly.
Up to that sign, correlation function computations would be identical:
if we integrate out the $X_{\mu \nu}$, then from the formal argument above,
for isolated critical points and worldsheet genus zero,
\begin{equation}
\langle f \rangle \: = \: \sum_{\rm vacua} \frac{f}{H} (\sigma_1 - \sigma_2)^2,
\end{equation}
whereas if we do not integrate out the $X_{\mu \nu}$ and take the
complete matrix of second derivatives, as we did earlier in 
section~\ref{sect:gkn:corr-fns}, then
\begin{equation}
\langle f \rangle \: = \: \sum_{\rm vacua} \frac{f}{H'},
\end{equation}
which matches (up to a sign).
Thus, we see that the formal results in section~\ref{sect:compare-corr-fn}
agree with concrete computations.

\subsection{$G(k,n) \cong G(n-k,n)$}
\label{sect:gkn:duality}

Mathematically, the Grassmannian $G(k,n)$ is identical to the
Grassmannian $G(n-k,n)$.  Physically, one is described by a two-dimensional
$U(k)$ gauge theory, the other by a two-dimensional $U(n-k)$ gauge
theory, which in the UV are different but
are related by a two-dimensional analogue of Seiberg
duality.

In this section we will discuss in what sense the mirrors proposed
here are compatible with the duality relating A-twisted sigma models
on the isomorphic spaces $G(k,n)$ and
$G(n-k,n)$.

Our proposed mirror to $G(k,n)$ is an $S_k$ orbifold of a
theory with $k(n+k)$ fields:  $k$ $\sigma_a$s, $kn$ $Y_{ia}$s, and
$k^2-k$ $Z_{\mu}$'s.  On the other hand, our proposed mirror to
$G(n-k,n)$ is an $S_{n-k}$ orbifold of a theory with
$(n-k)(2n-k)$ fields:  $n-k$ $\sigma_a$s,
$n(n-k)$ $Y_{ia}$s, and $(n-k)^2-(n-k)$ $Z_{\mu}$s.
In the UV, these are two very different theories.

Nevertheless, despite the fact that the orbifold groups are different
and the dimension of the space of matter fields is different, these
two B-twisted theories have the same critical loci,
since
\begin{displaymath}
\left( \begin{array}{c} n \\ k \end{array} \right) \: = \:
\left( \begin{array}{c} n \\ n-k \end{array} \right),
\end{displaymath}
and isomorphic
chiral rings (since each matches the chiral ring of the corresponding
A model, and $G(k,n) \cong G(n-k,n)$).  We have only checked that
B model correlation functions match A model correlation functions in
the $G(2,n)$ families, but if our conjecture is correct, then again since
$G(k,n) \cong G(n-k,n)$, the correlation functions  
in the two mirrors also match.

In short, these two theories, as B-twisted theories, appear to be
equivalent in the IR, just as the A-twisted theories to which
they are mirror also are equivalent in the IR.

We propose that these two theories, mirror to dual A-models,
are themselves Seiberg dual to one another, given that they appear to have
the same (B-twisted) IR limits, and in this sense, the mirror proposed
in this paper is compatible with this duality.

This particular realization of duality, this Seiberg duality for Toda duals,
is not shared by all other proposed mirrors.  We shall see in
section~\ref{sect:compare-mirrors} that the constructions of
Rietsch \cite{r1,r2,r3} have the property that the duality
$G(k,n) \cong G(n-k,n)$ is realized more simply on the mirror -- the two
Grassmannians have equivalent mirrors.

\subsection{Comparison to Rietsch's mirrors}
\label{sect:compare-mirrors}

Although there has not existed previously, 
to our knowledge, a proposal for mirrors for
general nonabelian GLSMs, there certainly exist proposals for
mirrors to Grassmannians, see {\it e.g.} \cite{ehx,r1,r2,r3} for
various formulations of one such.

Briefly, for any Grassmannian $G(k,n)$, the mirror in these proposals
can be described as a Landau-Ginzburg model on $G(n-k,n)$, with a 
superpotential with poles.  We outline details in a few
examples in appendix~\ref{app:rietsch}.
The resulting Landau-Ginzburg mirrors are explicitly invariant
under $k \leftrightarrow n-k$:  the superpotential and underlying
spaces are explicitly isomorphic.  By contrast, 
only the low-energy physics of
our proposed Landau-Ginzburg mirrors is invariant
under the duality $G(k,n) \leftrightarrow G(n-k,n)$, 
as discussed
previously in
section~\ref{sect:gkn:duality}.

As a result, the $G(k,n)$ mirror proposals of \cite{ehx,r1,r2,r3}
have a different form than the one presented here.  Nevertheless,
we are under the impression that 
they are known to also match physics, at least
in the sense that the quantum cohomology ring of the A-twisted theory
can be derived from the ring of functions on the critical loci of the
B-twisted Landau-Ginzburg model (see {\it e.g}
\cite{r2}[remark 8.10]).   

Briefly, we conjecture that the mirror to $G(k,n)$
presented in \cite{ehx,r1,r2,r3} is Seiberg-dual to the mirror presented here,
in the sense that they share the same number of critical loci and
Hessians at those loci, so that correlation functions and
quantum cohomology rings match.

\subsection{Comparison to Hori-Vafa's proposed mirror}

In \cite{Hori:2000kt}[appendix A], there is another short proposal
for a mirror.  Briefly, it is similar to ours, in that it has
mirrors $Y_{ia}$ to the matter fields, has a 
similar action, but crucially omits our $X_{\tilde{\mu}}$ fields,
and also speaks of taking Weyl-invariant fields
instead of a Weyl orbifold.  Furthermore, the proposal there
inserts
factors of
\begin{equation}
\prod_{a < b} (\sigma_a - \sigma_b)
\end{equation}
into period integrals such as \cite{Hori:2000kt}[equ'n (A.3)].
(For a mathematical treatment of the same proposal, see
{\it e.g.} \cite{bcfk}.)

We observe that in our proposed mirror, analogous factors arise by virtue of 
the $X_{\tilde{\mu}}$ fields.  For example, in computing correlation functions
in the mirror to $G(2,n)$ in section~\ref{sect:gkn:corr-fns}, 
we derived a factor of
\begin{displaymath}
\Pi_1 - \Pi_2,
\end{displaymath}
mirror to $\sigma_1-\sigma_2$.  This factor arose from the determinant of
matrix of second derivatives of the superpotential, and in fact can be
traced directly to the second derivatives with respect
to $X_{\tilde{\mu}}$.  In effect, in our mirror, the W bosons are
directly responsible for factors such as the one above.  Furthermore,
we saw such factors arise more explicitly when integrating out the
$X_{\tilde{\mu}}$ fields in section~\ref{sect:gkn:int-out}.

Phrased more simply, our proposal could be understood as a proposed
UV completion of the proposal in \cite{Hori:2000kt}[appendix A],
explicitly describing a set of fields (the $X_{\tilde{\mu}}$ fields),
together with an orbifold, that give a higher-energy description of the
integration measure factors that \cite{Hori:2000kt}[appendix A] utilize.

\subsection{Comparison to Gomis-Lee's proposed mirror}

In \cite{Gomis:2012wy}[section 4], 
a proposal was made for the mirror of the
the $U(k)$ gauge theory corresponding to
the Grassmannian $G(k,n)$, in the form of the $S^2$ partition function of
a Landau-Ginzburg model with a nontrivial measure factor.  
We discussed this partition function in
section~\ref{sect:justification}.  Briefly, the $S^2$ partition function
discussed there is in agreement with what we would expect from the
theory described here.  For example,
their Landau-Ginzburg theory does not explicitly include
the $X$ fields (associated with W bosons in the mirror), 
but it does have extra factors of $\sigma_a - \sigma_b$
in the partition function, which is what we would expect to see after 
integrating out the mirror W bosons $X_{\mu \nu}$ of this paper,
as discussed in section~\ref{sect:compare-corr-fn}.
Similarly, their proposed $S^2$ partition function has a
multiplicative factor of $1/|W|$, for $W$ the Weyl group, which is
consistent with the $S^2$ partition function of a Weyl orbifold.

In principle, \cite{Gomis:2012wy} only proposed a mirror for
$U(k)$ gauge theories, but analogous analyses are straightforward
for other Lie groups.  For example, \cite{Benini:2016qnm}[section 5.1]
rewrites the physical partition function of a GLSM in the form of
the partition function of a Landau-Ginzburg model, with an extra
measure factor that, following our section~\ref{sect:compare-corr-fn}, 
we can identify with
integrated-out $X$ fields.  To summarize, after integrating
out the fields $X_{\tilde{\mu}}$, our propoposal reduces to
a Landau-Ginzburg
model with modified integration measure of exactly the same form as predicted
for mirrors in \cite{Gomis:2012wy}[section 4] and
\cite{Benini:2016qnm}[section 5.1].

\section{Example:  Two-step flag manifold}
\label{sect:two-step}

In principle, the same proposal also applies to flag manifolds.
To be concrete, let us work out the proposed mirror to a two-step flag
manifold, and check that it describes the correct number of vacua.

Consider the two-step flag manifold $F(k_1,k_2,n)$, $k_1 < k_2 < n$,
which is
described in GLSMs as \cite{Donagi:2007hi} as
a $U(k_1) \times U(k_2)$ gauge theory with
\begin{itemize}
\item one set of chiral superfields in the $({\bf k_1}, {\bf \overline{k}_2})$
bifundamental representation,
\item $n$ chiral superfields in the $({\bf 1},{\bf k}_2)$ representation.
\end{itemize}
(In other words, a representation of a quiver.)

Following our proposal, the mirror is an orbifold of a Landau-Ginzburg
model with fields
\begin{itemize}
\item $Y_a^{\alpha}$, $a \in \{1, \cdots, k_1\}$, $\alpha \in \{1, \cdots,
k_2\}$, corresponding to the bifundamentals,
\item $\tilde{Y}_{i \alpha}$, $i \in \{1, \cdots, n\}$, corresponding
to the second set of matter fields,
\item $X_{\mu \nu} = \exp(-Z_{\mu \nu})$, 
$\mu, \nu \in \{1, \cdots k_1\}$, corresponding
to the W bosons from the $U(k_1)$,
\item $\tilde{X}_{\mu', \nu'} = \exp(-\tilde{Z}_{\mu' \nu'})$, 
$\mu', \nu' \in \{1, \cdots, k_2\}$,
corresponding to the W bosons from the $U(k_2)$,
\item $\sigma_h$, $\tilde{\sigma}_{h'}$, 
$h \in \{1, \cdots,
k_1\}$, $h' \in \{1, \cdots, k_2\}$,
\end{itemize}
and superpotential
\begin{eqnarray}
W & = &
\sum_{h=1}^{k_1} \sigma_h \left(
\sum_{a, \alpha} \rho_{a \alpha}^h Y_{a}^{\alpha} \: + \:
\sum_{\mu, \nu} \alpha^h_{\mu \nu} Z_{\mu \nu} \: - \: t \right)
\nonumber \\
& & \: + \:
\sum_{h'=1}^{k_2} \tilde{\sigma}_{h'} \left(
\sum_{a, \beta} \rho^{h'}_{a \beta} Y_{a}^{\beta} \: + \:
\sum_{i=1}^n \rho^{h'}_{i \alpha} \tilde{Y}_{i \alpha} \: + \:
\sum_{\mu', \nu'} \alpha^{h'}_{\mu' \nu'} \tilde{Z}_{\mu' \nu'}
\: - \: \tilde{t}
\right)
\nonumber \\
& & \: + \:
\sum_{a, \alpha} \exp\left( - Y_{a}^{\alpha} \right) \: + \:
\sum_{i, \alpha} \exp\left( - \tilde{Y}_{i \alpha} \right) \: + \:
\sum_{\mu, \nu} X_{\mu \nu} 
\: + \:
\sum_{\mu', \nu'} \tilde{X}_{\mu', \nu'}.
\end{eqnarray}
In the expression above,
\begin{equation}
\rho^h_{a \alpha} \: = \: \delta^h_a, \: \: \:
\rho^{h'}_{a \alpha} \: = \: - \delta^{h'}_{\alpha}, \: \: \:
\rho^{h'}_{i \alpha} \: = \: \delta^{h'}_{\alpha}, \: \: \:
\alpha^h_{\mu \nu} \: = \: - \delta^h_{\mu} + \delta^h_{\nu}, \: \: \:
\alpha^{h'}_{\mu' \nu'} \: = \: - \delta^{h'}_{\mu'} + \delta^{h'}_{\nu'},
\end{equation}
so we can rewrite the superpotential as
\begin{eqnarray}
W & = &
\sum_{h=1}^{k_1} \sigma_h \left(
\sum_{\alpha} Y^{\alpha}_h \: - \: \sum_{\nu \neq h} (Z_{h \nu} - Z_{\nu h})
\: - \: t \right)
\nonumber \\
& & \: + \:
\sum_{h'=1}^{k_2} \tilde{\sigma}_{h'} \left(
- \sum_{a} Y^{h'}_a \: + \:
\sum_i \tilde{Y}_{i h'} \: - \:
\sum_{\nu' \neq h'}( \tilde{Z}_{h' \nu'} - \tilde{Z}_{\nu' h'} )
\: - \: \tilde{t} \right)
\nonumber \\
& & \: + \:
\sum_{a, \alpha} \exp\left( - Y_{a}^{\alpha} \right) \: + \:
\sum_{i, \alpha} \exp\left( - \tilde{Y}_{i \alpha} \right) \: + \:
\sum_{\mu, \nu} X_{\mu \nu}
\: + \:
\sum_{\mu', \nu'} \tilde{X}_{\mu' \nu'} .
\end{eqnarray}

For reasons previously discussed, we focus on the untwisted sector of
the Weyl group orbifold.
Integrating out $\sigma_h$, $\tilde{\sigma}_{h'}$, we get the constraints
\begin{eqnarray}
\sum_{\alpha} Y^{\alpha}_h \: - \: \sum_{\nu \neq h} (Z_{h \nu} - Z_{\nu h})
& = & t, 
\\
- \sum_{a} Y^{h'}_a \: + \:
\sum_i \tilde{Y}_{i h'} \: - \:
\sum_{\nu' \neq h'}\left( \tilde{Z}_{h' \nu'} - \tilde{Z}_{\nu' h'} \right)
& = & \tilde{t},
\end{eqnarray}
which we can solve as
\begin{eqnarray}
Y^{k_2}_h & = & - \sum_{\alpha=1}^{k_2-1} Y^{\alpha}_h \: + \:
\sum_{\nu \neq h} (Z_{h \nu} - Z_{\nu h}) \: + \: t,
\\
\tilde{Y}_{n k_2} & = &
- \sum_{i=1}^{n-1} \tilde{Y}_{i k_2} \: + \:
\sum_{\nu' \neq k_2} \left( \tilde{Z}_{k_2 \nu'} - 
\tilde{Z}_{\nu' k_2} \right) \: + \: \tilde{t}
\nonumber \\
& &
\: + \: \sum_{a=1}^{k_1} \left[
- \sum_{\alpha=1}^{k_2-1} Y_a^{\alpha} \: + \:
\sum_{\nu \neq a} \left( Z_{a \nu} - Z_{\nu a} \right) 
\: + \: t \right]
\end{eqnarray}
and for $h' < k_2$,
\begin{eqnarray}
\tilde{Y}_{n h'} & = &
- \sum_{i=1}^{n-1} \tilde{Y}_{i h'} \: + \:
\sum_{a=1}^{k_1} Y^{h'}_a \: + \:
\sum_{\nu' \neq h'} \left( \tilde{Z}_{h' \nu'} - 
\tilde{Z}_{\nu' h'} \right) \: + \: \tilde{t}.
\end{eqnarray}

For later use, define
\begin{eqnarray}
\Pi_a & = & \exp\left( - Y_a^{k_2} \right), \\
& = & q \left( \prod_{\alpha=1}^{k_2-1} \exp\left( + Y_a^{\alpha}
\right) \right) \left( \prod_{\nu \neq a} 
\frac{X_{a \nu}}{ X_{\nu a} }
\right),
\\
\Gamma_{\alpha} & = & \exp\left( - \tilde{Y}_{n \alpha} \right)
\mbox{  for } \alpha < k_2, \\
& = & 
 \tilde{q}
\left( \prod_{i=1}^{n-1} \exp\left( + \tilde{Y}_{i \alpha} \right) \right)
\left( \prod_{a=1}^{k_1} \exp\left( - Y_a^{\alpha} \right) \right)
\left( \prod_{\nu' \neq \alpha} 
\frac{ \tilde{X}_{\alpha \nu'} }{ \tilde{X}_{\nu' \alpha} }
\right),
\\
T & = & \exp\left( - \tilde{Y}_{n k_2} \right), \\
& = &
\tilde{q}
q^{k_1}
\left( \prod_{i=1}^{n-1} \exp\left( + \tilde{Y}_{i k_2} \right) \right)
\left( \prod_{\nu' \neq k_2} 
\frac{ \tilde{X}_{k_2 \nu'} }{ \tilde{X}_{\nu' k_2} }
\right)
\cdot
\nonumber \\
& & \hspace*{1.0in} \cdot
\left( \prod_{a=1}^{k_1} \prod_{\alpha=1}^{k_2-1}
\exp\left( + Y_a^{\alpha} \right) \right)
\left( \prod_{a=1}^{k_1} \prod_{\nu \neq a}
\frac{ X_{\nu a} }{ X_{a \nu} }
\right),
\end{eqnarray}
for $q = \exp(-t)$, $\tilde{q} = \exp(-\tilde{t})$.

The superpotential then becomes
\begin{eqnarray}
W & = &
\sum_{a=1}^{k_1} \sum_{\alpha=1}^{k_2-1} \exp(-Y_a^{\alpha})
\: + \: 
\sum_{i=1}^{n-1} \sum_{\alpha=1}^{k_2} \exp\left( - \tilde{Y}_{i \alpha}
\right) \: + \:
\sum_{\mu, \nu} X_{\mu \nu}
\: + \:
\sum_{\mu', \nu'} \tilde{X}_{\mu' \nu'}
\nonumber \\
& & \: + \:
\sum_{a=1}^{k_1} \Pi_a 
\: + \:
\sum_{\alpha=1}^{k_2-1} \Gamma_{\alpha}
\: + \:
T.
\end{eqnarray}

We compute the critical locus as follows:
\begin{eqnarray*}
\frac{\partial W}{\partial Y_a^{\alpha}}: & &
 \exp\left( - Y_a^{\alpha}\right) \: = \: \Pi_a \: - \:
\Gamma_{\alpha} \: + \: T ,
\\
\frac{\partial W}{\partial \tilde{Y}_{i \alpha} }: & &
 \exp\left(- \tilde{Y}_{i \alpha} \right)
\: = \: \left\{ \begin{array}{cl}
\Gamma_{\alpha}, & \alpha \neq k_2, \\
 T , & \alpha=k_2,
\end{array} \right.
\\
\frac{\partial W}{\partial X_{ab}}: & &
X_{ab}  \: = \: - \Pi_a + \Pi_b , 
\\
\frac{\partial W}{\partial \tilde{X}_{\alpha \beta} }: & &
\tilde{X}_{\alpha \beta}  \: = \:
\left\{ \begin{array}{cl}
- \Gamma_{\alpha} + \Gamma_{\beta}, 
& \alpha \neq k_2, \beta \neq k_2, \\
 + \Gamma_{\beta} - T,
& \alpha = k_2, \beta \neq k_2, \\
- \Gamma_{\alpha}  + T,
& \alpha \neq k_2, \beta = k_2.
\end{array} \right.
\end{eqnarray*}

As discussed earlier in section~\ref{sect:gkn:excluded},
we must require that the
$X_{ab}$ and $\tilde{X}_{\alpha \beta}$ be nonzero.  Also using the
fact that $\exp(-Y) \neq 0$, we have
\begin{eqnarray}
\Pi_a & \neq & \Gamma_{\alpha} - T, 
\\
\Gamma_{\alpha} & \neq & 0, 
\\
T & \neq & 0,
\\
\Pi_a & \neq & \Pi_b \: \mbox{  for }a \neq b,
\\
\Gamma_{\alpha} & \neq & \Gamma_{\beta} \: \mbox{  for }\alpha \neq \beta,
\\
\Gamma_{\alpha} & \neq & T.
\end{eqnarray}
These guarantee that the critical locus does not intersect the
fixed point locus of the Weyl orbifold.

On the critical locus, from the definitions we then find
\begin{eqnarray}
\Pi_a & = & 
q (-)^{k_1-1} \left( \prod_{\alpha=1}^{k_2-1}
\frac{1}{\Pi_a - \Gamma_{\alpha} + T} \right),
\label{eq:flag:pi-result}
\\
(\Gamma_{\alpha})^n & = & \tilde{q} (-)^{k_2-1}
\left( \prod_{a=1}^{k_1} (\Pi_a - \Gamma_{\alpha} + T) \right),
\label{eq:flag:gamma-result}
\\
T^n & = & 
\tilde{q} 
q^{k_1}
(-)^{k_2-1 + k_1(k_1-1)} \left(
\prod_{a=1}^{k_1} \prod_{\alpha=1}^{k_2-1} \frac{1}{\Pi_a - \Gamma_{\alpha}
+ T} \right),
\\
& = &
\tilde{q} (-)^{k_2-1}
\left( \prod_{a=1}^{k_1} \Pi_a \right),
\label{eq:flag:t-result}
\end{eqnarray}
where the simplification in~(\ref{eq:flag:t-result}) was derived
using~(\ref{eq:flag:pi-result}).

It is useful to compare to the operator mirror map.
From equations~(\ref{eq:op-mirror-1}), (\ref{eq:op-mirror-2}), we
expect that the A and B model variables should be related as
\begin{eqnarray}
\exp\left( - Y_a^{\alpha} \right) & = & \sigma_a - \tilde{\sigma}_{\alpha},
\\
\exp\left( - \tilde{Y}_{i \alpha} \right) & = &
\tilde{\sigma}_{\alpha}, 
\\
X_{\mu \nu}  & = &
- \sigma_{\mu} + \sigma_{\nu},
\\
\tilde{X}_{\mu' \nu'}  & = &
- \tilde{\sigma}_{\mu'} + \tilde{\sigma}_{\nu'}.
\end{eqnarray}
These relations are consistent with the identities derived for the critical
locus above if we identify
\begin{eqnarray}
\sigma_a & = & \Pi_a + T, 
\\
\tilde{\sigma}_{\alpha} & = & \left\{ \begin{array}{cl}
\Gamma_{\alpha} & \alpha < k_2, \\
T & \alpha = k_2.
\end{array} \right.
\end{eqnarray}

Furthermore, applying the critical locus results~(\ref{eq:flag:pi-result}),
(\ref{eq:flag:gamma-result}), (\ref{eq:flag:t-result}),
we see that
\begin{eqnarray}
\prod_{\alpha=1}^{k_2} 
\left( \sigma_a - \tilde{\sigma}_{\alpha} \right)
& = & \Pi_a \prod_{\alpha=1}^{k_2-1} \left( \Pi_{\alpha} - \Gamma_{\alpha}
+ T \right), \\
& = & (-)^{k_1-1} q,
\\
\left( \tilde{\sigma}_{\alpha} \right)^n & = &
\left( \Gamma_{\alpha} \right)^n \mbox{  for } \alpha < k_2, \\
& = &
(-)^{k_2-1} \tilde{q} \left( \prod_{a=1}^{k_1} \left( \Pi_a - \Gamma_{\alpha} + T
\right) \right) \: = \:
(-)^{k_2-1} \tilde{q} \prod_{a=1}^{k_1} \left( \sigma_a - \tilde{\sigma}_{\alpha}
\right),
\\
\left( \tilde{\sigma}_{k_2} \right)^n & = &
T^n, \\
& = & (-)^{k_2-1} \tilde{q} \left( \prod_{a=1}^{k_1} \Pi_a \right)
\: = \: (-)^{k_2-1} \tilde{q} \prod_{a=1}^{k_1} \left( \sigma_a - 
\tilde{\sigma}_{k_2}
\right).
\end{eqnarray}

Now, let us compare to the A model.
The one-loop effective action for $F(k_1,k_2,n)$ on the Coulomb
branch was computed in \cite{Donagi:2007hi}[section 5.2],
where in the notation of that reference, it was shown that
\begin{eqnarray}
\prod_{\alpha=1}^{k_2} \left( \Sigma_{1a} - \Sigma_{2 \alpha} \right) & = & q_1
\mbox{ for each } a,
\\
(\Sigma_{2 \alpha})^n & = &
q_2 \prod_{a=1}^{k_1} \left( \Sigma_{1a} - \Sigma_{2 \alpha}
\right).
\end{eqnarray}

If we identify $\sigma_a = \Sigma_{1a}$, $\tilde{\sigma}_{\alpha} = 
\Sigma_{2 \alpha}$, $(-)^{k_1-1} q = q_1$, $(-)^{k_2-1} \tilde{q} = q_2$,
then we see that the algebraic equations for the proposed B model mirror match
the Coulomb branch relations derived from the A-twisted GLSM,
including the Weyl group $S_{k_1} \times S_{k_2}$ orbifold group action
which appears both here in the Landau-Ginzburg model and also on the
Coulomb branch of the A-twisted GLSM.  Since the critical loci here
match the critical loci of the one-loop twisted effective superpotential
of the original A-twisted GLSM for the flag manifold,
the number of vacua of the proposed
B model mirror necessarily
match those of the A model, and the quantum cohomology
ring of the A model matches the relations in the proposed B model mirror.

Let us conclude with a comment on dualities.
Flag manifolds have a duality analogous to the duality
$G(k,n) \cong G(n-k,n)$ of Grassmannians \cite{Donagi:2007hi}[section 2.4]:
\begin{equation}
F(k_1,k_2,n) \: \cong \: F(n-k_2, n-k_1, n).
\end{equation}
In principle, we expect this duality to be realized in the same
fashion as the symmetry $G(k,n) \cong G(n-k,n)$, namely as an IR
relation between two mirror Landau-Ginzburg orbifolds.  For example, 
from the analysis above for each of the two cases, 
the two mirrors are guaranteed to have
the same Coulomb branch relations and the same number of vacua.

\section{Example:  Adjoint-valued matter}
\label{sect:adj}

Consider an A-twisted GLSM with gauge group $U(k)$, $n$ chiral multiplets in
the fundamental representation, and one chiral multiplet in the
adjoint representation, and no superpotential.

Following our proposal, the mirror is an $S_k$-orbifold of a Landau-Ginzburg
model with fields
\begin{itemize}
\item $Y_{ia}$, $i \in \{1, \cdots, n\}$, $a \in \{1, \cdots, k\}$,
corresponding to the $n$ fundamentals,
\item $\tilde{Y}_{\mu \nu}$, $\mu, \nu \in \{1, \cdots, k\}$,
corresponding to the adjoint matter representation,
\item $X_{\mu \nu} = \exp(-Z_{\mu \nu})$, 
$\mu, \nu \in \{1, \cdots, k\}$, $\mu \neq \nu$,
corresponding to the W bosons,
\item $\sigma_a$,
\end{itemize}
and superpotential
\begin{eqnarray}
W & = & \sum_a \sigma_a \left(
\sum_{ib} \rho^a_{ib} Y_{ib} \: + \: \sum_{\mu \nu} \alpha^a_{\mu \nu}
\tilde{Y}_{\mu \nu} \: + \:
\sum_{\mu \nu} \alpha^a_{\mu \nu} Z_{\mu \nu} \: - \: t\right)
\nonumber \\
& & \: + \:
\sum_{i, a} \exp\left(-Y_{ia} \right)
\: + \: \sum_{\mu, nu} \exp\left( - \tilde{Y}_{\mu \nu} \right)
\: + \: \sum_{\mu \neq \nu} X_{\mu \nu}.
\end{eqnarray}
In the expression above,
\begin{equation}
\rho^a_{ib} \: = \: \delta^a_b, \: \: \:
\alpha^a_{\mu \nu} \: = \: - \delta^a_{\mu} + \delta^b_{\nu},
\end{equation}
so that we can rewrite the superpotential as
\begin{eqnarray}
W & = & \sum_{a=1}^k \sigma_a \left(
\sum_i Y_{ia} \: + \: \sum_{\nu \neq a} \left( - \tilde{Y}_{a \nu} +
\tilde{Y}_{\nu a} \right) \: + \:
\sum_{\nu \neq a} \left( - Z_{a \nu} + Z_{\nu a} \right) \: - \: t \right)
\nonumber \\
& & \: + \:
\sum_{ia} \exp\left( - Y_{ia} \right) \: + \:
\sum_{\mu} \exp\left( - \tilde{Y}_{\mu \mu} \right) \: + \:
\sum_{\mu \neq \nu} \exp\left( - \tilde{Y}_{\mu \nu} \right)
\: + \: \sum_{\mu \neq \nu} X_{\mu \nu}.
\end{eqnarray}

Note that in this construction
the fields $\tilde{Y}_{\mu \mu}$ are decoupled from the rest
of the fields, and appear in the superpotential above only as
\begin{displaymath}
\sum_i \exp\left(- \tilde{Y}_{\mu \mu}\right).
\end{displaymath}
This reflects the fact that generically on the Coulomb branch of 
the original gauge theory, they correspond to
uncharged free fields with respect to the Cartan subgroup we have
chosen.  (Of course, in the entire original gauge theory, they are not
decoupled, but rather appear decoupled at low energies on the Coulomb
branch, which is the reason for this artifact of our construction.)  
For purposes of comparison, the Hori-Vafa
mirror \cite{Hori:2000kt} of a single uncharged free chiral superfield consists
of a single field $Y$ with superpotential
$\exp(-Y)$.  This is exactly the structure we see above for the
fields $\tilde{Y}_{\mu \mu}$, as expected.

In the rest of this section, we will focus on the untwisted sector
of the orbifold, as the critical locus does not
intersect the fixed points of the orbifold.

Integrating out the fields $\sigma_a$, we get the constraints
\begin{equation}
\sum_{i=1}^n Y_{ia}  \: + \: \sum_{\nu \neq a} \left( - \tilde{Y}_{a \nu} +
\tilde{Y}_{\nu a} \right) \: + \:
\sum_{\nu \neq a} \left( - Z_{a \nu} + Z_{\nu a} \right)
\: = \: t,
\end{equation}
which we solve as
\begin{equation}
Y_{na} \: = \: 
- \sum_{i=1}^{n-1} Y_{ia} \: - \:
 \sum_{\nu \neq a} \left( - \tilde{Y}_{a \nu} +
\tilde{Y}_{\nu a} \right) \: - \:
\sum_{\nu \neq a} \left( - Z_{a \nu} + Z_{\nu a} \right)
\: + \: t.
\end{equation}
Define
\begin{eqnarray}
\Pi_a & = & \exp\left( - Y_{na} \right), \\
& = &
q \left( \prod_{i=1}^{n-1} \exp\left( + Y_{ia} \right) \right)
\left( \prod_{\nu \neq a} \exp\left( -\tilde{Y}_{a \nu} + \tilde{Y}_{\nu a}
\right) \right)
\left( \prod_{\nu \neq a} \frac{ X_{a \nu} }{ X_{\nu a} } \right),
\label{eq:adj:pi-defn}
\end{eqnarray}
where $q = \exp(-t)$.  Then, the superpotential becomes
\begin{eqnarray}
W & = & \sum_{i=1}^{n-1} \sum_a \exp\left( - Y_{ia} \right)
\: + \: \sum_{\mu} \exp\left( - \tilde{Y}_{\mu \mu} \right)
\: + \: \sum_{\mu \neq \nu} \exp\left( - \tilde{Y}_{\mu \nu} \right)
\: + \: \sum_{\mu \neq \nu} X_{\mu \nu}
\nonumber \\
& & \hspace*{3.5in}
\: + \: \sum_a \Pi_a.
\end{eqnarray}

We compute the critical locus as follows:
\begin{eqnarray*}
\frac{\partial W}{\partial Y_{ia} }: & &
\exp\left( - Y_{ia} \right) \: = \: \Pi_a,
\\
\frac{\partial W}{\partial \tilde{Y}_{\mu \nu}}: & & 
\exp\left( - \tilde{Y}_{\mu \nu} \right) \: = \:
\left\{ \begin{array}{cl}
- \Pi_{\mu} + \Pi_{\nu} & \mu \neq \nu, \\
0 & \mu=\nu,
\end{array} \right.
\\
\frac{\partial W}{\partial X_{\mu \nu}}: & &
X_{\mu \nu} \: = \: 
- \Pi_{\mu} + \Pi_{\nu}.
\end{eqnarray*}
We shall omit the decoupled fields $\tilde{Y}_{\mu \mu}$ from the
rest of our analysis.

As discussed in section~\ref{sect:gkn:excluded}, 
we exclude the loci $\{ X_{\mu \nu} = 0 \}$,
where the superpotential has poles,
which implies the constraint
\begin{equation}
\Pi_{\mu} \neq \Pi_{\nu}
\end{equation}
for $\mu \neq \nu$.
(As a result, the critical locus does not intersect any
orbifold fixed points.)  This can also be established from the fact
that $\exp(-\tilde{Y}) \neq 0$.  Furthermore, since
$\exp(-Y) \neq 0$, we also have that each $\Pi_a \neq 0$.

Then, plugging into definition~(\ref{eq:adj:pi-defn}), we find
\begin{equation}   \label{eq:adj:pi-reln}
\left( \Pi_{\mu} \right)^n \: = \: q.
\end{equation}

The operator mirror map~(\ref{eq:op-mirror-1}), (\ref{eq:op-mirror-2})
in this case states
\begin{eqnarray}
\exp\left( - Y_{ia} \right) & = & \sum_b \sigma_b \rho^b_{ia} 
\: = \: \sigma_a, 
\\
\exp\left( - \tilde{Y}_{\mu \nu} \right) & = &
\left\{ \begin{array}{cl}
\sum_{b} \sigma_a \alpha^b_{\mu \nu} \: = \:
- \sigma_{\mu} + \sigma_{\nu} & \mu \neq \nu, \\
0 & \mu=\nu,
\end{array} \right.
\\
X_{\mu \nu} & = & \sum_a \sigma_a \alpha^a_{\mu \nu}
\: = \: - \sigma_{\mu} + \sigma_{\nu}.
\end{eqnarray}
Comparing to the critical locus equations, we find immediately that
\begin{equation}
\Pi_a \: \leftrightarrow \: \sigma_a.
\end{equation}

Now, let us compare to corresponding A model Coulomb branch results.
From \cite{Morrison:1994fr}[equ'n (3.36)], the one-loop twisted
effective superpotential for the original gauge theory, for distinct,
large but
otherwise generic $\sigma_a$, has the form
\begin{eqnarray}
\tilde{W} & = & \sum_a \sigma_a \Biggl( it\: -\: \sum_{i=1}^n \sum_{c=1}^k
\rho^a_{ic} \ln \left( \sum_{b=1}^k \rho^b_{ic} \sigma_b \right)
\: - \: \sum_{\mu \neq \nu} \alpha^a_{\mu \nu}\ln \left(
\sum_{b=1}^k \alpha^b_{\mu \nu} \sigma_b \right) \Biggr),
\\
& = &
\sum_a \sigma_a \left( it \: - \:
n  \ln \sigma_a \: - \:
\sum_{\mu \neq \nu} \left( - \delta^a_{\mu} + \delta^a_{\nu} \right)
\ln \left(  - \sigma_{\mu} + \sigma_{\nu} 
\right) \right),
\\
& = &
\sum_a \sigma_a \left( it \: - \:
n \ln \sigma_a \right) \: - \:
\left( \begin{array}{c} k \\ 2 \end{array} \right) \pi i,
\end{eqnarray}
from which we derive the critical locus
\begin{equation}
\left( \sigma_a \right)^n \: = \: \tilde{q}
\end{equation}
for some constant $\tilde{q}$.
Given the operator mirror map, this clearly matches our mirror
equation~(\ref{eq:adj:pi-reln}), confirming that our proposed
mirror is functioning correctly in this case.

\section{Example:  Symmetric $m$-tensor-valued matter}
\label{sect:symmetric}

Consider an A-twisted GLSM with gauge group $U(k)$,
$n$ chiral multiplets in the fundamental representation,
and one chiral multiplet in the representation Sym$^m {\bf k}$.
(We assume $m < k$, for simplicity.)

Following our proposal, the mirror is an $S_k$-orbifold of a Landau-Ginzburg
model with fields
\begin{itemize}
\item $Y_{ia}$, $i \in \{1, \cdots, n\}$, $a \in \{1, \cdots, k\}$,
corresponding to the $n$ fundamentals,
\item $\tilde{Y}_{a_1 \cdots a_m}$, symmetric in its indices
$a_1, \cdots, a_m \in \{1, \cdots, k\}$, corresponding to the symmetric
matter representation,
\item $X_{\mu \nu} = \exp(-Z_{\mu \nu})$, 
$\mu, \nu \in \{1, \cdots, k\}$, corresponding to the
W bosons,
\item $\sigma_a$,
\end{itemize}
and superpotential
\begin{eqnarray}
W & = & \sum_a \sigma_a \left( \sum_{i b} \rho^a_{i b} Y_{i b} \: + \:
\sum_{a_1 \leq \cdots \leq a_m} \rho^a_{a_1 \cdots a_m} \tilde{Y}_{a_1 \cdots a_m}
\: + \: \sum_{\mu \nu} \alpha^a_{\mu \nu} Z_{\mu \nu} \: - \: t \right)
\nonumber \\
& & \: + \:
\sum_{i, a} \exp\left( - Y_{ia} \right) \: + \:
\sum_{a_1 \leq \cdots \leq a_m} \exp\left( - \tilde{Y}_{a_1 \cdots a_m} \right)
\: + \: 
\sum_{\mu \neq \nu} X_{\mu \nu}.
\end{eqnarray}
In the expression above,
\begin{equation}
\rho^a_{ib} \: = \: \delta^a_b, \: \: \:
\rho^a_{a_1 \cdots a_m} \: = \: \delta^a_{a_1} + \delta^a_{a_2} + \cdots +
\delta^a_{a_m}, \: \: \:
\alpha^a_{\mu \nu} \: = \: - \delta^a_{\mu} + \delta^a_{\nu},
\end{equation}
so that we can write the superpotential as
\begin{eqnarray}
W & = & \sum_{a=1}^k \sigma_a \left(
\sum_i Y_{ia} \: + \:
\sum_{j=1}^m \sum_{a_1\leq \cdots\leq \hat{a}_j\leq \cdots\leq a_m} 
\tilde{Y}_{a_1  \cdots a_{j-1} a a_{j+1} \cdots a_m}
\: + \: \sum_{\nu} \left(-Z_{a \nu} + Z_{\nu a} \right)
\: - \: t \right)
\nonumber \\
& & \: + \:
\sum_{i, a} \exp\left( - Y_{ia} \right) \: + \:
\sum_{a_1\leq \cdots\leq a_m} \exp\left( - \tilde{Y}_{a_1 \cdots a_m} \right)
\: + \: 
\sum_{\mu \neq \nu} X_{\mu \nu}.
\end{eqnarray}
Integrating out the $\sigma_a$, we get the constraints
\begin{equation}
\sum_i Y_{ia} \: + \:
\sum_{j=1}^m  \sum_{a_1\leq \cdots\leq \hat{a}_j\leq \cdots\leq a_m}
\tilde{Y}_{a_1 \cdots  a_{j-1} a a_{j+1}  \cdots  a_m}
\: + \: \sum_{\nu} \left(-Z_{a \nu} + Z_{\nu a} \right)
\: = \: t,
\end{equation}
which we can solve as
\begin{equation}
Y_{n a} \: = \:
- \sum_{i=1}^{n-1} Y_{ia} \: - \: 
\sum_{j=1}^m  \sum_{a_1\leq  \cdots\leq \hat{a}_j\leq \cdots\leq a_m}
\tilde{Y}_{a_1 \cdots a_{j-1} a a_{j+1} \cdots a_m}
\: - \:  \sum_{\nu \neq a} \left(-Z_{a \nu} + Z_{\nu a} \right)
\: + \: t.
\end{equation}
Define
\begin{eqnarray}
\Pi_a & = & \exp\left( - Y_{na} \right), \\
& = & q \left( \prod_{i=1}^{n-1} \exp\left( + Y_{ia} \right) \right)
\left( \prod_{j=1}^m \prod_{a_1\leq \cdots\leq \hat{a}_j\leq \cdots
\leq a_m}
\exp\left( + \tilde{Y}_{a_1 \cdots a_{j-1} a a_{j+1} \cdots a_m}
\right) \right)
\cdot \nonumber \\
& & \hspace*{1.0in} \cdot
\left( \prod_{\nu \neq a} 
\frac{ X_{a \nu} }{ X_{\nu a} }
\right),
\label{eq:symm-tensor:pi-defn}
\end{eqnarray} 
where $q = \exp(-t)$.
Then, the superpotential becomes
\begin{eqnarray}
W & = &
\sum_{i=1}^{n-1} \sum_a  \exp\left( - Y_{ia} \right) \: + \:
\sum_{a_1 \leq \cdots\leq a_m} \exp\left( - \tilde{Y}_{a_1 \cdots a_m} \right)
\: + \: 
\sum_{\mu \neq \nu} X_{\mu \nu} 
\nonumber \\
& & \hspace*{1.0in}
\: + \: \sum_a \Pi_a.
\end{eqnarray}

Excluding the loci $\{ X_{\mu \nu} = 0 \}$ as in 
section~\ref{sect:gkn:excluded}, we see the
vacua are restricted to $X_{\mu \nu} \neq 0$, and therefore do not
intersect the fixed points of the Weyl orbifold.  As a result, we do not
expect any contributions to vacua from twisted sectors.

We compute the critical locus as follows:
\begin{eqnarray*}
\frac{\partial W}{\partial Y_{ia} }: & &
\exp\left( - Y_{ia} \right) \: = \: \Pi_a,
\\
\frac{\partial W}{\partial \tilde{Y}_{a_1 \cdots a_m} }: & &
\exp\left( - \tilde{Y}_{a_1 \cdots a_m} \right) \: = \:
\Pi_{a_1} + \Pi_{a_2} + \cdots + \Pi_{a_m},
\\
\frac{\partial W}{\partial X_{\mu \nu}}: & &
X_{\mu \nu}  \: = \:
- \Pi_{\mu} + \Pi_{\nu}.
\end{eqnarray*}
Then, plugging into the definition~(\ref{eq:symm-tensor:pi-defn}), we find
\begin{equation} \label{eq:uksymm:pi-eqn}
\left( \Pi_a \right)^n \: = \: q 
\left[ \prod_{j=1}^m \prod_{a_1\leq \cdots\leq \hat{a}_j\leq \cdots\leq a_m}
\frac{1}{ \Pi_{a_1} + \cdots + \Pi_{a_{j-1}} + \Pi_a +
\Pi_{a_{j+1}} + \cdots + \Pi_{a_m} } 
\right]
(-)^{k-1},
\end{equation}
and as before, from the equation for $X_{\mu \nu}$ and the fact that the
potential diverges when $X_{\mu \nu}=0$, we find that the
$\Pi_a$ are all distinct.

The operator mirror map~(\ref{eq:op-mirror-1}), (\ref{eq:op-mirror-2}) 
in this case states
\begin{eqnarray}
\exp\left( - Y_{ia} \right) & = & \sum_a \sigma_b \rho^b_{ia}
\: = \: \sigma_a, 
\\
\exp\left( - \tilde{Y}_{a_1 \cdots a_m} \right) & = &
\sum_b \sigma_b \rho^b_{a_1 \cdots a_m} \: = \:
\sigma_{a_1} + \cdots + \sigma_{a_m},
\\
X_{\mu \nu} & = & \sum_a \sigma_a \alpha^a_{\mu \nu} \: = \:
-\sigma_{\mu} + \sigma_{\nu}.
\end{eqnarray}
Comparing to the critical locus equations, we find immediately that
\begin{equation}
\Pi_a \: \leftrightarrow \: \sigma_a.
\end{equation}

Now, let us compare to corresponding A model Coulomb branch results.
From \cite{Morrison:1994fr}[equ'n (3.36)], the one-loop twisted
effective superpotential for the original gauge theory, for distinct,
large but
otherwise generic $\sigma_a$, has the form
\begin{eqnarray}
\tilde{W} & = & \sum_a \sigma_a \Biggl( it\: -\: \sum_{i=1}^n \sum_{c=1}^k
\rho^a_{ic} \ln \left( \sum_{b=1}^k \rho^b_{ic} \sigma_b \right)
\nonumber \\
& & \hspace*{0.5in}
\: - \:
\sum_{a_1 \leq \cdots \leq a_m} \rho^a_{a_1 \cdots a_m} \ln \left(
\sum_{b=1}^k \rho^b_{a_1 \cdots a_m} \sigma_b \right)
\Biggr),
\\
& = & \sum_a \sigma_a \Biggl( it \: - \: n \sum_{c=1}^k \ln \sigma_c
\nonumber \\
& & \hspace*{0.5in}
\: - \:
\sum_{a_1 \leq \cdots \leq a_m} \left( \delta^a_{a_1} + \cdots + 
\delta^a_{a_m} \right) \ln\left( \sum_b \left( \delta^b_{a_1} + \cdots +
\delta^b_{a_m} \right) \sigma_b \right) \Biggr),
\end{eqnarray}
from which we derive the critical locus
\begin{equation}
 (\sigma_a)^n 
\left[ \prod_{j=1}^m 
\prod_{a_1 \leq \cdots \leq \hat{a}_j \leq \cdots \leq a_m}
\left( \sigma_{a_1} + \cdots + \sigma_{a_{j-1}} + \sigma_a +
\sigma_{a_j} + \cdots \sigma_{a_m} \right)
\right]
\: = \: \tilde{q},
\end{equation}
where $\tilde{q}$ is some constant.
Given the operator mirror map, this clearly matches our mirror 
equation~(\ref{eq:uksymm:pi-eqn}), confirming that our proposed mirror
is functioning correctly in this case.

\section{Example:  $SU(k)$ gauge theory with twisted masses}
\label{sect:suk-twisted}

In this section we will consider the mirror of an $SU(k)$ gauge theory
with $n$ fields in the fundamental representation.
We will also turn on generic twisted masses.

Following our proposal, the mirror is an $S_k$-orbifold of a Landau-Ginzburg
model with fields
\begin{itemize}
\item $Y_{i \alpha}$, $i \in \{1, \cdots, n\}$,
$\alpha \in \{1, \cdots, k\}$, corresponding to the $n$ fundamentals,
\item $X_{\mu \nu} = \exp(-Z_{\mu \nu})$, $\mu, \nu \in \{1, \cdots k\}$,
$Z_{\mu \nu} = 0$ for $\mu = \nu$ but not (anti)symmetric,
\item $\sigma_a$, $a \in \{1, \cdots, k-1\}$,
\end{itemize}
and superpotential
\begin{eqnarray}
W & = & \sum_{a=1}^{k-1} \sigma_a \left(
\sum_{i \alpha} \rho^a_{i \alpha} Y_{i \alpha} \: + \: 
\sum_{\mu \nu} \alpha^a_{\mu \nu} Z_{\mu \nu}  \right)
\nonumber \\
& & \: + \: 
\sum_{i, \alpha} \exp\left( - Y_{i \alpha} \right)
\: + \: \sum_{\mu \nu} X_{\mu \nu}
 \: - \:
\sum_{i, \alpha} \tilde{m}_i  Y_{i \alpha}.
\label{eq:suk:sup-1}
\end{eqnarray}
Between
$Z$ and $X$, we take $X$ to be the fundamental field.
In the expression above,
\begin{equation}
\rho^a_{i \alpha} \: = \: \delta^a_{\alpha} - \delta^{k}_{\alpha},
\: \: \:
\alpha^a_{\mu \nu} \: = \:
- \delta_{\mu, a} + \delta_{\mu, k} + \delta_{\nu, a}  -
\delta_{\nu, k},
\end{equation}
corresponding to Cartan generators of the form diag$(0, \cdots, 0, 1, 
0, \cdots, 0, -1)$.
Also, since this is an $SU$ gauge theory, there is no FI parameter,
so there is no $t$ in the mirror above.

It will sometimes be helpful to work with a different representation
of the same theory, in terms of $k$ fields
$\tilde{\sigma}_{\alpha}$ obeying the constraint
\begin{equation} \label{eq:suk:sigma-constr}
\sum_{\alpha=1}^k \tilde{\sigma}_{\alpha} \: = \: 0.
\end{equation}
We can understand this description and its relation to the Landau-Ginzburg
model above by adding an additional Lagrange multiplier $\lambda$
and working with the superpotential
\begin{eqnarray}
W & = & \sum_{\alpha=1}^{k} \tilde{\sigma}_{\alpha} \left(
\sum_{i \beta} \tilde{\rho}^{\alpha}_{i \beta} Y_{i \beta} \: + \: 
\sum_{\mu \nu} \tilde{\alpha}^{\alpha}_{\mu \nu} Z_{\mu \nu}  \right)
\: + \: 
\lambda \sum_{\alpha=1}^k \tilde{\sigma}_{\alpha}
\nonumber \\
& & \: + \: 
\sum_{i, \alpha} \exp\left( - Y_{i \alpha} \right)
\: + \: \sum_{\mu \nu} X_{\mu \nu}
 \: - \:
\sum_{i, \alpha} \tilde{m}_i  Y_{i \alpha},
\end{eqnarray}
where
\begin{equation}
\tilde{\rho}^{\alpha}_{i \beta} \: = \: \delta^{\alpha}_{\beta},
\: \: \:
\tilde{\alpha}^{\alpha}_{\mu \nu} \: = \: - \delta^{\alpha}_{\mu}
+ \delta^{\alpha}_{\nu},
\end{equation}
the same as for $U(k)$.
The Weyl group orbifold $S_k$ acts on this presentation
by permuting the $\alpha$ indices
in the obvious way.
Integrating out $\lambda$ gives the constraint~(\ref{eq:suk:sigma-constr}),
which we can use to eliminate $\tilde{\sigma}_k$:
\begin{equation}
\tilde{\sigma}_k \: = \: - \sum_{a=1}^{k-1} \tilde{\sigma}_a.
\end{equation}
Plugging this back in, and identifying $\sigma_a = \tilde{\sigma}_a$
for $a < k$, we recover the superpotential~(\ref{eq:suk:sup-1})
with 
\begin{equation}
\rho^a_{i \beta} \: = \: \tilde{\rho}^a_{i \beta} - \tilde{\rho}^k_{i \beta},
\: \: \:
\alpha^a_{\mu \nu} \: = \: \tilde{\alpha}^a_{\mu \nu} - 
\tilde{\alpha}^k_{\mu \nu}.
\end{equation} 

Since the other examples of Weyl group actions in this paper have been
simpler than this case, 
let us also briefly but explicitly describe the Weyl group orbifold
for the special 
case of $SU(3)$.  For three $\tilde{\sigma}$s, the Weyl group orbifold
is an obvious $S_3$ action, but once one integrates out $\tilde{\sigma}_3$,
it is a bit more complicated.  Three roots, which we can take to be the
positive roots, can be read off from the expressions above:
\begin{equation}
X_{21} \: \sim \: (1,-1), \: \: \:
X_{31} \: \sim \: (2,1), \: \: \:
X_{32} \: \sim \: (1,2),
\end{equation}
with $X_{12}$, $X_{13}$, and $X_{23}$ corresponding to root vectors of
opposite signs.  These statements are equivalent to writing
\begin{equation}
\langle \tilde{\sigma}, \alpha_{21} \rangle = \tilde{\sigma}_1 - \tilde{\sigma}_2,
\: \: \:
\langle \tilde{\sigma}, \alpha_{31} \rangle = 2 \tilde{\sigma}_1 + \tilde{\sigma}_2, 
\: \: \:
\langle \tilde{\sigma}, \alpha_{32} \rangle = \tilde{\sigma}_1 + 2 \tilde{\sigma}_2.
\end{equation}
The Weyl reflection defined by $21$ maps
\begin{equation}
X_{32} \leftrightarrow X_{31}, \: \: \:
X_{21} \leftrightarrow X_{12},
\end{equation}
the Weyl reflection defined by $31$ maps
\begin{equation}
X_{21} \leftrightarrow - X_{32} = X_{23}, \: \: \:
X_{31} \leftrightarrow X_{13},
\end{equation}
and
the Weyl reflection defined by $32$ maps
\begin{equation}
X_{21} \leftrightarrow X_{31}, \: \: \:
X_{32} \leftrightarrow X_{23},
\end{equation}
and these are represented by the following three matrices,
corresponding to $21$, $31$, and $32$, respectively:
\begin{equation}
\left[ \begin{array}{cc}
0 & 1 \\ 1 & 0 \end{array} \right],
\: \: \:
 \left[ \begin{array}{cc}
-1 & 0 \\ -1 & 1 \end{array} \right],
\: \: \:
\left[ \begin{array}{cc}
1 & -1 \\ 0 & -1 \end{array} \right].
\end{equation}
The action on $\tilde{\sigma}_1$, $\tilde{\sigma}_2$ is described by
the transpose matrices, so that under the three Weyl reflections
above,
\begin{equation}
\left[ \begin{array}{c} \tilde{\sigma}_1 \\ \tilde{\sigma}_2 \end{array} \right] \: \mapsto
\: \left[ \begin{array}{c} \tilde{\sigma}_2 \\ \tilde{\sigma}_1 \end{array} \right], 
\: \: \:
\left[ \begin{array}{c} -\tilde{\sigma}_1 - \tilde{\sigma}_2 \\ \tilde{\sigma}_2 \end{array} \right],
\: \: \:
\left[ \begin{array}{c} \tilde{\sigma}_1 \\ - \tilde{\sigma}_1 - \tilde{\sigma}_2 \end{array} \right].
\end{equation}
These (partially) define the Weyl orbifold action on the fields above,
and it is straightforward to check that the superpotential is invariant.
For example, it is straightforward to check that under the
Weyl reflection defined by $21$,
\begin{equation}
\langle \tilde{\sigma}, \alpha_{32} \rangle \: = \:
\tilde{\sigma}_1 + 2 \tilde{\sigma}_2 \: \mapsto \:
\tilde{\sigma}_2 + 2 \tilde{\sigma}_1 \: = \:
\langle \tilde{\sigma}, \alpha_{31} \rangle,
\end{equation}
consistent with the statement that this Weyl reflection maps
$X_{32}$ to $X_{31}$.

Now, let us return to expression~(\ref{eq:suk:sup-1}).
Integrating out the $\sigma_a$ gives the constraint
\begin{equation}
\sum_{i=1}^n \left( 
Y_{i, a} - Y_{i, k} \right) 
\: + \:
\sum_{\mu} \left( - Z_{a, \mu} + Z_{k, \mu} +
Z_{\mu, a} - Z_{\mu, k} \right)
\: = \: 0,
\end{equation}
which can be solved to eliminate $Y_{n a}$ for $a<k$:
\begin{eqnarray}
Y_{n, a} & = & - \sum_{i=1}^{n-1} Y_{i a} \: + \:
\sum_{i=1}^n Y_{i, k} \: + \: 
\sum_{|mu} \left( Z_{a \mu} - Z_{\mu a} - Z_{k, \mu} + Z_{\mu k} \right).
\end{eqnarray}

Define for $a \in \{1, \cdots, k-1\}$
\begin{eqnarray}  
\Pi_a & = & \exp\left( - Y_{n, a} \right), \\
& = & \left( \prod_{i=1}^{n-1} \exp\left( + Y_{i a} \right) \right)
\left( \prod_{i=1}^n \exp\left( - Y_{i k} \right) \right)
\left( \prod_{\mu \neq a} \frac{X_{a \mu} }{X_{\mu a} } \right)
\left( \prod_{\mu \neq k} \frac{ X_{\mu k} }{ X_{k \mu} } \right),
 \label{eq:sun:pi-defn}
\end{eqnarray}
for $X_{\mu \nu} = \exp(-Z_{\mu \nu})$.
The superpotential then becomes
\begin{eqnarray}
W & = &
\sum_{i=1}^{n-1} \sum_{\alpha=1}^k \exp\left( - Y_{i \alpha} \right)
\: + \: \exp\left( - Y_{n, k} \right)
\: + \: \sum_{\mu \nu} X_{\mu \nu} 
\: + \:
\sum_{a=1}^{k-1} \Pi_a
\nonumber \\
& & \: - \:
 \tilde{m}_n  Y_{n, k} 
\: - \:
\sum_{i=1}^{n-1} \sum_{\alpha=1}^k \tilde{m}_i 
Y_{i \alpha} 
\nonumber \\
& & \: - \:
\sum_{a=1}^{k-1} \tilde{m}_n  \Biggl(    
\: - \: \sum_{i=1}^{n-1}  Y_{i,a} \: + \:
\sum_{i=1}^n  Y_{i,k} 
\: + \: \sum_{\mu} \left( Z_{a,\mu} - Z_{k,\mu} - Z_{\mu,a} + Z_{\mu,k}
\right)
 \Biggr) .
\end{eqnarray}

The critical locus is determined by
\begin{eqnarray*}
\frac{\partial W}{\partial Y_{i \alpha}}: & &
\exp\left( - Y_{i \alpha} \right) \: = \: \left\{ \begin{array}{cl}
\Pi_{\alpha} - \tilde{m}_i + \tilde{m}_n & i < n, \alpha < k, \\
\Pi_{\alpha} & i=n, \alpha < k, \\
- (\sum_{a=1}^{k-1} \Pi_a) - \tilde{m}_i - (k-1) \tilde{m}_n  
& i < n, \alpha = k, \\
- (\sum_{a=1}^{k-1} \Pi_a) - k \tilde{m}_n &
i=n, \alpha=k,
\end{array} \right.
\\
\frac{\partial W}{\partial X_{\mu \nu} }: & &
X_{\mu \nu} \: = \:
\left\{ \begin{array}{cl}
-\Pi_{\mu} + \Pi_{\nu} & \mu < k, \nu < k, \\
 (\sum_{a=1}^{k-1} \Pi_{a}) + \Pi_{\nu} + k \tilde{m}_n & \mu=k, \nu<k, \\
- \Pi_{\mu} - (\sum_{a=1}^{k-1} \Pi_{a}) -k \tilde{m}_n  & \mu<k, \nu=k.
\end{array} \right.
\end{eqnarray*}
In the expressions above, $Y_{n \alpha}$ for $\alpha < k$ was
eliminated as a dynamical field using the constraints, 
but $\Pi_{\alpha}$ is defined
to be $\exp(-Y_{n \alpha})$ for $\alpha < k$ using the constraint
solution.

Note that since $\exp(-Y) \neq 0$, on the critical locus,
\begin{equation} \label{eq:suk:constr-1}
\Pi_{a} + \tilde{m}_n \: \neq \: \tilde{m_i},
\: \: \:
- \sum_{b=1}^{k-1} \left( \Pi_b + \tilde{m}_n \right) \: \neq \:
\tilde{m}_i,
\end{equation}
for all $a$, $i$, and as before excluding the locus
$\{ X_{\mu \nu} = 0\}$ ($\mu \neq \nu$),
\begin{equation} \label{eq:suk:constr-2}
\Pi_{a} + \tilde{m}_n \: \neq \: \Pi_{b} + \tilde{m}_n
\: \neq \: - \sum_{c=1}^{k-1} \left( \Pi_c + \tilde{m}_n \right),
\end{equation}
for all $a \neq b$.

Plugging into the definition~(\ref{eq:sun:pi-defn}) of $\Pi_a$, we find
\begin{eqnarray}
\Pi_a \left( \prod_{i=1}^{n-1} \left( \Pi_a + \tilde{m}_n - \tilde{m}_i 
\right) \right) & = &
 \left(  -  \sum_b \left(\Pi_b + \tilde{m}_n\right)  - \tilde{m}_n \right)
\cdot 
\nonumber \\
& & \hspace*{0.25in} \cdot
 \prod_{i=1}^{n-1} \left[  - \sum_b \left( \Pi_b + \tilde{m}_n \right) -
\tilde{m}_i   \right] .
\label{eq:sun:critical-loci-masses}
\end{eqnarray}

Next, we would like to compare to the results for the original
gauge theory described in \cite{Hori:2006dk}[section 3].
The conventions used there involve $k$ fields $\tilde{\sigma}_{\alpha}$,
subject to the constraint~(\ref{eq:suk:sigma-constr}).
In terms of such variables, using earlier work in this section,
the operator mirror map~(\ref{eq:op-mirror-mass-1}), 
(\ref{eq:op-mirror-mass-2}) 
says
\begin{eqnarray}
\exp\left( - Y_{i \alpha} \right) & = & - \tilde{m}_i \: + \:
\sum_{\beta}^k \tilde{\sigma}_{\beta} \rho^{\alpha}_{i \beta}, 
\\
& = & - \tilde{m}_i \: + \: \tilde{\sigma}_{\alpha}, 
\\
X_{\mu \nu}  & = &
\sum_{\alpha} \tilde{\sigma}_{\alpha} \tilde{\alpha}^{\alpha}_{\mu \nu}.
\\
& = & - \tilde{\sigma}_{\mu} + \tilde{\sigma}_{\nu}.
\end{eqnarray}
Comparing to the critical locus, we see that 
\begin{equation}
\tilde{\sigma}_{\alpha} \: = \: 
\left\{ \begin{array}{cl}
\Pi_{\alpha} + \tilde{m}_n, & \alpha < k , \\
- \sum_{b=1}^{k-1} \left( \Pi_b + \tilde{m}_n \right) & \alpha=k,
\end{array} \right.
\end{equation}
which satisfy
\begin{displaymath}
\tilde{\sigma}_1 + \cdots + \tilde{\sigma}_k \: = \: 0.
\end{displaymath}
Note in this language that since we excluded loci
where any $X_{\mu \nu} = 0$, the solutions for $\tilde{\sigma}_{\alpha}$
must be distinct, and also since $\exp(-Y_{i \alpha}) \neq 0$,
$\sigma_{\alpha} \neq \tilde{m}_i$ for all $i$ and $\alpha$,
rewriting the constraints~(\ref{eq:suk:constr-1}), 
(\ref{eq:suk:constr-2}).

In these variables, we can rewrite the critical
locus equation~(\ref{eq:sun:critical-loci-masses}) as
\begin{equation}
\prod_{i=1}^n \left( \tilde{\sigma}_{\alpha} - \tilde{m}_i \right) \: = \:
  \prod_{i=1}^n \left( \tilde{\sigma}_k - \tilde{m}_i
\right),
\end{equation}
meaning that
\begin{equation}  \label{eq:suk:ind-quant}
\prod_{i=1}^n \left( \tilde{\sigma}_{\alpha} - \tilde{m}_i \right)
\end{equation}
is independent of $\alpha$.

This result precisely matches that of \cite{Hori:2006dk}[section 3].
They work with $k$ $\tilde{\sigma}$'s obeying
\begin{displaymath}
\tilde{\sigma}_1 + \cdots + \tilde{\sigma}_k \: = \: 0,
\end{displaymath}
and furthermore, assume the $\sigma_a$ are all distinct and also
different from all twisted masses,
which we have derived algebraically in the proposed mirror
in equations~(\ref{eq:suk:constr-1}), (\ref{eq:suk:constr-2}).
They then compute the effective
superpotential
\begin{displaymath} 
\tilde{W} \: = \: - \sum_i \sum_{\alpha=1}^k 
\left( \tilde{\sigma}_{\alpha} - \tilde{m}_i
\right) \left( \ln\left( \tilde{\sigma}_{\alpha} - \tilde{m}_i \right) - 1 \right),
\end{displaymath} 
from which they derive the vacuum equations
\begin{eqnarray} 
\prod_i \left( \tilde{\sigma}_{\alpha} - \tilde{m}_i \right) & = & \exp(-\lambda), \\
\sum_{\alpha=1}^k \tilde{\sigma}_{\alpha} & = & 0,
\end{eqnarray}
for $\lambda$ a Lagrange multiplier enforcing the second condition
above, the constraint that the
sum of the $\tilde{\sigma}_{\alpha}$ vanish.
In particular, this means that the quantity~(\ref{eq:suk:ind-quant})
is independent of $\alpha$, just as we have derived algebraically
in our proposed mirror, and so the results from our proposed mirror
match the analysis of \cite{Hori:2006dk}[section 3] for the original
gauge theory.

\section{Example:  $SO(2k)$ gauge theory}
\label{sect:so2k}

For most of this paper we so far have considered $U(k)$ gauge theories.
In this section we will consider an $SO(2k)$ gauge theory,
with $n$ fields in the vector representation,
to explictily demonstrate
how
our mirror proposal can be applied to other
gauge groups, and to compare to results in the literature.

\subsection{Mirror proposal}

Proceeding as in {\it e.g.} \cite{Seiberg:1994pq},
although the vector representation is real, we will take the
matter of the theory to consist of $n$ (complex) chiral superfields,
implicitly complexifying the representation.

The original gauge theory is then an $SO(2k)$ gauge theory with 
$n$ chiral superfields $\phi_i^a$ in the vector representation.

Following our proposal, the mirror is a Weyl-group-orbifold
of
a Landau-Ginzburg model with fields
\begin{itemize}
\item $Y_{i\alpha}$, $i \in \{1, \cdots, n\}$, $\alpha \in \{1, \cdots, 2k\}$,
\item $X_{\mu \nu} = \exp( - Z_{\mu \nu})$,
$X_{\mu \nu} =  X_{\nu \mu}^{-1}$, $Z_{\mu \nu} = - Z_{\nu \mu}$,
$\mu, \nu \in \{1, \cdots, 2k\}$,
(excluding\footnote{
These correspond to elements of a Cartan subalgebra we use in defining
constraints associated to the fields $\sigma_a$.
} $X_{2a-1,2a}$, $Z_{2a-1,2a}$),
corresponding to the W bosons,
\item $\sigma_a$, $a \in \{1, \cdots, k\}$,
\end{itemize}
and superpotential
\begin{eqnarray}
W & = &
\sum_{a=1}^k \sigma_a \left(
\sum_{i \alpha \beta} \rho^a_{i \alpha \beta} Y_{i \beta} \: + \:
\sum_{\mu < \nu; \mu' , \nu'} \alpha^a_{\mu \nu, \mu' \nu'} Z_{\mu' \nu'}
\: - \: t \right)
\nonumber \\
& & \: + \:
\sum_{i \alpha} \exp\left( - Y_{i \alpha} \right) 
\: + \: 
\sum_{\mu < \nu} X_{\mu \nu}
\nonumber \\
& & \: - \:
\sum_{i, \alpha} \tilde{m}_i \left( Y_{i \alpha} \: - \:
\sum_{\beta} \rho^a_{i \alpha \beta} t \right).
\end{eqnarray}
Since the gauge group is semisimple, there is no continuous
FI parameter; however, as noted in \cite{Hori:2011pd}[section 4],
\cite{Hori:1994uf,Hori:1994nc,Witten:1978ka},
for $k > 1$, there is a possible\footnote{
See also \cite{Gaiotto:2010be,Aharony:2013hda} for four-dimensional
versions of these discrete theta angles.
} ${\mathbb Z}_2$ discrete theta angle
(due to the fact that $\pi_1(SO(m)) = {\mathbb Z}_2$ for
$m \geq 3$),
so we have included $t$, which can take the values $\{0, \pi i\}$.

In the expression above,
\begin{eqnarray}
\rho^a_{i \alpha \beta} & = &  \delta_{\alpha, 2a-1}
\delta_{\beta, 2a} - \delta_{\beta, 2a-1} \delta_{\alpha, 2a} ,
\label{eq:so2k:rho-defn}
\\
\alpha^a_{\mu \nu, \mu' \nu'} & = &
 \delta_{\nu \nu'} \left( \delta_{\mu, 2a-1} \delta_{\mu', 2a} -
\delta_{\mu, 2a} \delta_{\mu', 2a-1} \right)
\nonumber \\
& & \hspace*{0.5in}
\: + \:
 \delta_{\mu \mu'} \left( \delta_{\nu, 2a-1} \delta_{\nu', 2a} - 
\delta_{\nu, 2a} \delta_{\nu', 2a-1} \right).
\label{eq:so2k:alpha-defn}
\end{eqnarray}
The Lie algebra of $SO(2n)$ can be described as imaginary antisymmetric
matrices, and we have implicitly absorbed factors of $-i$ in typical
definitions of Cartan generators into the Lie algebra matrices.
For example, $\rho^a_{i \alpha \beta}$ is essentially taken from
\cite{georgi}[chapter 19.1], and
$\alpha^a_{\mu \nu, \mu' \nu'}$ can be computed as the
commutator of a block-diagonal matrix with $i$ times  Pauli matrix $\sigma_2$
in one block, with the matrix of $Z$'s.
As a consistency check, note that
\begin{eqnarray*}
\sum_{\mu',\nu'} \alpha^a_{\mu \nu, \mu' \nu'} Z_{\mu' \nu'} & = &
\delta_{\mu, 2a-1} Z_{2a, \nu} + \delta_{\nu, 2a-1} Z_{\mu,2a}
- \delta_{\mu, 2a} Z_{2a-1,\nu} - \delta_{\nu,2a} Z_{\mu,2a-1}
\end{eqnarray*}
is antisymmetric in $\mu$, $\nu$, as expected.

Now, let us describe the Weyl group orbifold explicitly, and check that the
superpotential is invariant.
The Weyl group $W$
of $SO(2k)$ is \cite{fultonharris}[section 18.1]
$K \rtimes S_k \subset ({\mathbb Z}_2)^k
\rtimes S_k$, equivalently
\begin{displaymath}
1 \: \longrightarrow \: K \: \longrightarrow \: W \: 
\longrightarrow \: S_k \: \longrightarrow \: 1,
\end{displaymath}
where $K$ is the subgroup of $({\mathbb Z}_2)^k$ with
an even number of nontrivial generators, and
$({\mathbb Z}_2)^k \rtimes S_k$ is the Weyl group of $SO(2k+1)$.
The $S_k$ permutes the $S^1$ factors in the maximal torus as well as
factors in $({\mathbb Z}_2)^k$, and generator of
the $i$th ${\mathbb Z}_2$ in
$({\mathbb Z}_2)^k$ multiplies the $i$th factor in the maximal torus by
$-1$.

The Weyl group action on the $\sigma_a$ is
identical to its action on the $\sigma_a$ of the A-model Coulomb
branch.  Specifically, the Weyl group acts on $\sigma_a$'s by
permutating the $a$'s and also by acting on the $\sigma_a$'s with
signs, as
(see {\it e.g.} \cite{Hori:2011pd}[equ'n (4.16)]
\begin{equation}
\sigma_a \: \mapsto \: \epsilon_a \sigma_a,
\end{equation} 
where for gauge group $SO(2k)$ the $\epsilon_a$ obey
\begin{displaymath}
\epsilon_1 \cdots \epsilon_k \: = \: 1,
\end{displaymath}
(The Weyl group for $SO(2k+1)$ is nearly identical, except that there
is no constraint on the product of the signs.)

The Weyl group action on the $Y$'s is to permute two-element blocks
defined by 
\begin{displaymath}
\{ Y_{i, 2a-1}, Y_{i, 2a} \}, 
\end{displaymath}
and the sign elements of
the Weyl group, that flip the signs of the $\sigma_a$'s,
act on the $Y$'s by
exchanging elements of a given two-element block, as $Y_{i, {\rm even}}
\leftrightarrow Y_{i, {\rm odd}}$.
The Weyl group action on $Z$'s is similar:  a combination of
permutations of $2 \times 2$ blocks, and interchanging odd and even
elements of a given block.

Now, let us check invariance of the superpotential.
The terms
\begin{equation}
\sum_{i \alpha} \exp\left( - Y_{i \alpha} \right) 
\: + \: 
\sum_{\mu < \nu} X_{\mu \nu}
 \: - \:
\sum_{i, \alpha} \tilde{m}_i 
 Y_{i \alpha}
\end{equation}
are invariant because the Weyl group simply interchanges $Y$'s with $Y$'s
and $X$'s with $X$'s.
The constraint terms
\begin{equation}
\sum_{a=1}^k \sigma_a \left(
\sum_{i \alpha \beta} \rho^a_{i \alpha \beta} Y_{i \beta} \: + \:
\sum_{\mu < \nu; \mu' , \nu'} \alpha^a_{\mu \nu, \mu' \nu'} Z_{\mu' \nu'}
\: - \: t \right)
\end{equation}
are also invariant, but less trivially.  The part of the Weyl group
that acts by permutations of $a$'s manifestly leaves this
expression intact.  The part of the Weyl group that acts by ${\mathbb Z}_2$'s
is less trivial.  From the expressions~(\ref{eq:so2k:rho-defn}),
(\ref{eq:so2k:alpha-defn})
for $\rho^a_{i \alpha \beta}$ and $\alpha^a_{\mu \nu, \mu' \nu'}$,
we see that under the sign part of the Weyl group, the terms
\begin{displaymath}
\sum_{i \alpha \beta} \rho^a_{i \alpha \beta} Y_{i \beta} \: + \:
\sum_{\mu < \nu; \mu' , \nu'} \alpha^a_{\mu \nu, \mu' \nu'} Z_{\mu' \nu'}
\end{displaymath}
are antisymmetric.  That antisymmetry is cancelled out by the fact that
it is multiplied by $\sigma_a$ which also picks up a sign.
The only remaining term is $\sigma_a t$, and since $t$ is only defined
mod 2 (as it is a ${\mathbb Z}_2$ discrete theta angle), we see that this term 
too is well-defined.  Thus, the superpotential is invariant under
all parts of the Weyl group.

As in previous examples, the critical locus of the
superpotential is located away from orbifold fixed points, so we do not
expect any twisted sector contributions to vacua.

\subsection{Critical loci}

Now that we have defined the model and checked that the
superpotential is invariant under the orbifold group action,
let us begin analyzing critical loci.
Integrating out the $\sigma_a$ gives the constraint
\begin{equation}
 \sum_{i=1}^n \left(Y_{i,2a} - Y_{i, 2a-1}\right)
\: + \:  \sum_{\nu > 2a} \left( Z_{2a, \nu} - Z_{2a-1,\nu} \right)
\: + \: \sum_{\mu < 2a-1}  \left( Z_{\mu,2a} - Z_{\mu,2a-1} \right)
\: = \: t,
\end{equation}
which we can use to eliminate $Y_{n, 2a}$ as
\begin{equation}
Y_{n, 2a} \: = \:
- \sum_{i=1}^{n-1} Y_{i,2a} \: + \: \sum_{i=1}^n Y_{i, 2a-1}
\: - \:
 \sum_{\nu > 2a} \left( Z_{2a, \nu} - Z_{2a-1,\nu} \right)
\: - \: \sum_{\mu < 2a-1}  \left( Z_{\mu,2a} - Z_{\mu,2a-1} \right)
\: + \: t.
\end{equation}
Define
\begin{eqnarray}
\Pi_a & = & \exp\left( - Y_{n, 2a} \right), \\
& = & q 
\left( \prod_{i=1}^{n-1} \exp\left( + Y_{i, 2a} \right) \right)
\left( \prod_{i=1}^n \exp\left( - Y_{i, 2a-1} \right) \right)
\cdot \nonumber \\
& & \hspace*{1in} \cdot
\left( \prod_{\nu > 2a}
\frac{X_{2a-1,\nu} }{ X_{2a,\nu} }
\right)
\left( \prod_{\mu < 2a-1}
\frac{ X_{\mu,2a-1} }{ X_{\mu,2a} } \right),
\label{eq:so2k:pi-defn}
\end{eqnarray}
where $q = \exp(- t)$.

The superpotential then reduces to
\begin{eqnarray}
W & = &
\sum_{i=1}^{n-1} \sum_{a=1}^k \exp\left( - Y_{i, 2a} \right)
\: + \:
\sum_{i=1}^n \sum_{a=1}^k \exp\left( - Y_{i, 2a-1} \right)
\: + \:
\sum_{\mu < \nu} X_{\mu \nu} 
\: + \:
\sum_{a=1}^k \Pi_a
\nonumber \\
& & \: - \:
\sum_{i=1}^{n-1} \sum_a \tilde{m}_i Y_{i, 2a} 
\: - \: \sum_{i=1}^n \sum_a \tilde{m}_i Y_{i, 2a-1}
\nonumber \\
& & \: - \:
\tilde{m}_n \sum_a \Biggl(
- \sum_{i=1}^{n-1} Y_{i,2a} \: + \: \sum_{i=1}^n Y_{i, 2a-1}
\: - \: 
 \sum_{\nu > 2a} \left( Z_{2a, \nu} - Z_{2a-1,\nu} \right)
\nonumber \\
& & \hspace*{1.5in}
\: - \:
\sum_{\mu < 2a-1}  \left( Z_{\mu,2a} - Z_{\mu,2a-1} \right)
\Biggr)
\end{eqnarray}
(omitting constant terms).
Note that $X_{2a,\nu} \neq 0$ for $a>1$ and
$X_{\mu,2a} \neq 0$, as the potential diverges at such points.

The critical locus is computed as follows:
\begin{eqnarray*}
\frac{\partial W}{\partial Y_{i, 2a}}: & &
\exp\left( - Y_{i, 2a} \right) \: = \: \Pi_a - \tilde{m}_i + \tilde{m}_n,
\\
\frac{\partial W}{\partial Y_{i, 2a-1} }: & & 
\exp\left( - Y_{i, 2a-1} \right) \: = \:
- \Pi_a - \tilde{m}_i - \tilde{m}_n,
\\
\frac{\partial W}{\partial X_{\mu \nu} }: & &
X_{2a,2b} \: = \: \Pi_a + \Pi_b + 2 \tilde{m}_n \mbox{   for } a < b, \\
& &
X_{2a,2b-1} \: = \: \Pi_a - \Pi_b \mbox{   for }a < b, \\
& &
X_{2a-1,2b} \: = \: - \Pi_a + \Pi_b \mbox{   for } a < b, \\
& & 
X_{2a-1,2b-1} \: = \: - \Pi_a - \Pi_b - 2 \tilde{m}_n \mbox{   for } a < b.
\end{eqnarray*}
The constraints on $X_{\mu \nu}$ can be summarized as
\begin{eqnarray}
X_{\mu \nu} & = & 
\sum_a \left( \delta_{\mu, 2a} - \delta_{\mu, 2a-1}\right) \Pi_a
\: + \:
 \sum_a \left( \delta_{\nu, 2a} - \delta_{\nu, 2a-1} \right)\Pi_a 
\nonumber \\
& & \hspace*{1.0in}
\: + \:
\tilde{m}_n \sum_a \left( \delta_{\mu, 2a} - \delta_{\mu, 2a-1} \right)
\: + \:
\tilde{m}_n \sum_a \left( \delta_{\nu, 2a} - \delta_{\nu, 2a-1} \right), 
\end{eqnarray}
for $\mu < \nu$.
The cases $X_{2a-1,2a}$ are omitted as these do not correspond to
propagating fields -- they correspond to elements of the pertinent Cartan,
and so there are no corresponding W bosons.

Since $\exp(-Y) \neq 0$, we see immediately from the first two lines
that 
\begin{equation}  \label{eq:so2n:constr-1}
\Pi_a + \tilde{m}_n \: \neq \: \pm \tilde{m}_i,
\end{equation}
for any $a$, $i$.  In addition, since as in section~\ref{sect:gkn:excluded}
we exclude the loci
$X_{2a,\nu} = 0$ and since for $a \neq b$,
\begin{displaymath}
X_{2a,2b} \: = \: \Pi_a + \Pi_b + 2 \tilde{m}_n,
\: \: \:
X_{2a,2b-1} \: = \: \Pi_a - \Pi_b,
\end{displaymath}
we see that the $\Pi_a$ must be distinct from one another, and in addition,
\begin{displaymath}
\Pi_a + \tilde{m}_n \: \neq \: - \left( \Pi_b + \tilde{m}_n \right)
\end{displaymath}
for $a \neq b$, hence in general for $a \neq b$,
\begin{equation}   \label{eq:so2n:constr-2}
\Pi_a + \tilde{m}_n \: \neq \: \pm \left( \Pi_b + \tilde{m}_n \right).
\end{equation}

On the critical locus, from the definition~(\ref{eq:so2k:pi-defn}) of
$\Pi_a$, we find
\begin{equation}  \label{eq:so2k:pi-equn}
\Pi_a \prod_{i=1}^{n-1} \left(  \Pi_a - \tilde{m}_i + \tilde{m}_n \right)
\: = \:
q  
\prod_{i=1}^{n} \left( - \Pi_a - \tilde{m}_i - \tilde{m}_n \right) .
\end{equation}

Let us now compare to the results for the A-twisted GLSM
described in \cite{Hori:2011pd}[section 4].
First, the operator mirror map~(\ref{eq:op-mirror-mass-1}),
(\ref{eq:op-mirror-mass-2}) is given by
\begin{eqnarray}
\exp\left( - Y_{i \alpha} \right) & = & - \tilde{m}_i \: + \:
\sum_{a=1}^k \sum_{i, \beta} \sigma_a \rho^a_{i \beta \alpha},
\\
& = & - \tilde{m}_i \: + \:  \sum_a \left( \delta_{\alpha, 2a} \sigma_a \: - \:
\delta_{\alpha, 2a-1} \sigma_a \right), \\
& = & - \tilde{m}_i \: + \:
\left\{ \begin{array}{cl}
 \sigma_a & \alpha = 2a, \\
- \sigma_a & \alpha = 2a-1,
\end{array} \right.
\\
X_{\mu \nu}  & = &
\sum_{a=1}^k \sum_{\mu'< \nu'} \sigma_a \alpha^a_{\mu' \nu', \mu \nu},
\\
& = &  \sum_a \left(
\delta_{\mu, 2a} - \delta_{\mu, 2a-1} + \delta_{\nu, 2a} - 
\delta_{\nu, 2a-1} \right) \sigma_a
\mbox{   for }\mu < \nu.
\end{eqnarray}

Comparing the operator mirror map above and the critical locus equations,
we find
\begin{eqnarray*}
\exp\left(-Y_{i, 2a} \right)
& = & \Pi_a - \tilde{m}_i + \tilde{m}_n, \\
& = & - \tilde{m}_i + \sigma_a, 
\\
\exp\left( -Y_{i, 2a-1} \right) & = &
- \Pi_a - \tilde{m}_i - \tilde{m}_n, \\
& = & - \tilde{m}_i - \sigma_a,
\end{eqnarray*}
from which we find
\begin{displaymath}
\sigma_a \: = \:  \Pi_a + \tilde{m}_n .
\end{displaymath}
(Note that $Y_{n, 2a}$ was eliminated as a dynamical field, but we
have used the definition of $\Pi_a$ to extend the expression
above for $\exp(-Y_{i, 2a})$ to the case $i=n$.)

Comparing expressions for $X$'s, we similarly find for $\mu < \nu$
\begin{eqnarray*}
X_{\mu \nu}  & = &
\sum_a \left( \delta_{\mu, 2a} - \delta_{\mu, 2a-1} + \delta_{\nu, 2a}
- \delta_{\nu, 2a-1} \right) \left( \Pi_a + \tilde{m}_n \right), 
\\
& = & \sum_a  \left( \delta_{\mu, 2a} - \delta_{\mu, 2a-1} + \delta_{\nu, 2a}
- \delta_{\nu, 2a-1} \right) \left(  \sigma_a \right),
\end{eqnarray*}
hence
\begin{displaymath}
\sigma_a \: = \:  \Pi_a + \tilde{m}_n ,
\end{displaymath}
consistent with the result above.

Applying the operator mirror map,
equation~(\ref{eq:so2k:pi-equn}) for $\Pi_a$ becomes
\begin{equation}  \label{eq:so2k:mirror-sigma-eqn}
\prod_{i=1}^n \left( \sigma_a - \tilde{m}_i \right)
\: = \:
q \prod_{i=1}^n \left( - \sigma_a - \tilde{m}_i \right).
\end{equation}

By comparison, reference \cite{Hori:2011pd}[equ'n (4.15)] gives for
this model for suitable discrete theta angle,
\begin{displaymath}
\prod_{i=1}^n \left( \sigma_a - \tilde{m}_i \right) \: = \:
(-)^{n+1} \prod_{i=1}^n \left( - \sigma_a - \tilde{m}_i\right),
\end{displaymath}
which matches~(\ref{eq:so2k:mirror-sigma-eqn}) for suitable $q$.
(Remember that since there is only a discrete ${\mathbb Z}_2$ theta angle
in this theory, $q = \pm 1$.)

Furthermore, reference \cite{Hori:2011pd}[equ'n (4.17)] gives the
following constraints on Coulomb vacua:
\begin{equation}
\sigma_a \: \neq \: \pm \tilde{m}_i, 
\: \: \:
\sigma_a \: \neq \: \pm \sigma_b \mbox{    for }a\neq b.
\end{equation}
Under the operator mirror map above, it is trivial to see that these match
the algebraic constraints~(\ref{eq:so2n:constr-1}),
(\ref{eq:so2n:constr-2}) that we derived in the proposed mirror B model.

As a consistency check, let us compare the Weyl group action on the
$\sigma$'s.  Briefly, the Weyl group for $SO(2k)$ acts by permuting the
$\sigma_a$ as well as by signs,
as (see {\it e.g.} \cite{Hori:2011pd}[equ'n (4.16)]
\begin{equation}
\sigma_a \: \mapsto \: \epsilon_a \sigma_a,
\end{equation}
where for gauge group $SO(2k)$ the $\epsilon_a$ obey
\begin{displaymath}
\epsilon_1 \cdots \epsilon_k \: = \: 1,
\end{displaymath}
(The Weyl group for $SO(2k+1)$ is nearly identical, except that there
is no constraint on the product of the signs.)
Here, we see that the Weyl group action on the $\sigma_a$ is
mirror to the Landau-Ginzburg model as follows:
\begin{itemize}
\item the signs on the $\sigma$'s act by exchanging $Y_{i, {\rm even}}$
with $Y_{i, {\rm odd}}$, and similarly on the $X$'s,
\item the permutation elements permute $a$'s.
\end{itemize}

In any event, since we now have the same equations 
as in \cite{Hori:2011pd}[section 4], the rest of the physics analysis
proceeds identically.
For example, consider the number of vacua.
Equation~(\ref{eq:so2k:mirror-sigma-eqn}) is symmetric under $\sigma \mapsto
- \sigma$, and since it is of degree $n$, if $n$ is even there are
$n/2$ pairs $\pm \sigma$ of roots.  Hence, in order to get $k$ distinct
vacua obeying $\sigma_a \neq \pm \sigma_b$ for $a \neq b$,
a necessary condition is that $n \geq 2k$.  If $n < 2k$, there are no
Coulomb branch vacua.  If there are also no mixed or Higgs branch vacua,
then supersymmetry is broken, as described in
\cite{Hori:2011pd}[section 4].

\section{Example:  $SO(2k+1)$ gauge theory}
\label{sect:so2kp1}

Next we will consider the mirror of an A-twisted $SO(2k+1)$ gauge theory
with $n$ chiral superfields in the vector representation.  Much of the
analysis will be very similar to the $SO(2k)$ case, so we will 
be brief.

Following our proposal, the mirror is a Weyl-group-orbifold of a
Landau-Ginzburg model with fields
\begin{itemize}
\item $Y_{i\alpha}$, $i \in \{1, \cdots, n\}$, $\alpha \in \{1, \cdots, 2k+1\}$,
\item $X_{\mu \nu} = \exp( - Z_{\mu \nu})$,
$X_{\mu \nu} =  X_{\nu \mu}^{-1}$, $Z_{\mu \nu} = - Z_{\nu \mu}$,
$\mu, \nu \in \{1, \cdots, 2k+1\}$,
(excluding\footnote{
These correspond to elements of a Cartan subalgebra we use in defining
constraints associated to the $\sigma_a$.
} $X_{2a-1,2a}$, $Z_{2a-1,2a}$),
corresponding to the W bosons,
\item $\sigma_a$, $a \in \{1, \cdots, k\}$,
\end{itemize}
and superpotential
\begin{eqnarray}
W & = &
\sum_{a=1}^k \sigma_a \left(
\sum_{i \alpha \beta} \rho^a_{i \alpha \beta} Y_{i \beta} \: + \:
\sum_{\mu < \nu; \mu' , \nu'} \alpha^a_{\mu \nu, \mu' \nu'} Z_{\mu' \nu'}
\: - \: t \right)
\nonumber \\
& & \: + \:
\sum_{i \alpha} \exp\left( - Y_{i \alpha} \right) 
\: + \: 
\sum_{\mu < \nu} X_{\mu \nu}
\nonumber \\
& & \: - \:
\sum_{i, \alpha} \tilde{m}_i \left( Y_{i \alpha} \: - \:
\sum_{\beta} \rho^a_{i \alpha \beta} t \right).
\end{eqnarray}
As before, there is a discrete ${\mathbb Z}_2$ theta angle,
and
\begin{eqnarray}
\rho^a_{i \alpha \beta} & = &  \delta_{\alpha, 2a-1}
\delta_{\beta, 2a} - \delta_{\beta, 2a-1} \delta_{\alpha, 2a} ,
\label{eq:so2k1:rho-defn}
\\
\alpha^a_{\mu \nu, \mu' \nu'} & = &
 \delta_{\nu \nu'} \left( \delta_{\mu, 2a-1} \delta_{\mu', 2a} -
\delta_{\mu, 2a} \delta_{\mu', 2a-1} \right)
\nonumber \\
& & \hspace*{0.5in}
\: + \:
 \delta_{\mu \mu'} \left( \delta_{\nu, 2a-1} \delta_{\nu', 2a} - 
\delta_{\nu, 2a} \delta_{\nu', 2a-1} \right),
\label{eq:so2k1:alpha-defn}
\end{eqnarray}
identical to the $SO(2k)$ case.  Note that the group action encoded
in $\rho$ leaves $Y_{i,2k+1}$ invariant -- this reflects the fact that the
vector representation has one component which is neutral under the Cartan
subalgebra \cite{georgi}[chapter 19.2].

The Weyl group action is nearly identical to the $SO(2k)$ case -- an extension
\begin{displaymath}
1 \: \longrightarrow \: ({\mathbb Z}_2)^k \: \longrightarrow \: W \: 
\longrightarrow \: S_k \: \longrightarrow \: 1,
\end{displaymath}
with the only difference being that the kernel is all of
$({\mathbb Z}_2)^k$, rather than the subgroup that leaves the overall
product invariant.  Its action on the fields is the same as before,
leaving $Y_{i,2k+1}$ invariant, for example.

Integrating out the $\sigma_a$ gives the same constraint as before,
\begin{equation}
 \sum_{i=1}^n \left(Y_{i,2a} - Y_{i, 2a-1}\right)
\: + \:  \sum_{\nu > 2a} \left( Z_{2a, \nu} - Z_{2a-1,\nu} \right)
\: + \: \sum_{\mu < 2a-1}  \left( Z_{\mu,2a} - Z_{\mu,2a-1} \right)
\: = \: t,
\end{equation}
which we can use to eliminate $Y_{n, 2a}$ as
\begin{equation}
Y_{n, 2a} \: = \:
- \sum_{i=1}^{n-1} Y_{i,2a} \: + \: \sum_{i=1}^n Y_{i, 2a-1}
\: - \: 
 \sum_{\nu > 2a} \left( Z_{2a, \nu} - Z_{2a-1,\nu} \right)
\: - \: \sum_{\mu < 2a-1}  \left( Z_{\mu,2a} - Z_{\mu,2a-1} \right)
\: + \: t. 
\end{equation}
As before, we define
\begin{eqnarray}
\Pi_a & = & \exp\left( - Y_{n, 2a} \right), \\
& = & q 
\left( \prod_{i=1}^{n-1} \exp\left( + Y_{i, 2a} \right) \right)
\left( \prod_{i=1}^n \exp\left( - Y_{i, 2a-1} \right) \right)
\cdot \nonumber \\
& & \hspace*{1in} \cdot
\left( \prod_{\nu > 2a}
\frac{X_{2a-1,\nu} }{ X_{2a,\nu} }
\right)
\left( \prod_{\mu < 2a-1}
\frac{ X_{\mu,2a-1} }{ X_{\mu,2a} } \right),
\label{eq:so2k1:pi-defn}
\end{eqnarray}
where $q = \exp(- t)$.

The superpotential then reduces to
\begin{eqnarray}
W & = &
\sum_{i=1}^{n-1} \sum_{a=1}^k \exp\left( - Y_{i, 2a} \right)
\: + \:
\sum_{i=1}^n \sum_{a=1}^k \exp\left( - Y_{i, 2a-1} \right)
\: + \:
\sum_{i=1}^n \exp\left( - Y_{i, 2k+1} \right)
\nonumber \\
& & \: + \:
\sum_{\mu < \nu} X_{\mu \nu} 
\: + \:
\sum_{a=1}^k \Pi_a
\: - \:
\sum_{i=1}^{n-1} \sum_a \tilde{m}_i Y_{i, 2a} 
\: - \: \sum_{i=1}^n \sum_a \tilde{m}_i Y_{i, 2a-1}
\: - \: \sum_{i=1}^n \tilde{m}_i Y_{i, 2k+1}
\nonumber \\
& & \: - \:
\tilde{m}_n \sum_a \Biggl(
- \sum_{i=1}^{n-1} Y_{i,2a} \: + \: \sum_{i=1}^n Y_{i, 2a-1}
\: - \: 
 \sum_{\nu > 2a} \left( Z_{2a, \nu} - Z_{2a-1,\nu} \right)
\nonumber \\
& & \hspace*{1.5in}
\: - \:
\sum_{\mu < 2a-1}  \left( Z_{\mu,2a} - Z_{\mu,2a-1} \right)
\Biggr)
\end{eqnarray}
(omitting constant terms).
For reasons described earlier, we take $X_{2a,\nu} \neq 0$ for $a>1$ and
$X_{\mu,2a} \neq 0$.

Note that in this construction
the fields $Y_{i,2k+1}$ are essentially decoupled from the
rest of the fields, and if we omit the twisted masses, appear in the
superpotential above only as
\begin{displaymath}
\sum_i \exp\left(- Y_{i,2k+1}\right).
\end{displaymath}
This reflects the fact that generically on the Coulomb branch of the
original gauge theory, they act
as uncharged free fields with respect to the Cartan subalgebra we have
chosen.  (Of course, in the entire original gauge theory, they are not
decoupled, but rather appear decoupled at low energies on the Coulomb
branch, which is the reason for this artifact of our construction.)  
For purposes of comparison, the Hori-Vafa
mirror \cite{Hori:2000kt} of a single uncharged free chiral superfield consists
of a single field $Y$ with superpotential
$\exp(-Y)$.  This is exactly the structure we see above for the
fields $Y_{i,2k+1})$, as expected.
(We saw the same phenomenon in the case of adjoint-valued fields
in section~\ref{sect:adj}.)

The critical locus is computed as before:
\begin{eqnarray*}
\frac{\partial W}{\partial Y_{i, 2a}}: & &
\exp\left( - Y_{i, 2a} \right) \: = \: \Pi_a - \tilde{m}_i + \tilde{m}_n,
\\
\frac{\partial W}{\partial Y_{i, 2a-1} }: & & 
\exp\left( - Y_{i, 2a-1} \right) \: = \:
- \Pi_a - \tilde{m}_i - \tilde{m}_n,
\\
\frac{\partial W}{\partial Y_{i,2k+1} }: & &
\exp\left( - Y_{i,2k+1} \right) \: = \:
- \tilde{m}_i,
\\
\frac{\partial W}{\partial X_{\mu \nu} }: & &
X_{2a,2b} \: = \: \Pi_a + \Pi_b + 2 \tilde{m}_n \mbox{   for } a < b, \\
& &
X_{2a,2b-1} \: = \: \Pi_a - \Pi_b \mbox{   for }a < b, \\
& &
X_{2a,2k+1} \: = \: \Pi_a + \tilde{m}_n,  \\
& &
X_{2a-1,2b} \: = \: - \Pi_a + \Pi_b \mbox{   for } a < b, \\
& & 
X_{2a-1,2b-1} \: = \: - \Pi_a - \Pi_b - 2 \tilde{m}_n \mbox{   for } a < b.
\\ & &
X_{2a-1,2k+1} \: = \:  - \Pi_a - \tilde{m}_n.
\end{eqnarray*}
The constraints on $X_{\mu \nu}$ can be summarized as
\begin{eqnarray}
X_{\mu \nu} & = & 
\sum_a \left( \delta_{\mu, 2a} - \delta_{\mu, 2a-1}\right) \left( \Pi_a
+ \tilde{m}_n \right) \: + \:
 \sum_a \left( \delta_{\nu, 2a} - \delta_{\nu, 2a-1} \right)
\left( \Pi_a + \tilde{m}_n \right),
\end{eqnarray}
for $\mu < \nu$.  As before, the cases $X_{2a-1,2a}$ are omitted as these
are not propagating fields, instead corresponding to elements of the
pertinent Cartan.

Since $\exp(-Y) \neq 0$, we derive from the first three lines that
\begin{equation} \label{eq:so2k1:constr-1}
\Pi_a + \tilde{m}_n \: \neq \: \pm \tilde{m}_i.
\end{equation}
(We omit from consideration the decoupled mirrors $Y_{i,2k+1}$ of
free fields in the original gauge theory.)

Since we exclude $X_{2a,\nu}=0$, for reasons discussed earlier
in section~\ref{sect:gkn:excluded},
we find for $a \neq b$,
\begin{equation} \label{eq:so2k1:constr-2}
\Pi_a + \tilde{m}_n \: \neq \: \pm \left( \Pi_b + \tilde{m}_n \right),
\: \: \:
\Pi_a + \tilde{m}_n \: \neq \: 0.
\end{equation}

On the critical locus, from the definition~(\ref{eq:so2k1:pi-defn}) of
$\Pi_a$, we find
\begin{equation} \label{eq:so2k1:pi-eqn}
\Pi_a \prod_{i=1}^{n-1} \left(  \Pi_a - \tilde{m}_i + \tilde{m}_n \right)
\: = \:
-q  
\prod_{i=1}^{n} \left( - \Pi_a - \tilde{m}_i - \tilde{m}_n \right) .
\end{equation}
This is nearly identical to~(\ref{eq:so2k:pi-equn}) for the $SO(2k)$
gauge theory, except for an extra
minus sign, arising from the contribution of $X_{\mu,2k+1}$.

We can compute the operator mirror map~(\ref{eq:op-mirror-mass-1}),
(\ref{eq:op-mirror-mass-2}) as before:
\begin{eqnarray}
\exp\left( - Y_{i \alpha} \right) & = & - \tilde{m}_i \: + \:
\sum_{a=1}^k \sum_{i, \beta} \sigma_a \rho^a_{i \beta \alpha},
\\
& = & - \tilde{m}_i \: + \:  \sum_a \left( \delta_{\alpha, 2a} \sigma_a \: - \:
\delta_{\alpha, 2a-1} \sigma_a \right), \\
& = & - \tilde{m}_i \: + \:
\left\{ \begin{array}{cl}
 \sigma_a & \alpha = 2a, \\
- \sigma_a & \alpha = 2a-1, \\
0 & \alpha=2k+1,
\end{array} \right.
\\
X_{\mu \nu}  & = &
\sum_{a=1}^k \sum_{\mu'< \nu'} \sigma_a \alpha^a_{\mu' \nu', \mu \nu},
\\
& = &  \sum_a \left(
\delta_{\mu, 2a} - \delta_{\mu, 2a-1} + \delta_{\nu, 2a} - 
\delta_{\nu, 2a-1} \right) \sigma_a
\mbox{   for }\mu < \nu.
\end{eqnarray}
Comparing with the critical locus, we find
\begin{equation}
\sigma_a \: = \: \Pi_a + \tilde{m}_n.
\end{equation}

Under the operator mirror map, equation~(\ref{eq:so2k1:pi-eqn}) for
$\Pi_a$ becomes
\begin{equation} \label{eq:so2k1:mirror-sigma-eqn}
\prod_{i=1}^n \left( \sigma_a - \tilde{m}_i \right)
\: = \:
- q \prod_{i=1}^n \left( - \sigma_a - \tilde{m}_i \right).
\end{equation}
Given that $q = \pm 1$, this matches
\cite{Hori:2011pd}[equ'n (4.15)], which is written for a particular
choice of discrete theta angle.

Furthermore, the constraints we derived~(\ref{eq:so2k1:constr-1}),
(\ref{eq:so2k1:constr-2}) are easily
seen to mirror the constraints
\begin{equation}
\sigma_a \: \neq \: \pm \tilde{m}_i,
\: \: \:
\sigma_a \: \neq \: \pm \sigma_b \mbox{  for } a \neq b,
\: \: \:
\sigma_a \: \neq \: 0
\end{equation}
given in \cite{Hori:2011pd}[equ'n (4.17)].

\section{Example:  $Sp(2k)$ gauge theory}
\label{sect:sp2k}

Consider a $Sp(2k) = USp(2k)$ gauge theory (in conventions in which the
gauge group has rank $k$) with $n$ chiral superfields
in the fundamental representation.

Following our proposal, the mirror is a Weyl-group-orbifold of a 
Landau-Ginzburg model with fields
\begin{itemize}
\item $Y_{i \mu}$, $i \in \{1, \cdots, n \}$,
$\mu \in \{1, \cdots, 2k\}$,
mirror to the $n$ chiral superfields in the fundamental representation,
\item $X_{\mu \nu} = \exp(-Z_{\mu \nu})$ for $\mu \leq \nu$,
$\mu, \nu \in \{1, \cdots, 2k\}$, excluding
$X_{2a-1,2a}$ (which would be mirror to the Cartan subalgebra),
\item $\sigma_a$, $a \in \{1, \cdots, k\}$,
\end{itemize}
and superpotential
\begin{eqnarray}
W & = & \sum_{a=1}^k \sigma_a \Biggl( \sum_i \rho^a_{i \mu} Y_{i \mu}
\: + \: \sum_{\mu} \alpha^a_{\mu \mu} Z_{\mu \mu}
\nonumber \\
& & \hspace*{0.5in} \: + \: \sum_{b<c} \left(
\alpha^a_{2b,2c} Z_{2b,2c} + \alpha^a_{2b-1,2c-1} X_{2b-1,2c-1}
+ \alpha^a_{2b-1,2c} X_{2b-1,2c} + \alpha^a_{2b,2c-1} X_{2b,2c-1}
\right) \Biggr)
\nonumber \\
& & 
\: + \:
\sum_{i \mu} \exp\left( - Y_{i \mu} \right)
 \: - \:  \sum_{i \mu} \tilde{m}_i Y_{i \mu}
\nonumber \\
& & 
\: + \: \sum_{\mu} X_{\mu \mu} \: + \: 
\sum_{a < b} \left( X_{2a,2b} + X_{2a-1,2b-1} +
X_{2a-1,2b} + X_{2a,2b-1} \right).
\end{eqnarray}
Since the gauge group is semisimple, there is no continuous FI
parameter, and in fact in this case there is also
no discrete FI parameter \cite{Hori:2011pd}[section 5.2].
(Here in the proposed mirror, 
the Weyl group action restricts possible FI-like terms to the values
$0, \pi i$, just as in the $SO$ groups.  Turning on $\pi i$ is not
explicitly incompatible with the orbifold,
but also does not\footnote{
The center of $Sp(2k)$ is ${\mathbb Z}_2$ 
(see {\it e.g.} \cite{Distler:2007av}[appendix A]), and an
$Sp(2k)/{\mathbb Z}_2$ gauge theory
would have a discrete FI parameter of the form above; however, 
the center acts nontrivially on fundamental matter, so in the A-twisted
gauge theory above, the gauge group cannot be $Sp(2k)/{\mathbb Z}_2$.
Nevertheless, this could be relevant for mirrors to $Sp(2k)$ gauge
theories with different matter content than the theory whose mirror
we compute above.
}, so far as we are aware, yield a result that is the mirror
of an $Sp(2k)$ gauge theory with fundamental matter.)

As a consistency check, $Sp(2k)$ has dimension $k(2k+1)$,
the same number of elements in a symmetric $2k \times 2k$ matrix,
matching the number of $X_{\mu \nu}$ plus omitted Cartan mirrors.

Following \cite{georgi}[section 26],
\begin{eqnarray}
\rho^a_{i \mu} & = & \delta_{\mu, 2a} - \delta_{\mu, 2a-1}, 
\\
\alpha^a_{\mu \nu} & = & 
\delta_{\mu, 2a} - \delta_{\mu, 2a-1} +
\delta_{\nu, 2a} - \delta_{\nu, 2a-1}.
\end{eqnarray}

The Weyl group of $Sp(2k)$ has the same form as that of 
$SO(2k+1)$:  it is an extension \cite{fultonharris}[section 16.1]
\begin{equation}
1 \: \longrightarrow \: ( {\mathbb Z}_2 )^k \: \longrightarrow \: W
\: \longrightarrow \: S_k \: \longrightarrow \: 1.
\end{equation}
Furthermore, its action is similar:  the permutation group
$S_k$ exchanges $a$'s, and the signs $( {\mathbb Z}_2 )^k$ act by
interchanging pairs ($2a-1,2a$) and flipping the sign of $\sigma_a$'s.
For example, under the $ {\mathbb Z}_2 )^k$,
the $Y$'s are interchanged with no signs 
($Y_{i,2a} \leftrightarrow Y_{i,2a-1}$),
but
\begin{displaymath}
\sum_{i,\mu} \rho^a_{i \mu} Y_{i\mu} \: \mapsto \:
- \sum_{i,\mu} \rho^a_{i \mu} Y_{i\mu}.
\end{displaymath}
Because this is multiplied by $\sigma_a$ in the superpotential, 
the superpotential term is
invariant.  Other terms can be checked similarly.

Now, integrating out the $\sigma_a$ gives the constraint
\begin{equation}
\sum_{i=1}^n \left( Y_{i, 2a} - Y_{i, 2a-1} \right)
\: + \: \sum_{\mu \leq \nu}  \left( \delta_{\mu, 2a} 
 - \delta_{\mu, 2a-1} +
\delta_{\nu, 2a} - \delta_{\nu, 2a-1}\right) Z_{\mu \nu} \: = \: 0.
\end{equation}
We can use this to eliminate $Y_{n,2a}$:
\begin{eqnarray}
Y_{n,2a} & = & - \sum_{i=1}^{n-1} Y_{i,2a} \: + \:
\sum_{i=1}^n Y_{i,2a-1} \: - \: \sum_{\nu \geq 2a} Z_{2a,\nu}
\: + \: \sum_{\nu \geq 2a-1} Z_{2a-1,\nu}
\nonumber \\
& & \hspace*{1.75in}  \: - \:
\sum_{\mu \leq 2a} Z_{\mu,2a} \: + \:
\sum_{\mu \leq 2a-1} Z_{\mu,2a-1}.
\end{eqnarray}
Define
\begin{eqnarray}
\Pi_a & = & \exp\left( - Y_{n,2a} \right), \\
& = &
\left( \prod_{i=1}^{n-1} \exp\left( + Y_{i,2a} \right) \right)
\left( \prod_{i=1}^n \exp\left( - Y_{i,2a-1} \right) \right)
\cdot 
\nonumber \\
& & \hspace*{1in} \cdot
\left( \prod_{\nu \geq 2a} \frac{X_{2a-1,\nu}}{X_{2a,\nu} } \right)
\left( \prod_{\mu \leq 2a-1} \frac{ X_{\mu,2a-1} }{ X_{\mu,2a} } \right)
\frac{ X_{2a-1,2a-1} }{ X_{2a,2a} }.
\label{eq:sp2k:pi-defn}
\end{eqnarray}
The superpotential then becomes
\begin{eqnarray}
W & = &
\sum_{i=1}^{n-1} \sum_{a=1}^k \exp\left( - Y_{i,2a} \right)
\: + \:
\sum_{i=1}^n \sum_{a=1}^k \exp\left( - Y_{i,2a-1} \right)
\: + \: \sum_{a=1}^k \Pi_a
\nonumber \\
& & 
\: + \: \sum_{\mu} X_{\mu \mu} \: + \: 
\sum_{a < b} \left( X_{2a,2b} + X_{2a-1,2b-1} +
X_{2a-1,2b} + X_{2a,2b-1} \right)
\nonumber \\
& & 
\: - \: \sum_{i=1}^{n-1} \sum_{a=1}^{k}  \tilde{m}_i Y_{i,2a }
\: - \: \sum_{i=1}^n \sum_{a=1}^k \tilde{m}_i Y_{i,2a-1}
\nonumber \\
& & 
\: - \: \tilde{m}_n \sum_{a=1}^k \Biggl(
- \sum_{i=1}^{n-1} Y_{i,2a} \: + \:
\sum_{i=1}^n Y_{i,2a-1} \: - \: \sum_{\nu \geq 2a} Z_{2a,\nu}
\: + \: \sum_{\nu \geq 2a-1} Z_{2a-1,\nu}
\nonumber \\
& & \hspace*{2in}  \: - \:
\sum_{\mu \leq 2a} Z_{\mu,2a} \: + \:
\sum_{\mu \leq 2a-1} Z_{\mu,2a-1}
\Biggr).
\end{eqnarray}

We compute the critical locus as follows:
\begin{eqnarray*}
\frac{\partial W}{\partial Y_{i,2a}}: & &
\exp\left( - Y_{i,2a} \right) \: = \:
\Pi_a - \tilde{m}_i + \tilde{m}_n,
\\
\frac{\partial W}{\partial Y_{i,2a-1}}: & &
\exp\left( - Y_{i,2a-1} \right) \: = \: - \Pi_a - \tilde{m}_i -
\tilde{m}_n,
\\
\frac{\partial W}{\partial X_{2a,2b}}: & &
X_{2a,2b} \: = \:  \Pi_a + \Pi_b + 2 \tilde{m}_n \mbox{  for } a \leq b,
\\
\frac{\partial W}{\partial X_{2a-1,2b-1}}: & &
X_{2a-1,2b-1} \: = \: - \Pi_a - \Pi_b - 2 \tilde{m}_n \mbox{  for } a \leq b,
\\
\frac{\partial W}{\partial X_{2a,2b-1}}: & &
X_{2a,2b-1} \: = \: \Pi_a - \Pi_b,
\\
\frac{\partial W}{\partial X_{2a-1,2b}}: & &
X_{2a-1,2b} \: = \: - \Pi_a + \Pi_b.
\end{eqnarray*}

Since $\exp(-Y) \neq 0$, we find from the first two critical locus
equations that
\begin{equation}  \label{eq:sp2k:constr-1}
\Pi_a + \tilde{m}_n \: \neq \: \pm \tilde{m}_i.
\end{equation}
Similarly, since we exclude $X_{2a,\nu} = 0$ and
$X_{\mu,2a} = 0$, for reasons discussed earlier 
in section~\ref{sect:gkn:excluded}, we have that
\begin{equation} \label{eq:sp2k:constr-2}
\Pi_a \neq \Pi_b \mbox{ for }a \neq b, \: \: \:
\Pi_a + \Pi_b + 2 \tilde{m}_n \neq 0.
\end{equation}

Plugging into equation~(\ref{eq:sp2k:pi-defn}), we find that on the critical
locus,
\begin{eqnarray}
\Pi_a & = &
\left( \prod_{i=1}^{n-1} \frac{1}{ \Pi_a - \tilde{m}_i + \tilde{m}_n }
\right)
\left( \prod_{i=1}^n \left( - \Pi_a - \tilde{m}_i - \tilde{m}_n \right)
\right)
\cdot \nonumber \\
& & \cdot
\left( \prod_{b > a} \frac{ \left( -\Pi_a + \Pi_b \right)
\left( -\Pi_a - \Pi_b - 2 \tilde{m}_n \right) }{
\left( \Pi_a + \Pi_b + 2 \tilde{m}_n \right)
\left( \Pi_a - \Pi_b \right) }
\right)
\left( \frac{1}{2 \Pi_a + 2 \tilde{m}_n } \right)
\cdot \nonumber \\
& & \cdot
\left( \prod_{b < a} \frac{ \left( \Pi_b - \Pi_a \right)
\left( - \Pi_b - \Pi_a - 2 \tilde{m}_n \right) }{
\left( \Pi_b + \Pi_a + 2 \tilde{m}_n \right)
\left( - \Pi_b + \Pi_a \right) } \right)
\left( - 2 \Pi_a - 2 \tilde{m}_n \right)
\cdot \nonumber \\
& & \hspace*{2in} \cdot
\left( \frac{ -2 \Pi_a - 2 \tilde{m}_n }{2 \Pi_a + 2 \tilde{m}_n } \right).
\end{eqnarray}
This simplifies to
\begin{equation} \label{eq:sp2k:pi-reln}
\Pi_a \left( \prod_{i=1}^{n-1} \left(  \Pi_a - \tilde{m}_i + \tilde{m}_n
\right) \right)
\: = \:
\prod_{i=1}^n \left( - \Pi_a - \tilde{m}_i - \tilde{m}_n \right).
\end{equation}

From equations~(\ref{eq:op-mirror-mass-1}), (\ref{eq:op-mirror-mass-2}), 
the operator mirror map takes the form
\begin{eqnarray}
\exp\left( - Y_{i\mu} \right) & = & \sum_a \sigma_a \rho^a_{i \mu}
- \tilde{m}_i,
\\
X_{\mu \nu} & = & \sum_a \sigma_a \alpha^a_{\mu \nu},
\end{eqnarray}
or more concretely,
\begin{eqnarray}
\exp\left( - Y_{i,2a} \right) & = & \sigma_a - \tilde{m}_i,
\\
\exp\left( - Y_{i,2a-1} \right) & = & - \sigma_a - \tilde{m}_i, 
\\
X_{2a,2b} & = & \sigma_a + \sigma_b, 
\\
X_{2a-1,2b-1} & = & - \sigma_a - \sigma_b,
\\
X_{2a,2b-1} & = & \sigma_a - \sigma_b, 
\\
X_{2a-1,2b} & = & - \sigma_a + \sigma_b.
\end{eqnarray}
Comparing to the critical locus, we see
\begin{equation}
\sigma_a \: \leftrightarrow \: \Pi_a + \tilde{m}_n,
\end{equation}
and hence from equation~(\ref{eq:sp2k:pi-reln}), our mirror predicts
\begin{equation}
\prod_{i=1}^n \left( \sigma_a - \tilde{m}_i \right) \: = \:
\prod_{i=1}^n \left( - \sigma_a - \tilde{m}_i \right).
\end{equation}
Up to a relative sign, this matches \cite{Hori:2011pd}[equ'n (5.8)].

Applying the operator mirror map to the constraints~(\ref{eq:sp2k:constr-1})
and (\ref{eq:sp2k:constr-2}), we get that
\begin{eqnarray}
\sigma_a & \neq & \pm \tilde{m}_i,
\\
\sigma_a & \neq & \sigma_b \mbox{ for }a \neq b,
\\
\sigma_a + \sigma_b & \neq & 0 \mbox{ for all }a, b,
\end{eqnarray}
which precisely match the constraints for this gauge theory
described in \cite{Hori:2011pd}[section 5.3, equ'n (5.9)].

\section{Example:  pure $SU(k)$ gauge theories}
\label{sect:pure}

In \cite{Aharony:2016jki}, it was suggested that a two dimensional
(2,2) supersymmetric pure $SU(k)$ gauge theory should flow to a theory
of $k-1$ free twisted chiral multiplets.  In this section, we will describe
how to check that result from our mirror, at least to the extent
possible with our topological field theory computations.  
We also describe the analogous computation in pure
$SO(3) = SU(2)/{\mathbb Z}_2$ theories, where we will see that the result
crucially depends upon the value of the discrete theta angle.
For one discrete theta angle, the pure $SO(3)$ theory behaves the
same as the pure $SU(2)$ theory; for the other discrete theta angle,
the mirror has no critical loci, no vacua, which we interpret as 
supersymmetry breaking in the original gauge theory.

The results for $SU(2)$ and $SO(3)$ are closely related, in terms of
nonabelian decomposition \cite{Sharpe:2014tca}.  This says that an
$SU(k)$ gauge theory with center-invariant matter decomposes into
a disjoint union of $SU(k)/{\mathbb Z}_k$ gauge
theories with the same matter and different discrete theta angles.
Here, schematically,
\begin{displaymath}
SU(2) \: = \: SO(3)_+ + SO(3)_-,
\end{displaymath}
using $SU(2)$ and $SO(3)$ to denote gauge theories, and so we see that
the behavior of the two $SO(3)$ gauge theories is closely linked to that
of the pure $SU(2)$ theory.

Before describing the general case, we will first discuss the
special cases of $SU(2)$ and $SU(3)$ theories.

\subsection{$SU(2)$, $SO(3)$ theories}

Let us begin by describing the mirror to a pure $SU(2)$ theory.
we will first describe the mirror to the $SU(2)$ theory itself,
then we will analyze the mirrors to the $SU(2)/{\mathbb Z}_2$ component
theories, and recover the same result.

As in section~\ref{sect:suk-twisted}, it is convenient to
describe this in terms of
two $\tilde{\sigma}$ fields obeying the constraint
\begin{displaymath}
\sum_a \tilde{\sigma}_a \: = \: 0.
\end{displaymath}
The mirror superpotential then takes the form
\begin{equation}
W \: = \:
\sum_a \tilde{\sigma}_a \sum_{\mu \neq \nu} \alpha^a_{\mu \nu} Z_{\mu \nu}
\: + \: \sum_{\mu \neq \nu} X_{\mu \nu},
\end{equation}
where
\begin{displaymath}
\alpha^a_{\mu \nu} \: = \: - \delta^a_{\mu} + \delta^a_{\nu}.
\end{displaymath}

For $SU(2)$, this becomes
\begin{eqnarray}
W & = &
\tilde{\sigma}_1 \left( - Z_{12} + Z_{21} \right) \: + \:
\tilde{\sigma}_2 \left( - Z_{21} + Z_{12} \right) \: + \:
X_{12} \: + \: X_{21},
\\
& = &
2 \tilde{\sigma}_1 \left( \ln X_{12} - \ln X_{21} \right) \: + \:
X_{12} \: + \: X_{21},
\end{eqnarray}
using the fact that $\tilde{\sigma}_2 = - \tilde{\sigma}_1$, and that
$Z_{\mu \nu} = - \ln X_{\mu \nu}$.
Solving for the critical loci, we find
\begin{eqnarray}
\frac{\partial W}{\partial \tilde{\sigma}_1}: & &
\left( \frac{X_{12} }{X_{21}} \right)^2 \: = \: 1,
\\
\frac{\partial W}{\partial X_{12}}: & &
X_{12} \: = \: - 2 \tilde{\sigma}_1,
\\
\frac{\partial W}{\partial X_{21}}: & &
X_{21} \: = \: + 2 \tilde{\sigma}_1.
\end{eqnarray}
In particular, on the critical locus, we see
$X_{12} = - X_{21}$, which is consistent with the first equation above.

If we integrate out $X_{12}$, $X_{21}$, then we see that the
superpotential becomes $W=0$, with one field ($\tilde{\sigma}_1$)
remaining.
This is certainly consistent with reducing to one free field,
precisely as predicted for this case by \cite{Aharony:2016jki}.

If we integrate out $\tilde{\sigma}_1$, the conclusion is the same but
the analysis is longer-winded.  Integrating out $\tilde{\sigma}_1$ gives
the constraint
\begin{equation}
X_{12}^2 \: = \: X_{21}^2,
\end{equation}
with solutions $X_{12} = \pm X_{21}$.  We have to sum over those possibilities.
If $X_{12} = + X_{21}$, then $W = 2 X_{12}$, which has no vacua and
breaks supersymmetry.  If $X_{12} = - X_{21}$, then $W=0$ and we have
one free field remaining, consistent with the previous analysis.

To do this carefully in a physical theory, one should also take
into account the K\"ahler potential -- which we omit in our discussion of
topological field theories, we have omitted.  As a result, we do not have
enough information about the physical mirror to verify all aspects of the
claim of \cite{Aharony:2016jki} for a physical theory, but certainly this
TFT computation is consistent with their conclusion.

Now, let us repeat the same analysis for pure $SO(3)$ theories.
A two-dimensional $SO(3)$ theory admits a ${\mathbb Z}_2$ discrete
theta angle, so there are two different $SO(3)$ theories to consider.
We shall examine each in turn.

Following our earlier
ansatz, the mirror of the $SO(3)$ theory with vanishing discrete
theta angle is defined by the superpotential
\begin{equation}
W \: = \:
(1/2) \sum_a \tilde{\sigma}_a \sum_{\mu \neq \nu} \alpha^a_{\mu \nu} Z_{\mu \nu}
\: + \: \sum_{\mu \neq \nu} X_{\mu \nu}, 
\end{equation}
which simplifies to
\begin{equation}
W \: = \: \tilde{\sigma}_1 \left( \ln X_{12} - \ln X_{21} \right) \: + \:
X_{12} \: + \: X_{21}.
\end{equation}
Solving for the critical loci as before, we find
\begin{eqnarray}
\frac{\partial W}{\partial \tilde{\sigma}_1}: & &
 \frac{X_{12} }{X_{21}}  \: = \: 1,
\\
\frac{\partial W}{\partial X_{12}}: & &
X_{12} \: = \: -  \tilde{\sigma}_1,
\\
\frac{\partial W}{\partial X_{21}}: & &
X_{21} \: = \: +  \tilde{\sigma}_1.
\end{eqnarray}
However, these equations are inconsistent:  the first requires
$X_{12} = + X_{21}$, the others require $X_{12} = - X_{21}$.
Therefore, this $SO(3)$ theory, with vanishing discrete theta angle,
has no supersymmetric vacua.  We interpret this to mean that the original
$SO(3)$ gauge theory with vanishing discrete theta angle 
breaks supersymmetry.
(One can also argue this directly in the A-twisted theory using
the one-loop twisted effective superpotential; the argument there is
very similar.)

Now, let us consider the second $SO(3)$ theory, with a nonzero discrete
theta angle.  The mirror to this theory is defined by the superpotential
\begin{equation}
W \: = \:
(1/2) \sum_a \tilde{\sigma}_a \sum_{\mu \neq \nu} \alpha^a_{\mu \nu} Z_{\mu \nu}
\: + \: \pi i \tilde{\sigma}_1
\: + \: \sum_{\mu \neq \nu} X_{\mu \nu}, 
\end{equation}
which simplifies to
\begin{equation}
W \: = \: \tilde{\sigma}_1 \left( \ln X_{12} - \ln X_{21} \right) \: + \:
\pi i \tilde{\sigma_1} \: + \: X_{12} \: + \: X_{21}.
\end{equation}
Solving for the critical loci as before, we now find
\begin{eqnarray}
\frac{\partial W}{\partial \tilde{\sigma}_1}: & &
 \frac{X_{12} }{X_{21}}  \: = \: -1,
\\
\frac{\partial W}{\partial X_{12}}: & &
X_{12} \: = \: -  \tilde{\sigma}_1,
\\
\frac{\partial W}{\partial X_{21}}: & &
X_{21} \: = \: +  \tilde{\sigma}_1.
\end{eqnarray}
Unlike the previous $SO(3)$ theory, these equations do have a solution,
the same solution as the original $SU(2)$ theory in fact, and so
(plausibly) describe a free field.

Thus, of the two $SO(3)$ theories, we have evidence from our TFT computations
that one flows to a theory of a single free twisted chiral superfield,
matching the $SU(2)$ result, whereas the other one has no supersymmetric
vacua at all.

These results for $SU(2)$ and $SO(3)$ theories are closely related
by nonabelian decomposition.  As outlined earlier, nonabelian
decomposition in this case says, schematically,
\begin{displaymath}
SU(2) \: = \: SO(3)_+ \: + \: SO(3)_-.
\end{displaymath}
Our results for the two $SO(3)$ theories match, in effect,
the result for the single $SU(2)$ theory.
Thus, we have not only verified
the claim of \cite{Aharony:2016jki} for $SU(2)$, but in addition refined
it, to show in which $SO(3)$ summand the free twisted chiral superfield
arises.

\subsection{$SU(3)$ theory}

Now, we will repeat the same analysis for the pure $SU(3)$ theory,
which has a few technical complications relative to $SU(2)$.

As before, we begin by writing the mirror in terms of three 
$\tilde{\sigma}$'s subject to the constraint
\begin{displaymath}
\sum_a \tilde{\sigma}_a \: = \: 0.
\end{displaymath}
The mirror superpotential then takes the form
\begin{eqnarray}
W & = &
\tilde{\sigma}_1 \left( - Z_{12} - Z_{13} + Z_{21} + Z_{31} \right)
\nonumber \\
& & \: + \:
\tilde{\sigma}_2 \left( - Z_{21} - Z_{23} + Z_{12} + Z_{32} \right)
\nonumber \\
& & \: + \:
\tilde{\sigma}_3 \left( - Z_{31} - Z_{32} + Z_{13} + Z_{23} \right)
\nonumber \\
& & \: + \:
X_{12} + X_{13} + X_{21} + X_{23} + X_{31} + X_{32},
\\ \\
& = &
\tilde{\sigma}_1 \left( - Z_{12} + Z_{21} - 2 Z_{13} + 2 Z_{31} +
Z_{32} -  Z_{23} \right)
\nonumber \\
& & \: + \:
\tilde{\sigma}_2 \left( - Z_{21} + Z_{12} - 2 Z_{23} + 2 Z_{32} +
Z_{31} - Z_{13} \right)
\nonumber \\
& & \: + \:
X_{12} + X_{13} + X_{21} + X_{23} + X_{31} + X_{32}.
\end{eqnarray}
Evaluating the critical locus, we find
\begin{eqnarray}
\frac{\partial W}{\partial \tilde{\sigma}_1}: & &
\frac{ X_{12} }{X_{21} }
\left( \frac{ X_{13} }{X_{31}} \right)^2
\frac{ X_{23} }{ X_{32} } \: = \: 1,   \label{eq:pure:su3:crit1}
\\
\frac{\partial W}{\partial \tilde{\sigma}_2}: & &
\frac{ X_{21} }{ X_{12} }
\left( \frac{ X_{23} }{ X_{32} } \right)^2
\frac{ X_{13} }{ X_{31} } \: = \: 1,  \label{eq:pure:su3:crit2}
\end{eqnarray}
\begin{eqnarray}
\frac{\partial W}{\partial X_{12}}: & &
X_{12} \: = \: \tilde{\sigma}_2 - \tilde{\sigma}_1,
\\
\frac{\partial W}{\partial X_{21}}: & &
X_{21} \: = \: \tilde{\sigma}_1 - \tilde{\sigma}_2 \: = \: - X_{12},
\\
\frac{\partial W}{\partial X_{13}}: & &
X_{13} \: = \: - \left( 2 \tilde{\sigma}_1 + \tilde{\sigma}_2 \right),
\\
\frac{\partial W}{\partial X_{31}}: & &
X_{31} \: = \: 2 \tilde{\sigma}_1 + \tilde{\sigma}_2 \: = \: - X_{13},
\\
\frac{\partial W}{\partial X_{23}}: & &
X_{23} \: = \: - \left( \tilde{\sigma}_1 + 2 \tilde{\sigma}_2 \right),
\\
\frac{\partial W}{\partial X_{32}}: & &
X_{32} \: = \: + \tilde{\sigma}_1 + 2 \tilde{\sigma}_2 \: = \: - X_{23}.
\end{eqnarray}
Furthermore, it is easy to see that the fields on the critical locus obey
\begin{equation}
X_{13} \: - \: X_{23} \: = \: X_{12}.
\end{equation}

As for $SU(2)$, there are two equivalent approaches we could take
in integrating out fields.  If one
integrates out the $X$s, one is left with $W=0$,
and two remaining fields ($\tilde{\sigma}_{1,2}$), of which one takes
Weyl-orbifold invariants.
Thus, at the level of these TFT computations, the mirror is consistent
with the original $SU(3)$ theory flowing to a theory of two free
chiral superfields, as predicted by \cite{Aharony:2016jki}.

Alternatively, if we integrate out the $\tilde{\sigma}$s, we get the
same result, after further algebra.  From equations~(\ref{eq:pure:su3:crit1}),
(\ref{eq:pure:su3:crit2}), it can be shown that
\begin{equation}
\left( \frac{ X_{23} }{ X_{32} } \right)^3 \: = \:
\left( \frac{ X_{31} }{ X_{13} } \right)^3,
\: \: \:
\frac{ X_{12} }{ X_{21} } \: = \:
\left( \frac{ X_{23} }{ X_{32} } \right)^2
\frac{ X_{13} }{ X_{31} },
\end{equation}
so for any $\xi$ such that $\xi^3 = 1$, 
\begin{equation}
\frac{ X_{23} }{ X_{32} } \: = \: \xi \frac{ X_{31} }{ X_{13} },
\: \: \:
\frac{ X_{12} }{ X_{21} } \: = \: \xi^2 \frac{ X_{31} }{ X_{13} }.
\end{equation}
We must sum over the critical loci, corresponding to possible values
of $\xi$.
Eliminating $X_{23}$ and $X_{12}$, we have
\begin{eqnarray}
W & = & X_{12} + X_{13} + X_{23} + X_{21} + X_{31} + X_{32},
\\
& = &
\xi^2 \frac{ X_{31} X_{21} }{ X_{13} }
+ X_{13} + \xi \frac{ X_{31} X_{32} }{ X_{13} } + X_{21} + X_{31} + X_{32}.
\end{eqnarray}
The critical locus equations are then
\begin{eqnarray}
\frac{\partial W}{\partial X_{13}} & = &
- \xi^2 \frac{ X_{31} X_{21} }{ X_{13}^2 } - \xi \frac{ X_{31} X_{32} }{
X_{13}^2 } + 1 \: = \: 0,
\\
\frac{\partial W}{\partial X_{31}} & = &
\xi^2 \frac{ X_{21} }{ X_{13} } + \xi \frac{ X_{32} }{ X_{13} } + 1
\: = \: 0,
\\
\frac{\partial W}{\partial X_{21} } & = & \xi^2 \frac{ X_{31} }{ X_{13} }
+ 1 \: = \: 0,
\\
\frac{\partial W}{\partial X_{32} } & = & \xi \frac{ X_{31} }{ X_{13} }
+ 1 \: = \: 0.
\end{eqnarray}
Note that from the last two equations
\begin{equation}
\frac{ X_{31} }{ X_{13} } \: = \: - \frac{1}{\xi^2} \: = \:
- \frac{1}{\xi},
\end{equation}
so we see that there are no vacua unless $\xi=+1$.  We therefore focus on
that case.  Rewriting the other two equations implies
\begin{equation}
X_{13} \: = \: - X_{21} - X_{32}
\end{equation}
on the critical locus.
Restricting to the critical locus above, it is straightforward to see
that $W=0$, and there are exactly two remaining free fields,
consistent with the prediction for IR physics given in
\cite{Aharony:2016jki}.

\subsection{General $SU(k)$ theories}

Now, let us turn to mirrors of pure $SU(k)$ theories for more general $k$.
Following the same pattern, and after solving for $\tilde{\sigma}_k$,
the superpotential has the form
\begin{eqnarray}
W & = &
 \tilde{\sigma}_1 \bigl[  - Z_{12} - Z_{13} - \cdots - Z_{1 (k-1)}
- 2 Z_{1k} - Z_{2k} - \cdots - Z_{(k-1)k}
\nonumber \\
& & \hspace*{0.75in}
+ Z_{21} + Z_{31} + \cdots + Z_{(k-1)1} + 2 Z_{k1} +
Z_{k2} + \cdots + Z_{k(k-1)} \bigr]
\nonumber \\
& & \: + \:
 \tilde{\sigma}_2 \bigl[ - Z_{21} - Z_{23} - Z_{24} - \cdots -
Z_{2(k-1)} - 2 Z_{2k} - Z_{1k} - Z_{3k} - \cdots - Z_{(k-1)k}
\nonumber \\
& & \hspace*{0.75in}
+ Z_{12} + Z_{32} + Z_{42} + \cdots + Z_{(k-1) 2} + 2 Z_{k2}
+ Z_{k1} + Z_{k3} + Z_{k4} + \cdots + Z_{k(k-1)} \bigr]
\nonumber \\
& & \: + \: \cdots
 \: + \: \sum_{\mu \neq \nu} X_{\mu \nu}.
\end{eqnarray}

It is straightforward to see that the critical locus equations take
the form
\begin{equation}
X_{\mu \nu} \: = \: - X_{\nu \mu},
\end{equation}
and
\begin{eqnarray}
\frac{ X_{12} }{X_{21}} \frac{X_{13}}{X_{31}} \cdots
\frac{ X_{1(k-1)} }{X_{(k-1)1}} \left( \frac{ X_{1k} }{X_{k1}} \right)^2
\frac{ X_{2k} }{ X_{k2} } \cdots \frac{ X_{(k-1)k} }{ X_{k(k-1)}}
& = & 1,
\\
\frac{ X_{21} }{X_{12}} \frac{X_{23}}{X_{32}} \frac{X_{24}}{X_{42}}
\cdots \frac{X_{2(k-1)}}{X_{(k-1)2}} \left( \frac{ X_{2k}}{X_{k2}} \right)^2
\frac{ X_{1k} }{X_{k1}} \frac{X_{3k}}{X_{k3}} \cdots 
\frac{X_{(k-1)k}}{X_{k(k-1)}} 
& = & 1,
\end{eqnarray}
and so forth.  Each of these equations is a product of $2(k-1)$ ratios
of $X$'s associated to positive and negative roots.  Since each
ratio is $-1$ on the critical locus and there are an even number of factors,
the product equations are automatically consistent, and do not generate
a contradiction.  

Integrating out the $X$s, leads to a theory with $W=0$ and $k-1$ independent
fields (the $\tilde{\sigma}$, or rather their Weyl-invariant combinations),
as predicted by \cite{Aharony:2016jki}.
We emphasize again that we have not checked any statements about
K\"ahler metrics -- we are only performing a consistency test of those
quantities which can be computed in topological field theories, nothing
more.

\section{Example:  pure $SO$, $Sp$ gauge theories}
\label{sect:puresosp}

In the previous section, we discussed the form of the mirror to
pure $SU(k)$ gauge theories, in order to test predictions made in
\cite{Aharony:2016jki}.  In this section we will briefly outline tetss of
an analogous conjecture for
pure $SO$ and $Sp$ theories. Specifically,
we conjecture that pure $Sp$ theories, as well as pure
$SO$ and $Sp/{\mathbb Z}_2$ theories for one discrete
theta angle, flow in the IR to theories of free
twisted chiral superfields, as many as the rank of the
gauge group, and pure $SO$ and $Sp/{\mathbb Z}_2$
theories with the other discrete theta angle break supersymmetry.
We only test this claim at the level of
topological field theory computations; we have not tested any claims about
the form of metrics or other properties of physical untwisted theories.

\subsection{Pure $SO(2k)$}

Let us briefly test this conjecture for pure $SO(2k)$ supersymmetric
gauge theories
in two dimensions.
Following section~\ref{sect:so2k}, the mirror superpotential takes the
form
\begin{eqnarray}
W & = &
\sum_{a=1}^k \sigma_a \left(
\sum_{\mu < \nu; \mu' , \nu'} \alpha^a_{\mu \nu, \mu' \nu'} Z_{\mu' \nu'}
\: - \: t \right)
\: + \: 
\sum_{\mu < \nu} X_{\mu \nu},
\\
& = &
\sum_{a=1}^k \sigma_a \left(
 \sum_{\nu > 2a} \left( Z_{2a, \nu} - Z_{2a-1,\nu} \right)
\: + \: \sum_{\mu < 2a-1}  \left( Z_{\mu,2a} - Z_{\mu,2a-1} \right)
\: - \: t \right)
\nonumber \\
& & \hspace*{2in}
\: + \: 
\sum_{\mu < \nu} X_{\mu \nu},
\end{eqnarray}
where $t$ is a discrete theta angle.
The critical locus equations are defined by
\begin{eqnarray}
\frac{\partial W}{\partial \sigma_a}: & &
\left( \prod_{\nu > 2a} \frac{ X_{2a-1,\nu} }{ X_{2a,\nu} } \right)
\left( \prod_{\mu < 2a-1} \frac{ X_{\mu,2a-1} }{ X_{\mu, 2a} } \right)
\: = \: q^{-1},
\label{eq:pure:so2k:crit1}
\\
\frac{\partial W}{\partial X_{2a,\nu} }: & & X_{2a,\nu} = + \sigma_a,
\\
\frac{\partial W}{\partial X_{2a-1,\nu} }: & & X_{2a-1,\nu} = - \sigma_a,
\\
\frac{\partial W}{\partial X_{\mu,2a} }: & & X_{\mu,2a} = + \sigma_a,
\\
\frac{\partial W}{\partial X_{\mu,2a-1}}: & & X_{\mu, 2a-1} = - \sigma_a.
\end{eqnarray}
Since there are $2(k-1)$ ratio factors in equation~(\ref{eq:pure:so2k:crit1}),
and each factor is $-1$,
we see that solutions for critical points will exist precisely when
$q^{-1} = 1$.  In that case, for that discrete theta angle,
integrating out $X$s leads to a vanishing superpotential and verifies
the claim for pure $SO(2k)$ theories (at least to the extent that the
claim can be checked in a topological field theory).  
For the other discrete theta angle,
there are no vacua, and hence supersymmetry is broken.

\subsection{Pure $SO(2k+1)$}

Next, let us outline the analogous computation for pure $SO(2k+1)$
supersymmetric gauge theories in two dimensions.
Following section~\ref{sect:so2kp1}, the superpotential takes almost exactly
the same form, except that the indices on $X_{\mu \nu}$ extend to $2k+1$:
\begin{eqnarray}
W & = &
\sum_{a=1}^k \sigma_a \left(
\sum_{\mu < \nu; \mu' , \nu'} \alpha^a_{\mu \nu, \mu' \nu'} Z_{\mu' \nu'}
\: - \: t \right)
\: + \: 
\sum_{\mu < \nu} X_{\mu \nu},
\\
& = &
\sum_{a=1}^k \sigma_a \left(
 \sum_{\nu > 2a} \left( Z_{2a, \nu} - Z_{2a-1,\nu} \right)
\: + \: \sum_{\mu < 2a-1}  \left( Z_{\mu,2a} - Z_{\mu,2a-1} \right)
\: - \: t \right)
\nonumber \\
& & \hspace*{2in}
\: + \: 
\sum_{\mu < \nu} X_{\mu \nu},
\end{eqnarray}
where $t$ is a discrete theta angle.

The analysis of the critical loci is nearly identical to that for
pure $SO(2k)$, except that in the analogue of 
equation~(\ref{eq:pure:so2k:crit1}),
there are now $2(k-1)+1$ ratio factors, so as each ratio is $-1$ on the
critical locus, we see that solutions for critical points will
exist precisely when $q^{-1} = -1$, {\it i.e.} for the nontrivial
discrete theta angle.  For that discrete theta angle, the result
is consistent with a theory of $k$ free chiral multiplets, as before,
to the extent that these TFT computations can check statements about the
physical theory.  For the other discrete theta angle, there are no
vacua and so supersymmetry is broken.
In passing, note that this matches earlier results for the $SO(3)$ theories.

\subsection{Pure $Sp(2k)$}

Finally, let us outline the analogous computation for pure
$Sp(2k)$ gauge theories.
We treat these gauge theories in a nearly identical fashion, so we will
be brief.

Following section~\ref{sect:sp2k},
the mirror superpotential is given by
\begin{eqnarray}
W & = &
 \sum_{a=1}^k \sigma_a \Biggl(
 \sum_{\mu \leq \nu}  \left( \delta_{\mu, 2a} 
 - \delta_{\mu, 2a-1} +
\delta_{\nu, 2a} - \delta_{\nu, 2a-1}\right) Z_{\mu \nu} 
 \Biggr)
\nonumber \\
& & 
\: + \: \sum_{\mu} X_{\mu \mu} \: + \: 
\sum_{a < b} \left( X_{2a,2b} + X_{2a-1,2b-1} +
X_{2a-1,2b} + X_{2a,2b-1} \right).
\end{eqnarray}
The critical locus is defined by
\begin{eqnarray}
\frac{\partial W}{\partial X_{2a,2a}}: & &
X_{2a,2a} \: = \: + 2 \sigma_a, 
\\
\frac{\partial W}{\partial X_{2a-1,2a-1}}: & &
X_{2a-1,2a-1} \: = \: - 2 \sigma_a,
\\
\frac{\partial W}{\partial X_{2a,2b}}: & & X_{2a,2b} \: = \: \sigma_1 +
\sigma_b \: \mbox{ for }a < b, 
\\
\frac{\partial W}{\partial X_{2a-1,2b-a}}: & &
X_{2a-1,2b-1} \: = \: - \left( \sigma_a + \sigma_b \right)
 \: \mbox{ for }a < b, 
\\
\frac{\partial W}{\partial X_{2a-1,2b}}: & &
X_{2a-1,2b} \: = \: - \sigma_a + \sigma_b
 \: \mbox{ for }a < b, 
\\
\frac{\partial W}{\partial X_{2a,2b-1}}: & &
X_{2a,2b-1} \: = \: \sigma_a - \sigma_b
 \: \mbox{ for }a < b.
\end{eqnarray}
In addition, $\partial W/\partial \sigma_a = 0$ implies
\begin{equation}
\left( \prod_{2a \leq \nu} \frac{1}{X_{2a,\nu}} \right)
\left( \prod_{2a-1 \leq \nu} X_{2a-1,\nu} \right)
\left( \prod_{\mu \leq 2a} \frac{1}{X_{\mu,2a} } \right)
\left( \prod_{\mu \leq 2a-1} X_{\mu, 2a-1} \right) \: = \: 1,
\end{equation}
which can be rewritten as
\begin{equation}
\left( \frac{ X_{2a-1,2a-1} }{ X_{2a,2a} } \right)^2
\left( \prod_{b > a} \frac{ X_{2a-1,2b-1} }{ X_{2a,2b} }
\frac{ X_{2a-1,2b} }{ X_{2a,2b-1} } \right)
\left( \prod_{b<a} \frac{ X_{2b-1,2a-1} }{ X_{2b,2a} }
\frac{ X_{2b,2a-1} }{ X_{2b-1,2a} } \right) \: = \: 1.
\end{equation}
Along the critical locus, each of the ratios appearing in the product
above is $-1$.  Since there are manifestly an even number of them,
this critical locus equation is trivially satisfied.

As before, along the critical locus,
the superpotential vanishes, and so at the level of these TFT computations,
for the correct discrete theta angle,
this is consistent with the IR limit being 
a set of $k$ free twisted chiral superfields.
As a consistency check, note this is consistent with earlier computations
for the pure $SU(2) = Sp(2)$ theory.

\section{Hypersurfaces and complete intersections in $G(k,n)$}
\label{sect:hyp}

In this section, we will consider the analysis of a degree $d$ hypersurface
and complete intersections 
in $G(k,n)$, and compare to results in \cite{Hori:2006dk,Closset:2015rna}.
To be clear, the proposal of this paper yields a B-twisted Landau-Ginzburg
model that reproduces A model correlation functions; however, we do not
know if the same Landau-Ginzburg model ever appears as a phase in any
GLSM describing a geometry birational to a mirror geometry, possibly after
integrating out some fields.
We cannot exclude the possibility, but neither do we have any examples.
In this section we will focus on setting up the models and comparing to
correlation functions, and aside from a few general remarks in one section,
we will leave potential geometric comparisons
to future work.

\subsection{Mirror proposal and Coulomb computations for a hypersurface}

Consider an A-twisted GLSM for a hypersurface of degree $d$ in
$G(k,n)$.
This is described by a $U(k)$ gauge theory with matter
\begin{itemize}
\item $n$ chiral multiplets $\phi_{ia}$ in the fundamental representation,
$i \in \{1, \cdots, n\}$, $a \in \{1, \cdots, k\}$,
\item one field $p$ of charge $-d$ under $\det U(k)$,
\end{itemize}
and a superpotential
\begin{equation}
W \: = \: p G(B),
\end{equation}
where $G$ is a polynomial of degree $d$ in the baryons $B_{ij}$,
\begin{equation}
B_{i_1 \cdots i_k} \: = \: \epsilon^{a_1 \cdots a_k}
\phi_{i_1 a_1} \cdots \phi_{i_k a_k}.
\end{equation}
We take the chiral superfields $\phi_{ia}$
to have R-charge zero, and $p$ to have R-charge two.

The mirror of this theory is an orbifold of the Landau-Ginzburg
model with fields
\begin{itemize}
\item $kn$ chiral superfields $Y_{ia}$, 
mirror to $\phi_{ia}$,
\item one chiral superfield $X_p = \exp(-Y_p)$, mirror to $p$,
\item $X_{\mu \nu} = \exp(-Z_{\mu \nu})$, $\mu, \nu \in \{1, \cdots, k\}$,
\item $\sigma_a$, $a \in \{ 1, \cdots, k\}$,
\end{itemize}
and superpotential
\begin{eqnarray}
W & = &
\sum_a \sigma_a \left( \sum_{ib} \rho^a_{ib} Y_{ib} \: - \: d Y_p
\: + \: \sum_{\mu \neq \nu} \alpha^a_{\mu \nu} Z_{\mu \nu} \: - \: t \right)
\nonumber \\
& & 
\: + \: \sum_{ia} \exp\left( - Y_{ia} \right)  \: + \: X_p 
\: + \: \sum_{\mu \neq \nu} X_{\mu \nu},
\label{eq:hyp:1st-sup}
\end{eqnarray}
where
\begin{equation}
\rho^a_{ib} \: = \: \delta^a_b, \: \: \:
\alpha^a_{\mu \nu} \: = \: - \delta^a_{\mu} + \delta^a_{\nu}.
\end{equation}

We then take an $S_k$ orbifold of this
theory, where $S_k$ acts as the Weyl group of $U(k)$ in the same 
fashion described elsewhere.

Integrating out the $\sigma_a$ gives constraints
\begin{equation}
\sum_{i=1}^n Y_{ia} \: - \: d Y_p \: + \: \sum_{\nu \neq a}
\left( - Z_{a \nu} + Z_{\nu a} \right) \: = \: t,
\end{equation}
which we can solve by eliminating $Y_{na}$:
\begin{equation}
Y_{na} \: = \: - \sum_{i=1}^{n-1} Y_{ia} \: + \:  d Y_p
\: - \: \sum_{\nu \neq a} \left( -Z_{a \nu} + Z_{\nu a} \right) \: + \: t.
\end{equation}

Define
\begin{eqnarray}
\Pi_a & = & \exp\left( - Y_{na} \right), 
\\
\label{eq:hyp:pi-defn}
& = & q \left( \prod_{i=1}^{n-1} \exp\left( + Y_{ia} \right) \right)
\left( X_p \right)^d
\left( \prod_{\nu \neq a} \frac{ X_{a \nu} }{ X_{\nu a} } \right).
\end{eqnarray}

The superpotential then becomes
\begin{eqnarray}  \label{eq:hyp:final-sup}
W & = &  \sum_{i=1}^{n-1} \sum_a  \exp\left( - Y_{ia} \right)
\: + \: \sum_a  \Pi_a  \: + \: X_p
\: + \: \sum_{\mu \neq \nu} X_{\mu \nu}.
\end{eqnarray}

The critical locus is given as follows:
\begin{eqnarray}
\frac{\partial W}{\partial Y_{ia}}: & &
\exp\left( -Y_{ia} \right) \: = \:  \Pi_a ,
\label{eq:hyp:crit-1}\\
\frac{\partial W}{\partial X_p}: & &
X_p \: = \: -  d \sum_a  \Pi_a,
\label{eq:hyp:crit-2}\\
\frac{\partial W}{\partial X_{\mu \nu} }: & &
X_{\mu \nu} \: = \:  -  \Pi_{\mu}  \: + \:
 \Pi_{\nu} .
\label{eq:hyp:crit-3}
\end{eqnarray}

Since we exclude $X_{\mu \nu} =0$, for reasons discussed earlier
in section~\ref{sect:gkn:excluded},
we see that
\begin{equation}  \label{eq:hyp:constr-1}
 \Pi_{\mu}  \: \neq \:
 \Pi_{\nu}  \: \mbox{  for }\mu \neq \nu.
\end{equation}
Since $\exp(-Y) \neq 0$, we see
that 
\begin{equation}  \label{eq:hyp:constr-2}
 \Pi_a  \: \neq \: 0.
\end{equation}

These constraints guarantee that the critical locus does not intersect
the fixed-point locus of the orbifold, and so for the purposes of
our computations we can ignore twisted sectors.

On the critical locus, from the definition~(\ref{eq:hyp:pi-defn}) of
$\Pi_a$, we find
\begin{displaymath}
 \Pi_a  \: = \: q
\left( \prod_{i=1}^{n-1} \frac{1}{ \Pi_a } \right)
\left( - d \sum_b  \Pi_b  \right)^{d}
(-)^{k-1},
\end{displaymath}
or more simply
\begin{equation}  \label{eq:hyp:pi-reln}
\left( \Pi_a \right)^{n } \: = \:
q
\left( - d \sum_b  \Pi_b  \right)^{d}
(-)^{k-1}.
\end{equation}

Note as a consequence that if 
\begin{displaymath}
\sum_a \Pi_a \: = \: 0,
\end{displaymath}
then from~(\ref{eq:hyp:pi-reln}), $\Pi_a = 0$, which is forbidden
from~(\ref{eq:hyp:constr-2}), hence we have the constraint that
\begin{equation}   \label{eq:hyp:constr-3}
\sum_a  \Pi_a  \: \neq \: 0.
\end{equation}

The operator mirror map~(\ref{eq:op-mirror-1}),
(\ref{eq:op-mirror-2}) says
\begin{eqnarray}
\exp\left( - Y_{ia} \right)  & = & \sum_b \sigma_b \rho^b_{ia} 
\: = \: \sigma_a,
\\
X_p & = &
-  d \sum_a \sigma_a,
\\
X_{\mu \nu} & = & \sum_a \sigma_a \alpha^a_{\mu \nu} \: = \:
- \sigma_{\mu} + \sigma_{\nu}.
\end{eqnarray}
Comparing with equations~(\ref{eq:hyp:crit-1}), (\ref{eq:hyp:crit-2}),
(\ref{eq:hyp:crit-3}), we see that on the critical
locus we can identify
\begin{equation}
\sigma_a \: \leftrightarrow \:  \Pi_a .
\end{equation}

Using the operator mirror map, equation~(\ref{eq:hyp:pi-reln}) becomes
\begin{equation} \label{eq:hyp:sigma-reln}
\left( \sigma_a \right)^n \: = \: (-)^{k-1} q \left( -  d \sum_b
\sigma_b \right)^{d}.
\end{equation}
Up to a sign, this matches \cite{Hori:2006dk} above their equation (2.20).

Applying the mirror map above to the three
constraints~(\ref{eq:hyp:constr-1}), (\ref{eq:hyp:constr-2}),
and (\ref{eq:hyp:constr-3}), we get the constraints
\begin{equation}
\sigma_a \: \neq \: \sigma_b \: \mbox{ for }a \neq b,
\: \: \:
\sigma_a \: \neq \: 0,
\: \: \:
\sum_a \sigma_a \: \neq \: 0,
\end{equation}
which match the constraints for this model given in
\cite{Hori:2006dk}[equ'n (2.15)].

In passing, we can also get a bit of insight into the $r \ll 0$ phase
of the original gauge theory.  In that phase, in the original theory
the $p$ field gets a vacuum expectation value which breaks the $U(k)$
gauge symmetry to an $SU(k)$ gauge symmetry.  Here in the
mirror, when $r \ll 0$,
$q \rightarrow + \infty$ (since $q \propto \exp(-r)$), and from
equation~(\ref{eq:hyp:sigma-reln}), as $q \rightarrow \infty$,
\begin{displaymath}
\sum_a \sigma_a \: \rightarrow \: 0.
\end{displaymath}
This is the same constraint that defines an $SU(k)$ gauge theory from
a $U(k)$ gauge theory, as we saw in section~\ref{sect:suk-twisted},
and here on the critical locus, also implies that $X_p \rightarrow 0$
from equation~(\ref{eq:hyp:crit-2}).
In that limit, we can integrate out $Y_p$ from the 
superpotential~(\ref{eq:hyp:1st-sup}), which enforces the constraint
\begin{displaymath}
\sum_a \sigma_a \: = \: 0,
\end{displaymath}
and then, 
this theory reduces to the same form as the
$SU(k)$ mirror discussed in section~\ref{sect:suk-twisted}.

\subsection{Poles in correlation functions in $G(2,4)[4]$}

Next, let us compute locations of poles in
correlation functions to compare to results in 
\cite{Closset:2015rna}[section 8.2].

We would like to
proceed in the same fashion as in section~\ref{sect:gkn:corr-fns}.
Given the superpotential~(\ref{eq:hyp:final-sup}) 
with $\sigma_a$'s integrated out,
one would like to compute correlation functions on $S^2$ in the form
\begin{equation}  \label{eq:hyp:corr-1}
\langle f \rangle \: = \: \sum_{\rm vacua} \frac{f}{H},
\end{equation}
where the vacua are defined by the critical locus of the
superpotential, and $H$ is the determinant of the matrix of
second derivatives.  (In principle, there is an orbifold here,
but as none of the critical loci intersect the orbifold fixed points,
we do not anticipate any contributions from twisted sectors, so the
effect of the $G$-orbifold is to multiply the correlation function
above by a factor of $1/|G|$.)

However, the critical loci are a bit more subtle in this case.
First, when critical loci exist, they are not isolated.
Instead, since
\begin{equation}  \label{eq:hyp:corr:crit}
\sigma_1^4 \: = \: \sigma_2^4 \: = \: - q (-4)^4 \left( \sigma_1 +
\sigma_2 \right)^4
\end{equation}
is homogeneous in $\sigma_1$, $\sigma_2$, there are lines of critical
points, given by rescaling any one solution.  Mechanically, in terms of
the correlation function computation outlined above, since there is
a flat direction, the Hessian $H=0$ (which can be confirmed by
direct computation on the critical loci).  Thus,
the formula~(\ref{eq:hyp:corr-1}) above does not seem to apply.
Instead, in principle, the correct computation would seem to involve
a Hessian of directions `orthogonal' to the flat direction, for example.

The second subtlety is that critical loci will only exist for
special values of $q$ (reflecting the fact that on the A model
side, one will only have a Coulomb branch at special K\"ahler moduli.
We can see this constraint as follows.
From equation~(\ref{eq:hyp:corr:crit}), although $\sigma_1$ and
$\sigma_2$ are required to be distinct, they are proportional to one
another and to their sum:
\begin{equation}
\sigma_1 \: \propto \: \sigma_1 + \sigma_2 \: \propto \: \sigma_2.
\end{equation}
Following the spirit of \cite{Hori:2006dk}[section 2.3],
write 
\begin{equation}
\sigma_a \: = \: \frac{ \omega^{n_a} }{Z} \left( \sigma_1 + \sigma_2 
\right),
\end{equation}
where $\omega^4 = 1$, and 
\begin{equation}   \label{eq:hyp:corr:q-constr-1}
\frac{1}{Z^4} \: = \: -q (-4)^4 = - 2^8 q.
\end{equation}
This should encapsulate the most general possible set of
solutions to~(\ref{eq:hyp:corr:crit}).
However, if we now require that the $\sigma_a$ sum correctly, we find
\begin{displaymath}
\sigma_1 \: + \: \sigma_2 \: = \: \frac{
\omega^{n_1} + \omega^{n_2} }{ Z} \left( \sigma_1 + \sigma_2 \right),
\end{displaymath}
hence
\begin{displaymath}
  \frac{
\omega^{n_1} + \omega^{n_2} }{ Z}  \: = \: 1,
\end{displaymath}
implying
\begin{equation}    \label{eq:hyp:corr:q-constr-2}
\left(  \frac{
\omega^{n_1} + \omega^{n_2} }{ Z} \right)^4 \: = \: 1,
\end{equation}
which we can see already from equation~(\ref{eq:hyp:corr:q-constr-1}) 
is going to constrain
$q$.  

Now, the possible fourth roots of unity are $\pm 1$, $\pm i$.
Some of the sums of pairs of fourth roots of unity vanish:
for example, $1-1=0$.  However, these values of $\omega^{n_a}$
contradict~(\ref{eq:hyp:corr:q-constr-2}), and so we omit them.
For the remaining possibilities, 
\begin{displaymath}
\left( \omega^{n_1} + \omega^{n_2} \right)^4 \: = \:
\omega^{4 n_1} \left( 1 + \omega^{n_2-n_1} \right)^4 \: = \:
\left( 1 + \omega^{n_2-n_1} \right)^4,
\end{displaymath}
and since we know the $\sigma_a$ are distinct and the sum nonzero, 
it suffices to
consider the case $\omega^{n_2-n-1} = \pm i$.
It is straightforward to compute that
\begin{displaymath}
(1 \pm i )^4 \: = \: -4,
\end{displaymath}
hence equation~(\ref{eq:hyp:corr:q-constr-2}) becomes
\begin{equation}
\frac{1}{Z^4} \: = \: - \frac{1}{4},
\end{equation}
and so comparing with equation~(\ref{eq:hyp:corr:q-constr-1}), we get
the constraint
\begin{equation}
\frac{1}{Z^4} \: = \:  - \frac{1}{4} \: = \: - 2^8 q,
\end{equation}
or more simply, $2^{10} q = 1$.  

This result is in complete agreement with \cite{Hori:2006dk}[section 2.4] and
\cite{bcfkvs}, who found that in the A-twisted GLSM for the
Calabi-Yau $G(2,4)[4] = 
{\mathbb P}^5[2,4]$, there is a Coulomb
branch at a single value of $q$, determined by the same formula above,
or equivalently, that A model correlation functions will have
poles at $q = 2^{-10}$.

\subsection{Comments on mirror geometries}
\label{sect:comments-mirror}

So far, we have focused on the mirror simply as a tool for
comparing Coulomb branch relations and correlation functions.
However, at least in special cases such as the quintic threefold,
it is known that abelian versions of the mirror construction above
can be related more directly to a mirror geometry,
see {\it e.g.} \cite{Aganagic:2004yh}.

The hypersurface $G(2,4)[4]$ is particularly interesting in this
regard -- the Grassmannian $G(2,4)$ can be described as a hypersurface
in a projective space, ${\mathbb P}^5[2]$, and is the only Grassmannian
which can be (nontrivially) realized in this fashion.  Thus,
$G(2,4)[4] = {\mathbb P}^5[2,4]$, and as there is an extensive
literature on mirror symmetry in abelian cases, one might hope to
compare to existing results.  The mirror to ${\mathbb P}^5[2,4]$ is
discussed explicitly in {\it e.g.} \cite{bb}[section 8.7, case 2],
\cite{fredrickson,Morrison:2015qca,bcfkvs,bvs,lt} (see \cite{boehm} for
related considerations for Pfaffian mirrors).  Briefly, the geometric
mirror is expressed as an intersection in a nontrivial toric
variety, not as a hypersurface in a projective space, so it is
unlikely that the mirror we have described above will correspond
to a Landau-Ginzburg model for the geometry in as simple a fashion
as happens for the quintic threefold, for example.  Furthermore, to
correctly identify a mirror geometry might require taking into account
blowups needed to properly regularize discontinuities in the superpotential,
as discussed in section~\ref{sect:regularization}.  In any event,
it is possible that the `full' mirror (possibly after integrating out
some fields) is birational to a simpler
model which corresponds more directly to the mirror construction above,
in a fashion analogous to that for the quintic.  We leave such
explorations for future work.

In passing, let us make an observation about central charges that
lends support to the idea that these mirrors may be related to
geometric mirrors.  To be specific, let us consider our proposed
mirror to a hypersurface in $G(k,n)$, as described by the
Landau-Ginzburg model with superpotential~(\ref{eq:hyp:final-sup}) 
we computed previously.  In the Calabi-Yau case, this superpotential
is quasi-homogeneous, so one expects there should be a nontrivial
IR limit, for which we can compute the central charge (of the physical
untwisted theory).  
Define $q_i = n_i/d$ where
\begin{displaymath}
W\left( \lambda^{n_i} \Phi_i \right) \: = \: \lambda^d W\left( \Phi_i \right).
\end{displaymath}
If we were to take the fundamental
fields to be as before, namely
the $X_{\mu \nu}$, $X_p$, and $Y_{ia}$, then $q(X_{\mu \nu}) = 1$
and $q(X_p) = 1$.   If we take into account the translation symmetry
of the $Y$'s, then $q(\exp(-Y)) = 1$ but, given that there is no
multiplicative component, $q(Y) = 0$.
If we use the central charge formula of
\cite{Vafa:1989xc}[equ'n (15)], and assume that it also applies to
cases with fields with translation symmetries, then the central charge
is given by
\begin{eqnarray}
\frac{c}{3} & = & \sum_i \left( 1 - 2 q_i \right), \\
& = & (1)(1-2) + (k^2-k)(1-2) + k(n-1)(1-0),
\\
& = & k(n-k)-1,
\end{eqnarray}
matching the dimension of the Calabi-Yau hypersurface.  
(See also \cite{Hori:2006dk}[section 4.1], where this is computed
in the original theory as an exercise in $SU(k)$ gauge theories.)
(This gives a different R charge assignment than is used in the
Landau-Ginzburg point of the orbifold mirror, though here we
have implicitly integrated out some $Y$'s instead of the mirror to the
$p$ field as is traditionally done in writing the quintic mirror, which
presumably accounts for the difference.)
Finally, the reader should
note that we are not claiming that this implies our proposed mirror necessarily
matches the Landau-Ginzburg point of some GLSM for the mirror geometry;
rather, we are merely listing a suggestive computation.  Perhaps the
lesson here is that to get a geometry from a nonabelian mirror,
one must integrate out the $X$ fields first, so as a result, the
mirror may look like a Landau-Ginzburg model but with extra factors in
the measure.  We leave all such considerations of mirror geometries to
future work.

\subsection{Complete intersections in $G(k,n)$}

Consider a GLSM describing a complete intersection of $S$
hypersurfaces in $G(k,n)$ of degrees $Q_{\alpha}$, 
$\alpha \in \{1, \cdots, S\}$.
This is a $U(k)$ gauge theory with $n$ chiral multiplets in the fundamental
(of R charge zero) and $S$ chiral multiplets $p_{\alpha}$ (of R charge two)
charged under $\det U(k)$ with charge
$-Q_{\alpha}$.
This theory was considered in \cite{Closset:2015rna}, which computed
a general expression for correlation functions of $\sigma$'s in the
A-twisted gauge theory, a series of residues given in
\cite{Closset:2015rna}.  In this section we will study the mirror of this
theory, and outline how correlation functions in our mirror reproduce
the expression \cite{Closset:2015rna}[equ'n (8.32)] for correlation
functions given in the A-twisted
theory.

Following our proposal, the mirror is defined by
\begin{itemize}
\item chiral superfields $Y_{ia}$, $i \in \{ 1, \cdots, n \}$,
$a \in \{1, \cdots, k\}$,
\item chiral superfields $X_{\alpha} = \exp( - Y_{\alpha})$,
\item chiral superfields $X_{\mu \nu}$, mirror to the W bosons,
\item $\sigma_a$,
\end{itemize}
with superpotential
\begin{eqnarray}
W & = & \sum_a \sigma_a \left( \sum_{i b} \rho^a_{ib} Y_{ib} 
\: + \: \sum_{\alpha} Q_{\alpha} \ln X_{\alpha} \: - \:
\sum_{\mu \neq \nu} \alpha^a_{\mu \nu} \ln X_{\mu \nu} \: - \: t \right)
\nonumber \\
& & \: + \: \sum_{ia} \exp\left( - Y_{ia} \right) \: + \:
\sum_{\alpha} X_{\alpha} \: + \: \sum_{\mu \neq \nu} X_{\mu \nu}.
\end{eqnarray}
Integrating out the $X_{\mu \nu}$ and simplifying as in
section~\ref{sect:gkn:int-out}, the superpotential becomes
\begin{eqnarray}
W & = & \sum_a \sigma_a \left( \sum_{i b} \rho^a_{ib} Y_{ib} 
\: + \: \sum_{\alpha} Q_{\alpha} \ln X_{\alpha} \: - \:
\tilde{t} \right)
\nonumber \\
& & \: + \: \sum_{ia} \exp\left( - Y_{ia} \right) \: + \:
\sum_{\alpha} X_{\alpha},
\label{eq:ci:hyp2}
\end{eqnarray}
where $\tilde{t}$ is potentially shifted from $t$,
with the proviso that correlation functions will contain a factor of
\begin{displaymath}
\sum_{\mu < \nu} \left( \sigma_{\mu} - \sigma_{\nu} \right)^2.
\end{displaymath}
In the expressions above, as is usual for $U(k)$, we take
\begin{equation}
\rho^a_{ib} \: = \: \delta^a_b, \: \: \:
\alpha^a_{\mu \nu} \: = \: - \delta^a_{\mu} + \delta^a_{\nu}.
\end{equation}

It will later be useful to recall the operator mirror map~(\ref{eq:op-mirror-1}),
(\ref{eq:op-mirror-2}):
\begin{eqnarray}
\exp\left( - Y_{ia} \right) & \leftrightarrow & \sigma_a,
\label{eq:ci:mirr-1}
\\
X_{\alpha} & \leftrightarrow & - Q_{\alpha} \sum_a \sigma_a.
\label{eq:ci:mirr-2}
\end{eqnarray}

Integrating out the $\sigma_a$, we get constraints
\begin{equation}
\sum_{i=1}^n Y_{ia} \: + \: \sum_{\alpha} Q_{\alpha} \ln X_{\alpha} 
\: = \: \tilde{t},
\end{equation}
which we can use to eliminate some variables.
First, let us eliminate $\ln X_1$:
\begin{equation}
\ln X_1 \: = \: \frac{1}{Q_1} \left[ \tilde{t} \: - \: \sum_{i=1}^n Y_{ik}
\: - \: \sum_{\alpha=2}^S Q_{\alpha} \ln X_{\alpha} \right],
\end{equation}
where in the equation above, we have taken $a = k$ to be specific.
Eliminating $X_1$ in this fashion leaves $k-1$ independent constraints.
If we plug $X_1$ back into the constraints above for $a < k$, we find
that they reduce to
\begin{equation}
\sum_{i=1}^n \left( Y_{ia} - Y_{ik} \right) \: = \: 0,
\end{equation}
for $a < k$.  Define, for $a < k$,
\begin{eqnarray}
\hat{Y}_{ia} & \equiv & Y_{ia} - Y_{ik}, 
\end{eqnarray}
so that the remaining $k-1$ constraints can be written as
\begin{equation}
\sum_{i=1}^{n} \hat{Y}_{ia} \: = \: 0,
\end{equation}
for $a < k$.  We can use these constraints to
eliminate $\hat{Y}_{na}$:
\begin{equation}
\hat{Y}_{na} \: = \: - \sum_{i=1}^{n-1} \hat{Y}_{ia},
\label{eq:ci:hat-y-constr-1}
\end{equation}
for $a < k$.

In passing, in terms of these new variables, the 
operator mirror map~(\ref{eq:ci:mirr-1})
is
modified as follows:
\begin{eqnarray}
\exp\left( - \hat{Y}_{ia} \right) & \leftrightarrow &
\frac{\sigma_a}{\sigma_k},
\\
\exp\left( - \hat{Y}_{na} \right) & \leftrightarrow & \left(
\frac{\sigma_k}{\sigma_a} \right)^{n-1}.
\end{eqnarray}
The second expression is derived from the 
constraint~(\ref{eq:ci:hat-y-constr-1}).
Note that since the first expression is defined for $i$ including $i=n$,
self-consistency requires
\begin{equation}
\left( \sigma_a \right)^n \: = \: \left( \sigma_k \right)^n.
\label{eq:ci:mirr-constr-1}
\end{equation}
We will see this arise as a partial description of the critical locus,
later.

Plugging these expressions for $\hat{Y}_{ia}$
into the superpotential~(\ref{eq:ci:hyp2}),
we find that it becomes
\begin{eqnarray}
W & = & \sum_{i=1}^{n-1} \sum_{a=1}^{k-1} \exp\left( - \hat{Y}_{ia} \right)
\exp\left( - Y_{ik} \right) \: + \:
\sum_{i=1}^{n} \exp\left( - Y_{ik} \right)
\nonumber \\
& &  \: + \:
\sum_{a=1}^{k-1} \left[ \exp\left( - Y_{nk} \right)
 \prod_{i=1}^{n-1} \exp\left( + \hat{Y}_{ia} \right) \right]
\nonumber \\
& & \: + \:
\tilde{q}^{-1/Q_1} \left( \prod_{i=1}^n \exp\left( - Y_{ik}/Q_1 \right) \right)
\left( \prod_{\alpha=2}^S X_{\alpha}^{-Q_{\alpha}/Q_1} \right)
\: + \:
\sum_{\alpha=2}^S X_{\alpha}.
\label{eq:ci:sup-final}
\end{eqnarray}

Now, our goal is to outline how to reproduce the general
expression for correlation functions in the A-twisted theory given in
\cite{Closset:2015rna}[equ'n (8.32)], here in the mirror B-twisted
Landau-Ginzburg theory.  The first observation to make is that
the poles at which the residues are computed in \cite{Closset:2015rna},
will correspond to the critical loci of the B-twisted theory for which
we derived equations above.  In particular, in the B-twisted theory,
correlation functions can be expressed in the form
of residues
\begin{equation}
\langle {\cal O} \rangle \: = \: 
\oint \frac{dx_1 \wedge \cdots dx_n}{\partial_1 W \partial_2 W \cdots
\partial_n W} {\cal O},
\end{equation}
where $W = W(x_1, \cdots, x_n)$,
so by inspection, the poles at which the residue are computed in the B-twisted
theory correspond to critical loci of the superpotential.  (Strictly speaking,
the result above is for a flat metric on the target, and we have not
specified the K\"ahler metric in our proposal.  See
\cite{Guffin:2008kt} for analogous expressions for B-twisted Landau-Ginzburg
correlation functions for more
general cases.)  We will outline
how the residue here in the B-twisted theory matches the residue given in
\cite{Closset:2015rna}[equ'n (8.32)].  The first step will be to
argue that the poles in the integrand of \cite{Closset:2015rna}[equ'n (8.32)],
defined by the denominator
\begin{equation}
\left[ \prod_{a=1}^{k-1} \left( x_a^n - 1 \right) \right]
\left[ 1 + (-)^{k-n} q \left( \prod_{\alpha=1}^S Q_{\alpha}^{ Q_{\alpha} }
\right) \left( \sum_{a=1}^k x_a \right)^n \right],
\end{equation}
where $x_a$ is defined in \cite{Closset:2015rna} such that $x_a^n=1$,
correspond to the critical loci of the B-twisted mirror here.

In principle, to do this computation carefully, one should also
check whether the fundamental fields have changed, following analyses
described elsewhere in this paper, but to simplify and shorten the
argument, we will omit this step.  
(Part of the price we pay is that this loses information about
R-charges, so the mirror behaves somewhat more nearly like that
of a noncompact model, as we shall see later.)

Now, in principle one ought to check whether the fundamental fields may
have changed 
as a result of these variable changes, but instead of a lengthy rigorous
computation, we will
opt in this section to merely give a quick argument to formally
compare to 
results for correlation functions in \cite{Closset:2015rna}[section 8].
In the same spirit, in computations below, we will work with
$Y_{\alpha} = - \ln X_{\alpha}$ as a fundamental field
instead of $X_{\alpha}$, adding
suitable factors of 
\begin{displaymath}
\prod_{\alpha} \exp\left( - Y_{\alpha} \right)
\end{displaymath}
(from zero mode integration measures) to correlation functions.

From the superpotential~(\ref{eq:ci:sup-final}), the critical locus is
computed from
\begin{eqnarray}
\frac{\partial W}{\partial \hat{Y}_{ia}} & = &
- \exp\left( - \hat{Y}_{ia} \right) \exp\left( - Y_{ik} \right)
\: + \: \exp\left( - Y_{nk} \right) 
\prod_{j=1}^{n-1} \exp\left( + \hat{Y}_{ja} \right),
\label{eq:ci:crita-1}
\\
\frac{\partial W}{\partial Y_{ik} } & = &
\sum_{a=1}^{k-1} \exp\left( - \hat{Y}_{ia} \right) \exp\left( - Y_{ik} \right)
\: + \: \exp\left( - Y_{ik} \right)
\nonumber \\
& &  \: + \: \frac{1}{Q_1}
\tilde{q}^{-1/Q_1} \left( \prod_{j=1}^n \exp\left( - Y_{jk}/Q_1 \right) \right)
\left( \prod_{\alpha=2}^S X_{\alpha}^{-Q_{\alpha}/Q_1} \right)
\mbox{ for } i < n,
\label{eq:ci:crita-2}
\\
\frac{\partial W}{\partial Y_{ik} } & = &
\exp\left( - Y_{nk} \right) \: + \:
\sum_{a=1}^{k-1}  \left[ \exp\left( - Y_{nk} \right)
 \prod_{i=1}^{n-1} \exp\left( + \hat{Y}_{ia} \right) \right]
\nonumber \\
& & \: + \: \frac{1}{Q_1}
\tilde{q}^{-1/Q_1} \left( \prod_{j=1}^n \exp\left( - Y_{jk}/Q_1 \right) \right)
\left( \prod_{\alpha=2}^S X_{\alpha}^{-Q_{\alpha}/Q_1} \right)
\mbox{ for } i =n,
\label{eq:ci:crita-3}
\\
\frac{\partial W}{\partial Y_{\alpha} } & = &
X_{\alpha}
 \: - \: \frac{Q_{\alpha}}{ Q_1  }
\tilde{q}^{-1/Q_1} \left( \prod_{j=1}^n \exp\left( - Y_{jk}/Q_1 \right) \right)
\left( \prod_{\alpha=2}^S X_{\alpha}^{-Q_{\alpha}/Q_1} \right).
\label{eq:ci:crita-4}
\end{eqnarray}

From equation~(\ref{eq:ci:crita-1}), we see that on the critical locus
\begin{displaymath}
\exp\left( - \hat{Y}_{ia} \right) \exp\left( - Y_{ik} \right)
\end{displaymath}
is independent of $i$.  From equation~(\ref{eq:ci:crita-2}), we see that
\begin{displaymath}
\sum_{a=1}^{k-1} \left( 1 + \exp\left( - \hat{Y}_{ia} \right) \right)
\exp\left( - Y_{ik} \right)
\end{displaymath}
is independent of $i$, hence
\begin{displaymath}
\exp\left( - Y_{ik} \right)
\end{displaymath}
is independent of $i$, so cancelling out factors in 
equation~(\ref{eq:ci:crita-1}),
we see that on the critical locus
\begin{equation}
\exp\left( - \hat{Y}_{ia} \right) \: = \: 
\prod_{j=1}^{n-1} \exp\left( + \hat{Y}_{ja} \right).
\label{eq:ci:crita-res1}
\end{equation}
As a result,
\begin{displaymath}
\exp\left( - \hat{Y}_{ia} \right)
\end{displaymath}
is independent of $i$.  On the critical locus, define
\begin{equation}
z_a \: \equiv \: \exp\left( - \hat{Y}_{ia} \right)
\end{equation}
for any $i$, then from equation~(\ref{eq:ci:crita-res1}) we have
\begin{equation}
\left( z_a \right)^n \: = \: 1
\label{eq:ci:z-crit1}
\end{equation}
on the critical locus.
If we identify the $z_a$ here with the $x_a$ of \cite{Closset:2015rna},
then we see that the critical locus equation above defines some of the
locations of poles in the residue computation of the correlation function
for the A-twisted complete intersection in
\cite{Closset:2015rna}[equ'n (8.32)], as one would expect.

In passing, note that the operator mirror map~(\ref{eq:ci:mirr-1}) implies
\begin{equation}
z_a \: \leftrightarrow \: \frac{ \sigma_a }{\sigma_k}.
\label{eq:ci:z-op-mirr}
\end{equation}
The critical locus equation~(\ref{eq:ci:z-crit1}) above requires
\begin{displaymath}
\left( \sigma_a \right)^n \: = \: \left( \sigma_k \right)^n,
\end{displaymath}
which matches the self-consistency condition~(\ref{eq:ci:mirr-constr-1}) which
we derived earlier.  Thus, the critical locus equation above is compatible
with the mirror map.

Next, we shall compute the rest of the critical locus,
and see that it matches the rest of the poles of the integrand of
\cite{Closset:2015rna}[equ'n (8.32)].
From equation~(\ref{eq:ci:crita-4}), we see that on the critical locus,
\begin{displaymath}
\frac{ Q_1 X_{\alpha} }{ Q_{\alpha} } \: = \:
\tilde{q}^{-1/Q_1} \left( \prod_{i=1}^n \exp\left( - Y_{ik}/Q_1 \right) \right)
\left( \prod_{\beta=2}^S X_{\beta}^{-Q_{\beta}/Q_1} \right),
\end{displaymath}
independent of $\alpha$.
Define this to be a constant $C$:
\begin{equation}
C \: = \: 
\frac{ Q_1 X_{\alpha} }{ Q_{\alpha} } \: = \:
\tilde{q}^{-1/Q_1} \left( \prod_{i=1}^n \exp\left( - Y_{ik}/Q_1 \right) \right)
\left( \prod_{\beta=2}^S X_{\beta}^{-Q_{\beta}/Q_1} \right),
\end{equation}
Then,
\begin{equation}
\prod_{\alpha=2}^S X_{\alpha}^{- Q_{\alpha} / Q_1} \: = \:
\left[ \prod_{\alpha} \left( \frac{Q_{\alpha} }{ Q_1} \right)^{- Q_{\alpha}/ Q_1}
\right] C^{ 1 - \sum_{\alpha} Q_{\alpha} / Q_1 }.
\label{eq:ci:misc-1}
\end{equation}
From equation~(\ref{eq:ci:crita-4}), we know this also equals
\begin{equation}
C \tilde{q}^{+1/Q_1} \prod_{j=1}^{n} \exp\left( + Y_{jk}/Q_1 \right).
\label{eq:ci:misc-2}
\end{equation}
Taking a $Q_1$ power of both sides, we find
\begin{displaymath}
\tilde{q} C^{Q_1} \prod_{j=1}^n \exp\left( + Y_{jk} \right) \: = \:
C^{Q_1 - \sum_{\alpha} Q_{\alpha}} \left[
\prod_{\alpha} \left( \frac{ Q_{\alpha} }{ Q_1 } \right)^{- Q_{\alpha} }
\right],
\end{displaymath}
or more simply, on the critical locus,
\begin{displaymath}
\tilde{q} \left( \exp\left( + Y_{nk} \right) \right)^n \: = \:
C^{- \sum_{\alpha} Q_{\alpha} } \left[
\prod_{\alpha} \left( \frac{ Q_{\alpha} }{ Q_1 } \right)^{- Q_{\alpha} }
\right].
\end{displaymath}
From equation~(\ref{eq:ci:crita-2}), we also know
\begin{displaymath}
\frac{ Q_1 X_{\alpha} }{ Q_{\alpha} } \: = \: C \: = \: 
- Q_1 \left[ 1 \: + \: \sum_{a=1}^{k-1} z_a \right] \exp\left( - Y_{nk} \right),
\end{displaymath}
hence
\begin{displaymath}
\left( \exp\left(+ Y_{nk} \right) \right)^n \: = \:
(-)^n C^{-n} Q_1^n \left[ 1 \: + \: \sum_{a=1}^{k-1} z_a \right]^n.
\end{displaymath}
Combining these equations, we find
\begin{equation}
(-)^n \tilde{q} C^{-n} Q_1^n  \left[ 1 \: + \: \sum_{a=1}^{k-1} z_a \right]^n
\: = \:
C^{- \sum_{\alpha} Q_{\alpha} } \left[
\prod_{\alpha} \left( \frac{ Q_{\alpha} }{ Q_1 } \right)^{- Q_{\alpha} }
\right].
\end{equation}
Using the Calabi-Yau condition
\begin{equation}
\sum_{\alpha} Q_{\alpha} \: = \: n,
\end{equation}
we find
\begin{equation}
(-)^n \tilde{q} \left( \prod_{\alpha} Q_{\alpha}^{Q_{\alpha}}
\right) \left[ 1 \: + \: \sum_{a=1}^{k-1} z_a \right]^n
 \: = \: 1.
\end{equation}
Since $\tilde{q} = (-)^{k-1} q$, this can be rewritten as
\begin{equation}
1 \: + \: (-)^{k-n} q \left( \prod_{\beta} 
Q_{\beta}^{ Q_{\beta} } \right)
\left[ 1 \: + \:
 \sum_{a=1}^{k-1} z_a \right]^n \: = \: 0,
\end{equation}
which precisely matches the second condition for a pole in
the integrand of \cite{Closset:2015rna}[equ'n (8.32)].

So far, we have demonstrated that the critical loci of our proposed 
Landau-Ginzburg mirror match the poles of the integrand of the residue
formula for A model correlation functions in 
\cite{Closset:2015rna}[equ'n (8.32)].

Next, we will give a brief outline of how
the rest of the B model correlation function matches the A model
computations in \cite{Closset:2015rna}[section 8].
We emphasize that the analysis we describe here is designed to be
brief and hopefully readable, but not rigorous.

For later use, let us also compute $C$.
Combining equations~(\ref{eq:ci:misc-1}) and (\ref{eq:ci:misc-2}),
we have that 
\begin{displaymath}
\left[ \prod_{\alpha} \left( \frac{ Q_{\alpha} }{ Q_1 } \right)^{- Q_{\alpha} /
Q_1 } \right] C^{- \sum_{\alpha} Q_{\alpha}/Q_1} \: = \:
\tilde{q}^{+1/Q_1} \prod_{j=1}^n \exp\left( + Y_{jk}/Q_1 \right).
\end{displaymath}
Simplifying, using the Calabi-Yau condition, and solving for $C$,
we find
\begin{equation}
C \: = \: \tilde{q}^{-1/n} Q_1 \left[ \prod_{\alpha}   
 Q_{\alpha}^{-Q_{\alpha} } \right]^{1/n} \prod_{j=1}^n \exp\left( - Y_{jk}/n
\right).
\end{equation}

Next, we shall formally integrate out the $\hat{Y}_{ia}$.  It is
straightforward to compute that
\begin{equation}
\frac{\partial^2 W}{\partial \hat{Y}_{ia} \partial \hat{Y}_{jb} } \: = \:
\delta_{ij} \delta_{ab} \exp\left( - \hat{Y}_{ia} \right)
\exp\left( - Y_{ik} \right) \: + \: \delta_{ab}
\exp\left( - Y_{nk} \right)
\prod_{\ell=1}^{n-1} \exp\left( + \hat{Y}_{\ell a} \right).
\end{equation}
To compute the determinant of the resulting matrix of second derivatives,
it is useful to block-diagonalize with blocks of the same $a$.
Each such block takes the form
\begin{displaymath}
\left[ \begin{array}{cccc}
A+B & A & A & \cdots \\
A & A+B & A & \cdots \\
A & A & A+B & \cdots \\
\vdots & & & 
\end{array}
\right],
\end{displaymath} 
and the determinant of an $n \times n$ block of this form can be shown
to be
\begin{displaymath}
B^n \: + \: n A B^{n-1}.
\end{displaymath}
Here, on the critical locus,
\begin{eqnarray}
A & = & \exp\left( - Y_{nk} \right) 
\prod_{\ell=1}^{n-1} \exp\left( + \hat{Y}_{\ell a} \right)
\: = \: \left( z_a \right)^{-n+1} \exp\left( - Y_{nk} \right),
\\
B & = &  \exp\left( - \hat{Y}_{ia} \right)
\exp\left( - Y_{ik} \right) \: = \:
z_a \exp\left( - Y_{nk} \right),
\end{eqnarray}
so the determinant of the entire matrix (the product of the determinants
of $k-1$ blocks \cite{powell}, each an $(n-1) \times (n-1)$ matrix)
is given by
\begin{equation}
\exp\left( - (n-1)(k-1) Y_{nk} \right) \prod_{a=1}^{k-1} \left[
 z_a^{n-2} \left( z_a \: + \: (n-1) (z_a)^{-n+1} \right) \right].
\end{equation}
On the critical locus, we can use equation~(\ref{eq:ci:z-crit1}),
namely $z_a^n=1$, to write this as
\begin{equation}
\exp\left( - (n-1)(k-1) Y_{nk} \right) \prod_{a=1}^{k-1} \left[
z_a^{-2} \left( n z_a \right) \right].
\end{equation}
This will be a factor appearing in correlation functions, which we will
utilize later.

The superpotential~(\ref{eq:ci:sup-final}) becomes
\begin{eqnarray}
W & = & \sum_{i=1}^{n-1} \sum_{a=1}^{k-1} z_a
\exp\left( - Y_{ik} \right) \: + \:
\sum_{i=1}^{n} \exp\left( - Y_{ik} \right)
\nonumber \\
& &  \: + \:
\sum_{a=1}^{k-1} \left[ \exp\left( - Y_{nk} \right)
\left( z_a \right)^{1-n}
\right]
\nonumber \\
& & \: + \:
\tilde{q}^{-1/Q_1} \left( \prod_{i=1}^n \exp\left( - Y_{ik}/Q_1 \right) \right)
\left( \prod_{\alpha=2}^S X_{\alpha}^{-Q_{\alpha}/Q_1} \right)
\: + \:
\sum_{\alpha=2}^S X_{\alpha}.
\label{eq:ci:sup-final2}
\end{eqnarray}

Next, we similarly integrate out the $Y_{\alpha}$.
From the superpotential above, we compute
\begin{eqnarray*}
\frac{\partial W}{\partial Y_{\alpha} } & = &
X_{\alpha} \: - \: \frac{ Q_{\alpha} }{ Q_1 }
\tilde{q}^{-1/Q_1} \left( \prod_{i=1}^n \exp\left( - Y_{ik}/Q_1 \right) \right)
\left( \prod_{\gamma=2}^S X_{\gamma}^{-Q_{\gamma}/Q_1} \right),
\\
\frac{\partial^2 W}{\partial Y_{\alpha} \partial Y_{\beta} } & = &
\delta_{\alpha \beta} X_{\alpha} + 
\frac{Q_{\alpha} Q_{\beta} }{ Q_1^2}
\tilde{q}^{-1/Q_1} \left( \prod_{i=1}^n \exp\left( - Y_{ik}/Q_1 \right) \right)
\left( \prod_{\gamma=2}^S X_{\gamma}^{-Q_{\gamma}/Q_1} \right).
\end{eqnarray*}
On the critical locus,
\begin{eqnarray}
\frac{\partial^2 W}{\partial Y_{\alpha} \partial Y_{\beta} } & = &
\delta_{\alpha \beta} X_{\alpha} \: + \:
\frac{ Q_{\alpha} Q_{\beta} }{ Q_1^2 } C,
\\
& = & C \left( \delta_{\alpha \beta} \frac{Q_{\alpha}}{Q_1} \: + \:
\frac{Q_{\alpha} Q_{\beta}}{Q_1^2} \right).
\end{eqnarray}
The determinant of the matrix of second derivatives is then
proportional to
\begin{equation}
C^{S-1 } \: \propto \:
\tilde{q}^{- (S-1)/n}
\prod_{j=1}^n \exp\left( - Y_{jk} (S-1)/n \right)
\end{equation}
(up to irrelevant constant factors).

The superpotential~(\ref{eq:ci:sup-final2}) becomes
\begin{eqnarray}
W & = & \sum_{i=1}^{n-1} \sum_{a=1}^{k-1} z_a
\exp\left( - Y_{ik} \right) \: + \:
\sum_{i=1}^{n} \exp\left( - Y_{ik} \right)
\nonumber \\
& &  \: + \:
\sum_{a=1}^{k-1} \left[ \exp\left( - Y_{nk} \right)
\left( z_a \right)^{n-1}
\right]
\nonumber \\
& & \: + \:
C
\: + \:
\sum_{\alpha=2}^S \frac{ Q_{\alpha} C}{Q_1},
\label{eq:ci:sup-final3}
\\
& = & \sum_{i=1}^{n} \exp\left( - Y_{ik} \right)
\left[ 1 + \sum_{a=1}^{k-1} z_a \right]
\nonumber \\
& & 
 \: + \:
n \,
\tilde{q}^{-1/n} \left[ \prod_{\alpha}  Q_{\alpha}^{-Q_{\alpha} }
 \right]^{1/n} \prod_{j=1}^n \exp\left( - Y_{jk}/n
\right).
\label{eq:ci:sup-final3a}
\end{eqnarray}

In passing, if we were to 
identify $\exp(-Y_{jk}/n)$ as the fundamental fields,
then the expression above would look like the Landau-Ginzburg mirror 
of a degree $n$ hypersurface in ${\mathbb P}^{n-1}$.  That observation
glosses over the fact that correlation functions have extra factors
of determinants computed earlier, so we are certainly not claiming that
this is the same as the mirror to a Calabi-Yau hypersurface
in ${\mathbb P}^{n-1}$; nevertheless, the resemblance of the superpotential
above is striking.

Now, let us start collecting factors.
From integrating out the $\hat{Y}_{ia}$, we derived a determinant $H$,
and hence a factor of $1/H^{1-g}$ in correlation functions, for
\begin{displaymath}
H \: \propto \: \exp\left( - (n-1)(k-1) Y_{nk} \right),
\end{displaymath}
and similarly from integrating out the $X_{\alpha}$, we derived
a determinant proportional to
\begin{displaymath}
\prod_{j=1}^n \exp\left( - Y_{jk} (S-1)/n \right).
\end{displaymath}
On the critical locus, where $\exp(-Y_{ik} )$ is independent of $i$,
this is an overall factor of $1/H'^{1-g}$ for
\begin{equation}
H' \: \propto \: \exp\left[ - Y_{nk} \left( (n-1)(k-1) + (S-1) \right)
\right].
\end{equation}
Finally, from integrating out the $X_{\mu \nu}$, correlation functions will
have a factor of $H''^{1-g}$ for
\begin{equation}
H'' \: = \: \sum_{\mu < \nu} \left( \sigma_{\mu} - \sigma_{\nu} \right)^2
\: \propto \: \sum_{\mu < \nu} \left( \frac{\sigma_{\mu}}{\sigma_k} 
- \frac{\sigma_{\nu}}{\sigma_k} \right)^2 \sigma_k^2,
\end{equation}
which contributes factors of $\sigma_k^2 = \exp(-2Y_{nk})$ to our analysis.
Putting this together,
when computing correlation functions,
and writing everything with the operator mirror map in terms of $\sigma$'s,
we get factors of
\begin{eqnarray}
\left(
\frac{ \exp\left( - (k^2-k) Y_{nk} \right) }{
\exp\left( - \left( (n-1)(k-1) + (S-1) \right) Y_{nk} \right) }
\right)^{1-g}
& = &
\left(
\frac{1}{
\sigma_k^{ nk - n + S - k^2 }
} \right)^{1-g},
\end{eqnarray}
together of course with a factor of
\begin{equation}
\sum_{\mu < \nu} \left( \frac{\sigma_{\mu}}{\sigma_k} 
- \frac{\sigma_{\nu}}{\sigma_k} \right)^2 
\: = \:
\sum_{\mu < \nu} \left( z_{\mu} - z_{\nu} \right)^2,
\end{equation}
using the operator mirror map~(\ref{eq:ci:z-op-mirr}).

Now, let us collect other factors.
From the original zero mode integration measure, we have factors of
\begin{displaymath}
\prod_{\alpha} X_{\alpha} \: = \:
\prod_{\alpha} \exp\left( - Y_{\alpha} \right).
\end{displaymath}
We eliminated $X_1$ as an independent field earlier, but there is
still a factor of $\exp(-Y_1)$.  It will be simplest to write these
using the operator mirror map~(\ref{eq:ci:mirr-2}) in terms of $\sigma$'s:
\begin{displaymath}
\exp\left( - Y_{\alpha} \right) \: = \: - Q_{\alpha} \sum_a \sigma_a,
\end{displaymath}
hence the factor in correlation functions from zero mode integrals is
\begin{eqnarray}
\prod_{\alpha} \left( - Q_{\alpha} \sum_a \sigma_a\right) & = &
(-)^S \left( \sum_a \sigma_a \right)^S \prod_{\alpha} Q_{\alpha}^{Q_{\alpha}},
\\
& = &
(-)^S \sigma_k^S \left( \sum_a \frac{\sigma_a}{\sigma_k} \right)^S
\prod_{\alpha} Q_{\alpha}^{Q_{\alpha}},
\\
& = &
(-)^S \sigma_k^S \left( \sum_a z_a \right)^S
\prod_{\alpha} Q_{\alpha}^{Q_{\alpha}},
\end{eqnarray}
where we have used the mirror map relation~(\ref{eq:ci:z-op-mirr}).

Putting all of the factors discussed together, and working at genus zero
for simplicity, in correlation functions we have an overall factor 
proportional to
\begin{equation}
\left( \sum_a z_a \right)^S
\left[ \sum_{\mu < \nu} \left( z_{\mu} - z_{\nu} \right)^2 \right]
\left[ \prod_{a=1}^{k-1} nz_a^{n-1} \right]
\sigma_k^{- (k(n-k) - n) } .
\label{eq:ci:final-facts}
\end{equation}

To get a nonzero correlation function, we will want as many factors
of $\sigma_k$ in the operators as in the denominator in the expression
above.  The reader should note that with the choices of fundamental
variables we are using, we are effectively taking the $p$ fields of
the original $A$ model not to have R-charge, hence we are working in
an analogue of the $V_+$ model.  One therefore expects that
nonzero correlation functions
will therefore have operators of the same dimension as the ambient
space ($k(n-k)$), plus an extra factor of $\sigma_k^n$ arising from
the obstruction sheaf, so as to reproduce correlation functions on the
compact complete intersection.  This many factors of $\sigma_k$ exactly
matches the number of factors in the denominator of the expression
above.

Now, let us briefly outline how the computation above should
match the expression in 
\cite{Closset:2015rna}[equ'n (8.32)], which gives the corresponding
correlation functions in the A model as
\begin{eqnarray}
\langle {\cal O} \rangle & = & \left( \prod_{\alpha} Q_{\alpha} \right)
{\rm Res}\, \frac{
\left[ \prod_{\mu < \nu} \left( z_{\mu} - z_{\nu} \right)^2 \right]
\left( \sum_{a=1}^k z_a \right)^S {\cal O}
}{
\left[ \prod_{a=1}^{k-1} \left( z_a^n - 1 \right) \right]
\left[ 1 + (-)^{k-n} q \left( \prod_{\alpha} Q_{\alpha}^{Q_{\alpha}} 
\right) \left( \sum_{a=1}^k z_a \right)^n
\right]
},
\end{eqnarray}
up to overall constants.
We have already explained the origin of the factors
\begin{displaymath}
\sigma_k^{- (k(n-k) - n) }.
\end{displaymath}
The factors in the numerator
\begin{equation}
\left[ \prod_{\mu < \nu} \left( z_{\mu} - z_{\nu} \right)^2 \right]
\left( \sum_{a=1}^k z_a \right)^S 
\end{equation}
trivially match factors in~(\ref{eq:ci:final-facts}).
To write the result as a residue, roughly, we would take our
factors and divide by equations for the critical locus, which should
define the location of the poles of the residue.
The factor of 
\begin{displaymath}
\left[ \prod_{a=1}^{k-1} nz_a^{n-1} \right]
\end{displaymath}
in~(\ref{eq:ci:final-facts}) serves the correctly renormalize factors
arising when computing the residue of
\begin{displaymath}
\left[ \prod_{a=1}^{k-1} \left( z_a^n - 1 \right) \right].
\end{displaymath}

To summarize, we have checked that the critical loci of the proposed
mirror to a GLSM for a complete intersection in $G(k,n)$ match
the poles of residue expressions for correlation functions in
\cite{Closset:2015rna}[section 8], and we have sketched how the
B model correlation functions should match the A model correlation functions
given in \cite{Closset:2015rna}[section 8].

\section{Conclusions}

In this paper we have proposed a mirror construction for
two-dimensional A-twisted gauge theories, addressing an old unsolved
issue 
in the literature.  We have explicitly
compared our proposal to results in the literature for Coulomb branch
relations and constraints and, when appropriate, quantum cohomology
rings for two-dimensional gauge theories with a variety of gauge
groups, matter representations, and twisted masses, ranging from
GLSM descriptions of Grassmannians to various $SO$ and $Sp$ gauge theories with
chiral superfields in vector representations with twisted masses.
We have also tested and refined predictions for pure $SU$ theories and also
tested analogous conjectures for pure $SO$ and $Sp$ theories.

One can imagine several future directions for this work.
One direction
we wish to highlight are analogues for nonabelian
(0,2) supersymmetric
gauge theories, generalizing work on mirrors to abelian 
(0,2) supersymmetric gauge theories in 
\cite{Chen:2016tdd,Chen:2017mxp,Gu:2017nye}.

\section{Acknowledgements}

We would like to thank K.~Cho, C.~Closset, J.~Guo, K.~Hori, S.~Katz, H.~Kim,
J.~Knapp, Z.~Lu, I.~Melnikov, J.~Park, T.~Pantev, and C.~Vafa
for useful conversations.
E.S. was partially supported by NSF grant PHY-1720321.

\appendix

\section{Rietsch's mirror to $G(k,n)$}
\label{app:rietsch}

In this appendix we will describe Eguchi-Hori-Xiong and 
Rietsch's proposed mirror to
Grassmannians \cite{ehx,r1,r2,r3}
in more detail.  As mentioned in the text, this is also
a Landau-Ginzburg model with critical loci duplicating, to our
knowledge, both correlation functions and quantum cohomology relations,
hence we conjecture that it is `Seiberg-dual' to the proposal in the main
text.  

There are several different-looking but essentially\footnote{
Some different presentations may also differ in having slightly different
(partial) compactifications of the underlying space of the Landau-Ginzburg
model.  For our purposes, so long as the critical loci are unchanged
by such compactifications,
the low-energy physics should also be unchaged.
}
equivalent formulations
of this mirror.  Reference \cite{r2}
in particular
lays out the equivalence between various formulations of its proposal
as well as with \cite{ehx}[section 6.3],
so for simplicitly we shall
focus here on reviewing just one of the forms in \cite{r2}.

To compare to \cite{r2}, we first need to give an alternative
description of Pl\"ucker coordinates.  Recall that Pl\"ucker coordinates
are `baryons' formed in the GLSM for a Grassmannian $G(k,n)$. 
The GLSM is a $U(k)$ gauge theory with $n$ fields $\phi^i$ in the
fundamental representation of $U(k)$.  The baryons are then
\begin{displaymath}
B^{i_1 \cdots i_k} \: = \: \phi^{i_1}_{a_1} \cdots \phi^{i_k}_{a_k}
\epsilon^{a_1 \cdots a_k}.
\end{displaymath}
Mathematically, these give homogeneous coordinates on a projective
space of dimension
\begin{displaymath}
\left( \begin{array}{c}
n \\ k \end{array} \right) - 1,
\end{displaymath}
and so define a projective embedding of the Grassmannian.

These same baryons are in one-to-one correspondence with
Young tableaux sitting inside a $k \times (n-k)$ box.
Take a decreasing sequence $(\lambda_1, \cdots, \lambda_k)$
where $0 \leq \lambda_i \leq n-k$ for each $i$,
then $(\lambda_1+k, \lambda_2+k-1, \cdots, \lambda_k + 1)$ is a
decreasing sequence with each entry $\leq n$.
Identify
\begin{displaymath}
i_1 = \lambda_1 + k, \: \: \:
i_2 = \lambda_2 + k-1, \: \: \:
\cdots, \: \: \:
i_k = \lambda_k + 1.
\end{displaymath}
In this fashion, we can associate a baryon.
(Note that these conventions may yield Young tableaux which are transposes
of those of \cite{r2}.)

Some examples of the above may be helpful.
Consider for example the case $k=2$.
Then Young tableaux can be associated to baryon indices as follows:
\begin{center}
\begin{tabular}{cl}
$(i_1, i_2)$ & Young tableaux \\ \hline
$(2,1)$ & $\emptyset$ $(\lambda_1=\lambda_2=0)$ \\
$(3,2)$ & $\tiny\yng(1,1)$ $(\lambda_1 = \lambda_2 = 1)$ \\
$(4,3)$ & $\tiny\yng(2,2)$ $(\lambda_1=\lambda_2=2)$ \\
$(3,1)$ & $\tiny\yng(1)$  $(\lambda_1=1, \lambda_2=0)$
\end{tabular}
\end{center}
The special set of Young tableaux corresponding to
the sequential set
\begin{displaymath}
(i_1, \cdots, i_k) = (i+k, i+k-1, i+k-2, \cdots, i+1)
\end{displaymath}
(with indices interpreted mod $n$)
will be denoted $\mu_{i}$, following \cite{r2}.
For example, for the case $k=2$, $n=4$, we have
\begin{center}
\begin{tabular}{cccc}
$\mu$ & $(i_1,i_2)$ & $(\lambda_1,\lambda_2)$ & Young tableau \\ \hline
$\mu_1$ & $(3,2)$ & $(1,1)$ & $\tiny\yng(1,1)$ \\
$\mu_2$ & $(4,3)$ & $(2,2)$ & $\tiny\yng(2,2)$ \\
$\mu_3$ & $(1,4) = (4,1)$ & $(2,0)$ & $\tiny\yng(2)$ \\
$\mu_4$ & $(2,1)$ & $(0,0)$ & $\emptyset$
\end{tabular}
\end{center}
Similarly, the Young tableaux corresponding to the sequential set
\begin{displaymath}
(i_1, \cdots, i_k) = (i+k+1, i+k-1, i+k-2, \cdots, i+1)
\end{displaymath}
will be denoted $\hat{\mu}_{i}$.

Then, the Landau-Ginzburg
mirror to the Grassmannian $G(n-k,n)$ is constructed in \cite{r2} as
a Landau-Ginzburg model over (an open\footnote{
So long as our open subset contains all the critical points, we are free
to restrict to any convenient open subset of the original space without
changing the B model correlation functions or chiral ring.
} subset of) $G(k,n)$ with superpotential
\cite{r2}[equ'n (6.4)]
\begin{displaymath}
W \: = \:
\sum_{i=1}^n \frac{ p_{ \hat{\mu}_i } }{ p_{ \mu_i } } q^{\delta_{i, n-k} }
\: = \:
\sum_{i \neq n-k}  \frac{ p_{ \hat{\mu}_i } }{ p_{ \mu_i } }
+ q \frac{ p_{ \hat{\mu}_{n-k} } }{ p_{ \mu_{n-k} } }.
\end{displaymath}
Moreover, if we fix the open set by taking one of the $p \neq 0$,
then the resulting open set is isomorphic to ${\mathbb C}^{k(n-k)}$
\cite{gh}[chapter 1.5], as we shall see explicitly in examples below.

\subsection{${\mathbb P}^2$}

Let us first apply this to the special case of ${\mathbb P}^2$, realized
in two different ways, to explicitly verify that we recover the
known Toda mirror:
\begin{itemize}
\item First, let us construct the mirror Landau-Ginzburg model to
$G(1,3)$.  Here, $k-1$, $n=3$.  We compute
\begin{center}
\begin{tabular}{cccc}
$\mu$ & $i$ & $\lambda$ & Young tableau \\ \hline
$\mu_1$ & $2$ & $1$ & $\tiny\yng(1)$ \\
$\mu_2$ & $3$ & $2$ & $\tiny\yng(2)$ \\
$\mu_3$ & $1$ & $0$ & $\emptyset$\\
$\hat{\mu}_1$ & $3$ & $2$ & $\tiny\yng(2) = \mu_2$ \\
$\hat{\mu}_2$ & $1$ & $0$ & $\emptyset = \mu_3$ \\
$\hat{\mu}_3$ & $2$ & $1$ & $\tiny\yng(1) = \mu_1$
\end{tabular}
\end{center}
so the mirror superpotential is given by
\begin{eqnarray*}
W & = & \frac{p_{ \hat{\mu}_1 } }{ p_{\mu_1} } \: + \:
q \frac{ p_{ \hat{\mu}_2 } }{ p_{ \mu_2 } } \: + \:
\frac{ p_{ \hat{\mu}_3} }{ p_{\mu_3} }, \\
& = & \frac{ p_{\tiny\yng(2)} }{ p_{\tiny\yng(1) } } \: + \:
q \frac{ p_{\emptyset} }{p_{\tiny\yng(2)}} \: + \:
\frac{ p_{\tiny\yng(1)} }{ p_{\emptyset} }.
\end{eqnarray*}
The three Pl\"ucker coordinates coincide with homogeneous coordinates
on ${\mathbb P}^2$, and
\begin{displaymath}
{\mathbb P}^2 - \{ p_{\emptyset} = 0 \} - \{ p_{\tiny\yng(1) }=0 \} -
\{ p_{\tiny\yng(2) } \} \: = \: ({\mathbb C}^{\times})^2.
\end{displaymath}
Working in the patch $p_{\emptyset} \neq 0$, say, and identifying
$x = p_{\tiny\yng(1)}$, $y = p_{\tiny\yng(2)} / p_{\tiny\yng(1)}$,
then we see that the mirror superpotential can be written as
\begin{displaymath}
W \: = \: x + y + \frac{q}{xy},
\end{displaymath}
matching other expressions for the Toda dual of ${\mathbb P}^2$.
\item Next, we construct the mirror Landau-Ginzburg model to
$G(2,3)$.  Here, $k=2$, $n=3$.  We compute
\begin{center}
\begin{tabular}{cccc}
$\mu$ & $(i_1,i_2)$ & $(\lambda_1,\lambda_2)$ & Young tableau \\ \hline
$\mu_1$ & $(3,2)$ & $(1,1)$ & $\tiny\yng(1,1)$ \\
$\mu_2$ & $(1,3)=(3,1)$ & $(1,0)$ & $\tiny\yng(1)$ \\
$\mu_3$ & $(2,1)$ & $(0,0)$ & $\emptyset$ \\
$\hat{\mu}_1$ & $(1,2) = (2,1)$ & $(0,0)$ & $\emptyset = \mu_3$ \\
$\hat{\mu}_2$ & $(2,3) = (3,2)$ & $(1,1)$ & $\tiny\yng(1,1) = \mu_1$ \\
$\hat{\mu}_3$ & $(3,1)$ & $(1,0)$ & $\tiny\yng(1) = \mu_2$
\end{tabular}
\end{center}
The mirror superpotential is
\begin{eqnarray*}
W & = & q \frac{ p_{ \hat{\mu}_1 } }{ p_{ \mu_1 } } \: + \:
\frac{ p_{ \hat{\mu}_2 } }{ p_{ \mu_2 } } \: + \:
\frac{ p_{ \hat{\mu}_3 } }{ p_{ \mu_3 } }, \\
& = & q \frac{ p_{\emptyset}}{p_{\tiny\yng(1,1) } } \: + \:
\frac{ p_{\tiny\yng(1,1)} }{ p_{\tiny\yng(1) } } \: + \:
\frac{ p_{\tiny\yng(1) } }{ p_{\emptyset} }.
\end{eqnarray*}
Modulo transposing the Young tableaux, this mirror is identical to that
for $G(1,3)$.
\end{itemize}

In the mirror to ${\mathbb P}^2$ computed above,
it happened that each $\hat{\mu}_i$ coincided with $\mu_j$ for some
$j$.  This reflects that fact that $k=1$ or $n-1$, and does not happen
in general.

\subsection{$G(2,4)$}

Consider the Grassmannian $G(2,4)$.  Its mirror is constructed as a
Landau-Ginzburg model over (an open subset of) $G(2,4)$.
We compute
\begin{center}
\begin{tabular}{cccc}
$\mu$ & $(i_1,i_2)$ & $(\lambda_1,\lambda_2)$ & Young tableau \\ \hline
$\mu_1$ & $(3,2)$ & $(1,1)$ & $\tiny\yng(1,1)$ \\
$\mu_2$ & $(4,3)$ & $(2,2)$ & $\tiny\yng(2,2)$ \\
$\mu_3$ & $(1,4)=(4,1)$ & $(2,0)$ & $\tiny\yng(2)$ \\
$\mu_4$ & $(2,1)$ & $(0,0)$ & $\emptyset$ \\ \hline
$\hat{\mu}_1$ & $(4,2)$ & $(2,1)$ & $\tiny\yng(2,1)$ \\
$\hat{\mu}_2$ & $(1,3)=(3,1)$ & $(1,0)$ & $\tiny\yng(1)$ \\
$\hat{\mu}_3$ & $(2,4)=(4,2)$ & $(2,1)$ & $\tiny\yng(2,1)$ \\
$\hat{\mu}_4$ & $(3,1)$ & $(1,0)$ & $\tiny\yng(1)$
\end{tabular}
\end{center}
It has superpotential
\begin{eqnarray*}
W & = & \frac{ p_{ \hat{\mu}_1 } }{ p_{ \mu_1 } } \: + \:
q \frac{ p_{ \hat{\mu}_2 } }{ p_{ \mu_2 } } \: + \:
\frac{ p_{ \hat{\mu}_3 } }{ p_{ \mu_3 } } \: + \: 
\frac{ p_{ \hat{\mu}_4 } }{ p_{ \mu_4 } }, \\
& = & 
\frac{ p_{ \tiny\yng(2,1) } }{ p_{ \tiny\yng(1,1) } } \: + \:
q \frac{ p_{ \tiny\yng(1) } }{ p_{\tiny\yng(2,2) } } \: + \:
\frac{ p_{ \tiny\yng(2,1) } }{ p_{\tiny\yng(2) } } \: + \:
\frac{ p_{ \tiny\yng(1) } }{ p_{\emptyset} },
\end{eqnarray*}
where the $p$'s are Pl\"ucker coordinates associated to Young tableaux,
as above, and the open subset of $G(2,4)$ is defined by excluding the points
where $W$ has poles.

Now, we need to coordinatize $G(2,4)$.  Points of $G(2,4)$ can be represented
by matrices
\begin{displaymath}
\left[ \begin{array}{cccc}
a_1 & a_2 & a_3 & a_4 \\
b_1 & b_2 & b_3 & b_4
\end{array} \right],
\end{displaymath}
and the Pl\"ucker coordinates are the minors determined by columsn of the
matrix above.
The open subset $\{ p_{\emptyset} \neq 0 \}$ can be identified with
${\mathbb C}^4$ as matrices of the form
\begin{displaymath}
\left[ \begin{array}{cccc}
1 & 0 & a_3 & a_4 \\
0 & 1 & b_3 & b_4
\end{array} \right].
\end{displaymath}
We now define coordinates on ${\mathbb C}^4$ as follows:
\begin{eqnarray*}
x_1 & \equiv & p_{\tiny\yng(1,1)} = p_{ \mu_1 } = - a_3 =
p_{ \hat{\mu}_3 }, \\
x_2 & \equiv & p_{\tiny\yng(2,1)} = p_{ \hat{\mu}_1 } = -a_4, \\
x_3 & \equiv & p_{\tiny\yng(1)} = p_{ \hat{\mu}_4 } = + b_3 =
p_{ \hat{\mu}_2 }, \\
x_4 & \equiv & p_{\tiny\yng(2)} = p_{ \mu_3 } = b_4,
\end{eqnarray*}
so that
\begin{displaymath}
p_{\tiny\yng(2,2)} = p_{ \mu_2 } = a_3 b_4 - a_4 b_3 = 
- x_1 x_4 + x_2 x_3.
\end{displaymath}
Then, in these variables, the mirror superpotential can be written
\begin{displaymath}
W \: = \:
\frac{ x_2 }{ x_1 } \: + \:
q \frac{ x_3 }{ x_2 x_3 - x_1 x_4} \: + \:
\frac{ x_2 }{x_4 } \: + \:
x_3.
\end{displaymath}

\subsection{$G(2,5)$}

Another example of a Landau-Ginzburg 
mirror to a Grassmannian in \cite{r2} is given as follows.
Consider the Grassmannian $G(2,5)$.  Its mirror is constructed as a
Landau-Ginzburg model over (an open subset of) $G(3,5)$.
We compute
\begin{center}
\begin{tabular}{cccc}
$\mu$ & $(i_1,i_2)$ & $(\lambda_1,\lambda_2)$ & Young tableau \\ \hline
$\mu_1$ & $(3,2)$ & $(1,1)$ & $\tiny\yng(1,1)$ \\
$\mu_2$ & $(4,3)$ & $(2,2)$ & $\tiny\yng(2,2)$ \\
$\mu_3$ & $(5,4)$ & $(3,3)$ & $\tiny\yng(3,3)$ \\
$\mu_4$ & $(1,5) = (5,1)$ & $(3,0)$ & $\tiny\yng(3)$ \\
$\mu_5$ & $(2,1)$ & $(0,0)$ & $\emptyset$ \\  \hline
$\hat{\mu}_1$ & $(4,2)$ & $(2,1)$ & $\tiny\yng(2,1)$ \\
$\hat{\mu}_2$ & $(5,3)$ & $(3,2)$ & $\tiny\yng(3,2)$ \\
$\hat{\mu}_3$ & $(1,4)=(4,1)$ & $(2,0)$ & $\tiny\yng(2)$ \\
$\hat{\mu}_4$ & $(2,5)=(5,2)$ & $(3,1)$ & $\tiny\yng(3,1)$  \\
$\hat{\mu}_5$ & $(3,1)$ & $(1,0)$ & $\tiny\yng(1)$
\end{tabular}
\end{center}
It has superpotential
\begin{eqnarray*}
W & = & \frac{ p_{ \hat{\mu}_1 } }{ p_{ \mu_1 } } \: + \:
\frac{ p_{ \hat{\mu}_2 } }{ p_{ \mu_2 } } \: + \:
q \frac{ p_{ \hat{\mu}_3 } }{ p_{ \mu_3 } } \: + \:
\frac{ p_{ \hat{\mu}_4 } }{ p_{ \mu_4 } } \: + \:
\frac{ p_{ \hat{\mu}_5 } }{ p_{ \mu_5 } }, \\
& = &
\frac{ p_{\tiny\yng(2,1)} }{ p_{ \tiny\yng(1,1) } } \: + \:
\frac{ p_{\tiny\yng(3,2)} }{ p_{\tiny\yng(2,2) } } \: + \:
q \frac{ p_{ \tiny\yng(2) } }{ p_{ \tiny\yng(3,3) } } \: + \:
\frac{ p_{ \tiny\yng(3,1) } }{ p_{\tiny\yng(3) } } \: + \:
\frac{ p_{ \tiny\yng(1) } }{ p_{ \emptyset } }.
\end{eqnarray*}
where the $p$'s are Pl\"ucker coordinates associated to Young tableaux,
as above, and the open subset of $G(3,5)$ is defined by excluding the points
where $W$ has poles.

Now, to understand this mirror, we need to understand the
intersection of $G(3,5)$ with its image in
\begin{displaymath}
{\mathbb P}^9 - \{ p_{\emptyset}=0\} - \{ p_{\tiny\yng(1,1)}=0\} -
\{ p_{\tiny\yng(2,2)}=0 \} - \{ p_{\tiny\yng(3,3)}=0 \} -
\{ p_{\tiny\yng(3)}=0 \}.
\end{displaymath}
We can study this space as follows\footnote{We would like to thank Z.~Lu for
explaining this to us.}.
Let the points of $G(2,5)$ be represented by the matrix
\begin{equation}  \label{eq:matrix}
\left[ \begin{array}{ccccc}
a_1 & a_2 & a_3 & a_4 & a_5 \\
b_1 & b_2 & b_3 & b_4 & b_5
\end{array} \right].
\end{equation}
Consider five of the homogeneous coordinates on
${\mathbb P}^9$ (the image of the
Pl\"ucker embedding), corresponding to the minors of the following
columns:
\begin{displaymath}
(1,2), \: \: \:
(1,3), \: \: \:
(1,4), \: \: \:
(1,5), \: \: \:
(2,3).
\end{displaymath}
(This is not the case we are interested in, but will be instructive to
consider.)  Restricting to the locus on which the first minor is nonzero
puts~(\ref{eq:matrix}) in the form
\begin{displaymath}
\left[ \begin{array}{ccccc}
1 & 0 & a_3 & a_4 & a_5 \\
0 & 1 & b_3 & b_4 & b_5 \end{array} \right],
\end{displaymath}
(after a possible $U(2)$ rotation)
parametrizing ${\mathbb C}^6$.  Then, the remaining minors determine the open sets $b_3 \neq 0$,
$b_4 \neq 0$, $b_5 \neq 0$, and
$a_3 \neq 0$.  Thus, in this case, the intersection of
$G(3,5)$ with the open locus in ${\mathbb P}^9$ is
$({\mathbb C}^{\times})^2 \times {\mathbb C}^2$.

However, the case we are interested in is different.  Specifically,
we are interested in the case that the $x_i$ correspond to minors of the
following columns:
\begin{displaymath}
(1,2), \: \: \:
(2,3), \: \: \:
(3,4), \: \: \:
(4,5), \: \: \:
(5,1).
\end{displaymath}
Restricting to the locus on which the first minor is nonzero
gives the same result as before,
but now requiring the other minors to be nonzero implies
\begin{displaymath}
a_3 \neq 0, \: \: \: 
 a_3 b_4 - a_4 b_3 \neq 0, \: \: \:
 a_4 b_5 - a_5 b_4 \neq 0, \: \: \:
 b_5 \neq 0.
\end{displaymath}
This describes the complement of two quadrics in
${\mathbb C}^4 \times ({\mathbb C}^{\times})^2$.
It is not clear whether this can be further simplified.

Now, to compute correlation functions, we need to compute the critical
locus of the superpotential and the Hessian at those critical loci.
To do this, we need to identify the `fundamental' fields over which the
path integral sums, in order to take derivatives of the superpotential
with respect to those fields.

To that end, let us identify the $p$'s as follows:
\begin{eqnarray*}
x_1 & \equiv & p_{\tiny\yng(1)} = p_{ \hat{\mu}_5 } = b_3, \\
x_2 & \equiv & p_{\tiny\yng(2,1)} = p_{ \hat{\mu}_1 } = -a_4, \\
x_3 & \equiv & p_{\tiny\yng(2)} = p_{\hat{\mu}_3} = b_4, \\
x_4 & \equiv & p_{\tiny\yng(3,1)} = p_{ \hat{\mu}_4 } = -a_5, \\
x_5 & \equiv & p_{\tiny\yng(1,1)} = p_{ \mu_1 } = -a_3, \\
x_6 & \equiv & p_{\tiny\yng(3)} = p_{ \mu_4 } = b_5,
\end{eqnarray*}
so that
\begin{eqnarray*}
p_{\tiny\yng(3,2)} & = & p_{\hat{\mu}_2 } = a_3 b_5 - a_5 b_3 =
- x_5 x_6 + x_4 x_1, \\
p_{\tiny\yng(2,2)} & = & p_{ \mu_2 } = a_3 b_4 - a_4 b_3 
=  - x_5 x_3 + x_2 x_1, \\
p_{\tiny\yng(3,3)} & = & p_{ \mu_3 } = a_4 b_5 - b_4 a_5 =
-x_2 x_6 + x_3 x_4.
\end{eqnarray*}
Then, in these variables, the mirror superpotential can be written
\begin{displaymath}
W \: = \: \frac{x_2}{x_5} + \frac{ x_4 x_1 - x_5 x_6 }{ x_1 x_2 - x_3 x_5 }
+ q \frac{x_3}{x_3 x_4 - x_2 x_6} +
\frac{ x_4 }{x_6} + x_1.
\end{displaymath}
It is straightforward to check that
the critical locus of the superpotential above is ten points,
exactly right to match A model results as the Euler characteristic is
\begin{displaymath}
\left( \begin{array}{c} 5 \\ 2 \end{array} \right) \: = \: 10.
\end{displaymath}

\section{Notes on mirrors to fields with R charges}
\label{app:r-charges}

In this section we will outline, at a formal level, how fields in the
A-twisted theory with nonzero R charges modify the mirror, as outlined
in the proposal of the previous section.

To understand this, let us first consider the action of T-duality
on a complex scalar field $\phi = \rho \exp(i \varphi)$,
closely following \cite{Hori:2000kt}[section 3.1].
If we take $\phi$ to have charge $Q$ under a gauged $U(1)$ with
gauge field $A_{\mu}$, and also couple it to a background (classical)
$U(1)_R$ gauge field $A_{\mu}^R$, giving it $U(1)_R$ charge $q$,
then the bosonic kinetic term would have the form\footnote{
The coupling we are describing is different from the coupling to a
$U(1)_R$ that occurs when one takes worldsheet curvature corrections
into account.  For example, to define a (2,2) rigidly supersymmetric
chiral multiplet on $S^2$ with a worldsheet superpotential,
one must also specify a Killing vector $X$ on the target space,
interpreted as the action of $U(1)_R$, under 
which the superpotential has charge $2$.  Curved-space supersymmetry
then requires that one add couplings such as, for example, 
\begin{displaymath}
- \frac{1}{4 r^2} g_{i \overline{\jmath}} X^i X^{\overline{\jmath}}
\: + \:
\frac{i}{4r^2} K_i X^i \: - \:
\frac{i}{4 r^2} K_{\overline{\imath}} X^{\overline{\imath}},
\end{displaymath}
where we have assumed a round metric on $S^2$, of radius $r$.
(See {\it e.g.} \cite{Jia:2013foa} and references therein.) 
In the expression above, we are doing something different -- we are not
adding curvature corrections, but rather we are coupling to a background
gauge field for $U(1)_R$.
}
\begin{equation}
L_{\varphi} \: = \: -\rho^2 \left( \partial_{\mu} \varphi + 
q A_{\mu}^R + Q A_{\mu} \right)^2.
\end{equation}
Following the same procedure as in \cite{Hori:2000kt}[section 3.1],
we then introduce a vector field $B_{\mu}$ and define the new
Lagrangian
\begin{equation}
L' \: = \: - \frac{1}{4 \rho^2} B_{\mu} B^{\mu} \: + \:
\epsilon^{\mu \nu} B_{\mu} \left( \partial_{\nu} \varphi + 
q A_{\nu}^R + Q A_{\mu} \right).
\end{equation}
Integrating out $B_{\mu}$ returns $L_{\varphi}$.  If instead
we integrate out $\varphi$, we get the constraint that $B_{\mu} = 
\partial_{\mu} \theta$ for some $\theta$ and the new Lagrangian
\begin{equation}
L_{\theta} \: = \: - \frac{1}{4 \rho^2} \partial_{\mu} \theta
\partial^{\mu} \theta \: - \: q \theta \epsilon^{\mu \nu} \frac{1}{2}
F^R_{\mu \nu} \: - \: Q \theta \epsilon^{\mu \nu} \frac{1}{2} 
F_{\mu \nu},
\end{equation}
where, since in the topological field theory,
$A_{\mu}^R$ equals the spin connection, so $F_{\mu \nu}^R$ should be
the worldsheet curvature.  For simplicity, here we will merely take
$F_{\mu \nu}^R$ to be $1/2 \epsilon_{\mu \nu}$, for simplicity.
Following \cite{Hori:2000kt}[section 3.1], in the supersymmetrization,
the bosonic term
\begin{displaymath}
-Q \theta \epsilon^{\mu \nu} \frac{1}{2} F_{\mu \nu}
\end{displaymath}
becomes a term $Q Y \Sigma$ in the mirror superpotential,
where $Y$ is the supersymmetrization of $\theta$ and $\Sigma$ is the
adjoint-valued superfield associated with the vector multiplet.
Here, by analogy, one would expect the supersymmetrization of the
term
\begin{displaymath}
- q \theta \epsilon^{\mu \nu} \frac{1}{2}
F^R_{\mu \nu}
\end{displaymath}
to become a term $(q/2) Y$ in the mirror superpotential (utilizing
our assumption about $F_{\mu \nu}^R$).

Following the pattern of \cite{Gomis:2012wy},
if we write the mirror B model partition function schematically in the
form\footnote{
To clarify, this is the form of the partition function for a theory of
twisted chiral superfields with (twisted)
superpotential on worldsheet $S^2$.  The path integral over nonzero modes
has been folded into a suppressed constant; shown is the integral over
bosonic zero modes and the contribution from the remaining bosonic
superpotential terms.  As discussed in {\it e.g.}
\cite{Jia:2013foa}, for a twisted chiral superfield, the only curvature
terms are of the form $i W/r - i \overline{W}/r$, for $W$ the superpotential
and $r$ the radius of $S^2$.  Restricting to constant bosonic modes,
\begin{displaymath}
\int_{S^2} d^2 z \, i \frac{W}{r} \: = \: \mbox{(area of $S^2$)} (i) \frac{W}{r}
\: = \: (4 \pi r^2) (i) \frac{W}{r} \: = \: 4 \pi i r W,
\end{displaymath}
hence the weighting shown.
}
\begin{equation}
Z \: = \: \int d^2 Y \exp(- 4 \pi i r W(Y) + 4 \pi i r \overline{W}(\overline{Y}))
\end{equation}
(where the integral over $Y$ is an ordinary integral, not a path integral),
then as described in \cite{Gomis:2012wy}[section 4.3], formally,
adding a term
\begin{equation}  \label{eq:r-charge-term}
\frac{1}{4 \pi i r} \frac{q}{2} Y
\end{equation}
to the superpotential is equivalent to modifying
the integral over bosonic zero modes:
\begin{eqnarray}
Z & = & \int d^2 Y e^{- (q/2) Y} e^{- (q/2) \overline{Y}} \exp(- 4 \pi i r W+ 
4 \pi i r \overline{W}), 
\\
& \propto & \int d^2 X \exp( - 4 \pi i r W(X) + 4 \pi i r \overline{W}(\overline{X}) ),
\end{eqnarray}
where $X \equiv \exp(- (q/2) Y )$.  In effect, so far as the B model is
concerned, adding the term~(\ref{eq:r-charge-term})
to the superpotential is consistent with changing
the fundamental field from $Y$ to $X = \exp(- (q/2) Y)$.
(The reader should note that this chain of reasoning is only possible
because the integral above is an ordinary integral, not a path
integral.  The same reasoning in a path integral would, on its face,
be more suspect, as for example the exponentials would become
nontrivial composite operators.  We use this argument here only because the
B model localizes onto constant maps.)

Later in this paper we will occasionally use this style of
argument to outline motivations for changing fundamental fields.

Next, let us turn to the dependence of the A-twisted theory on R charges.
If there is a superpotential, the R charge of the superpotential must
add up to two.  Other constraints also exist.  For example, we have
already observed that the A-twist requires R charges to be integral.
For another example,
it was observed in \cite{Hori:2013ika}[section 3.4] that positive-definiteness of
the potential in the Lagrangian density requires
every R-charge $r_i$ obey $0 \leq r_i \leq 2$.  

Let us assume we have multiple possible R-charges obeying the
constraints above.  (As a practical matter, this is only likely in an
untwisted theory, so our comments here can be read as formal 
observations on physical theories.)  Let us also assume that we have a nonzero
superpotential, describing a Calabi-Yau geometry,
and vanishing twisted masses.
In such cases, there is a formal argument that the closed\footnote{
In the open string A model, R charges appear in defining the matrix
factorization data, see {\it e.g.} \cite{Hori:2013ika}[section 3.4],
so this statement is less clear.  We restrict to
closed strings here.
} string A model
is independent of the (consistent) choices of R charges, at least
for fields appearing in the superpotential.
We will outline one argument below
in an example.
(The following argument
is only intended to be suggestive, not in any sense rigorous in detail.
We also emphasize that the claim is only for closed string theories, and the
argument given would only make sense for topological field theories.)

Consider the special case of the GLSM for the quintic hypersurface
in ${\mathbb P}^4$.
The A-twisted theory is a $U(1)$ gauge theory with matter
\begin{itemize}
\item five chiral superfields $\phi_i$ of charge $1$,
corresponding to homogeneous coordinates on ${\mathbb P}^4$,
\item one chiral superfield $p$ of charge $-5$,
\end{itemize}
and superpotential
\begin{equation}
W \: = \: p G(\phi),
\end{equation}
where $G(\phi)$ is a homogeneous polynomial of degree five,
corresponding to the quintic hypersurface.

Suppose each $\phi_i$ is assigned R-charge $r$,
and $p$, R-charge $2-5r$.  Setting aside questions of orbifolds,
we claim that the mirror is formally independent of $r$.

From \cite{Hori:2000kt,Aganagic:2004yh}, the mirror is a Landau-Ginzburg
orbifold with superpotential
\begin{equation}
W \: = \: \sigma \left( \sum_i Y_i - 5 Y_p + t \right)
\: + \: \sum_i \exp(- Y_i ) \: + \: \exp(-Y_p),
\end{equation}
and where the fundamental fields are
\begin{displaymath}
\exp\left( - (r/2) Y_i \right), \: \: \:
\exp\left( - ((2-5r)/2) Y_p \right).
\end{displaymath}

Now, proceeding as usual, we integrate out $\sigma$ to get a constraint
on $Y_p$:
\begin{equation}
Y_p \: = \: \frac{1}{5} \sum_i Y_i \: + \: t/5.
\end{equation}

Now, let us formally plug this into the integration over the
bosonic integrals over zero modes (meaning, constant maps, since we
are in the B model):
\begin{eqnarray}
\lefteqn{
\int \left( \prod_i d^2 \exp( - (r/2) Y_i ) \right)
d^2 \exp( - ( (2-5r)/2 ) Y_p ) \,
\delta\left( Y_p - (1/5) \sum_i Y_i + t/5 \right)
} 
\nonumber \\
 & \hspace*{1in} \propto &
\int \left( \prod_i d^2 Y_i \exp( - (r/2) Y_i ) \right)
d^2 Y_p \exp( - ( (2-5r)/2 ) Y_p ) \,
\nonumber \\
& & \hspace*{1in}
\delta\left( Y_p - (1/5) \sum_i Y_i + t/5 \right),
\\
& \hspace*{1in} \propto &
\int \left(  \prod_i d Y_i e^{- (r/2) Y_i } \right)
\exp\left(  - \frac{2-5r}{2} \left(  \frac{1}{5} \sum_i Y_i \: + \: t/5
\right)
\right)
 \\
& \hspace*{1in} \propto & 
\int \prod_i d Y_i \exp\left( - \frac{5r + 2-5r}{10} Y_i \right)
\\
& \hspace*{1in} = &
\int \prod_i d Y_i \exp\left( - \frac{1}{5} Y_i \right)
\\
& \hspace*{1in} \propto &
\int \prod_i d \exp\left( - \frac{1}{5} Y_i \right),
\end{eqnarray}
which hints that the fundamental field is $x_i \equiv
\exp(- Y_i/5 )$.  (In principle, there would also be a ${\mathbb Z}_5$
orbifold associated with each $x_i$, but as our discussion here
is meant to be a formal outline, not a rigorous argument, we shall
largely gloss over that detail.)
Now, to be convincing, one would need to repeat this discussion in
full path integrals, not just the zero mode piece, but as the B model
localizes on constant maps, the hints above are suggestive.

In the new variables, the Landau-Ginzburg orbifold has superpotential
\begin{equation}
W \: = \: \sum_i x_i^5 \: + \: q^{-1/5} x_1 x_2 x_3 x_4 x_5,
\end{equation}
independent of $r$, as advertised.  In addition,
this is the same result for the Hori-Vafa quintic
mirror as obtained in \cite{Aganagic:2004yh}.
In passing, in computations of Gromov-Witten invariants in
\cite{Jockers:2012dk}, results were independent of R-charges chosen
(within a fixed range).

\end{document}